\documentclass[journal,10pt]{IEEEtran}
\IEEEoverridecommandlockouts
\usepackage{cite}
\usepackage{longtable}
\usepackage{amsmath,amssymb,amsfonts}
\usepackage{mathtools} 

\usepackage{graphicx}
\usepackage{textcomp}
\usepackage{xcolor}
\usepackage{epstopdf}
\usepackage{algorithm,algpseudocode}
\usepackage{url}
\usepackage[normalem]{ulem}
\usepackage{hyperref}
\usepackage{multirow}
\usepackage{longtable}
\usepackage{rotating} 
\usepackage{booktabs}
\usepackage{multirow}
\usepackage{siunitx}
\usepackage{booktabs}
\usepackage{pifont}
\newcommand{\cmark}{\ding{51}}%
\newcommand{\xmark}{\ding{55}}%
\usepackage{ulem} 
\usepackage{tikz}
\usepackage{adjustbox}
\usepackage{bbding}
\usepackage{cancel}
\usepackage{array, rotating}
\usepackage{multirow, makecell}
\usepackage{afterpage}
\usepackage{placeins}
\usepackage{longtable}
\usepackage{xtab}
\usepackage{tabularx}

\DeclareUnicodeCharacter{03BC}{\ensuremath{\mu}}

\newcommand{\notcheckmark}{{$\surd$}\textsuperscript{\textcolor{black}{\kern-0.35em{\bf--}}}}
\makeatletter
\newcommand*{\rom}[1]{\expandafter\@slowromancap\romannumeral #1@}
\makeatother

\newcolumntype{P}[1]{>{\centering\arraybackslash}p{#1}}

\def\BibTeX{{\rm B\kern-.05em{\sc i\kern-.025em b}\kern-.08em
		T\kern-.1667em\lower.7ex\hbox{E}\kern-.125emX}}

\DeclareMathAlphabet      {\mathbfit}{OML}{cmm}{b}{it}

\usepackage{tikz}
\usetikzlibrary{shapes.geometric}
\usepackage{scalerel}
\usetikzlibrary{svg.path}
\definecolor{orcidlogocol}{HTML}{A6CE39}
\tikzset{
  orcidlogo/.pic={
    \fill[orcidlogocol] svg{M256,128c0,70.7-57.3,128-128,128C57.3,256,0,198.7,0,128C0,57.3,57.3,0,128,0C198.7,0,256,57.3,256,128z};
    \fill[white] svg{M86.3,186.2H70.9V79.1h15.4v48.4V186.2z}
                 svg{M108.9,79.1h41.6c39.6,0,57,28.3,57,53.6c0,27.5-21.5,53.6-56.8,53.6h-41.8V79.1z M124.3,172.4h24.5c34.9,0,42.9-26.5,42.9-39.7c0-21.5-13.7-39.7-43.7-39.7h-23.7V172.4z}
                 svg{M88.7,56.8c0,5.5-4.5,10.1-10.1,10.1c-5.6,0-10.1-4.6-10.1-10.1c0-5.6,4.5-10.1,10.1-10.1C84.2,46.7,88.7,51.3,88.7,56.8z};
  }
}

\newcommand\orcidicon[1]{\href{https://orcid.org/#1}{\mbox{\scalerel*{
\begin{tikzpicture}[yscale=-1,transform shape]
\pic{orcidlogo};
\end{tikzpicture}
}{|}}}}

\begin{document}
	
	\title{\textbf{From Centralized RAN to Open RAN: A Survey on the Evolution of Distributed Antenna Systems}}

    \author{
		\IEEEauthorblockN{Mahmoud A. Hasabelnaby\textsuperscript{\orcidicon{0000-0003-1456-5734}}, \textit{Member, IEEE}, Mohanad Obeed\textsuperscript{\orcidicon{0000-0001-6774-255X}}$^*$, \textit{Member, IEEE}, Mohammed Saif\textsuperscript{\orcidicon{0000-0001-6633-0799}}$^*$, \textit{Member, IEEE}, Anas Chaaban\textsuperscript{\orcidicon{0000-0002-8713-5084}}, \textit{Senior Member, IEEE}, and M. J. Hossain\textsuperscript{\orcidicon{0000-0002-3377-7831}}, \textit{Senior Member, IEEE} } 
		
		\thanks {
           $^*$Equal contribution. Listing order is random.
           
           Mahmoud A. Hasabelnaby and Mohanad Obeed are with the Advanced Wireless Technology Lab, Huawei Technologies Canada Co. Ltd, ON K2K 3J1, Canada (e-mail: mahmoud.hasabelnaby@huawei.com, mhndbaz@gmail.com). This work was done while M. Hasabelnaby and M. Obeed were a PhD candidate and a postdoctoral fellow, respectively, at the School of Engineering, University of British Columbia, Canada.

			Mohammed Saif is with the Department of Electrical and Computer Engineering, University of Toronto, Toronto, ON M5S, Canada (e-mail: mohammed.saif@utoronto.ca). This work was done while M. Saif was a postdoctoral fellow at the School of Engineering, University of British Columbia, Canada.

			M. J. Hossain and Anas Chaaban are with the School of Engineering, the University of British Columbia, Kelowna, BC V1V 1V7, Canada
			(e-mail: anas.chaaban@ubc.ca, jahangir.hossain@ubc.ca).

		}
		
	}
	\maketitle

\begin{abstract}
Next-generation mobile networks require evolved radio access network (RAN) architectures to meet the demands of high capacity, massive connectivity, reduced costs, and energy efficiency, and to realize communication with ultra-low latency and ultra-high reliability. 
{Meeting such} requirements for both mobile users and vertical industries in the next decade {requires novel solutions. One of the potential solutions that attracted significant research attention in the past 15 years} is to redesign the radio access network (RAN). In this survey, we present a comprehensive survey on distributed antenna system (DAS) architectures that address these challenges and improve network performance. We cover the transition from traditional decentralized RAN to DAS, including cloud radio-access networks (C-RAN), fog radio-access networks (F-RAN), virtualized radio-access networks (V-RAN), cell-free massive multiple-input multiple-output (CF-mMIMO), and {the most recent advances manifested in} open radio-access network (O-RAN). In the process, we discuss the benefits and limitations of these architectures, including the impact of limited-capacity fronthaul links, various cooperative uplink and downlink coding strategies, cross-layer optimization, and techniques to optimize the performance of DAS. Moreover, we review key enabling technologies for next-generation RAN systems, such as multi-access edge computing, network function virtualization, software-defined networking, and network slicing; in addition to some crucial radio access technologies, such as millimeter wave, massive multi-input multi-output, device-to-device communication, and massive machine-type communication. Last but not least, we discuss the major research challenges in DAS and identify several possible directions for future research.
\end{abstract}




\section{Introduction}
\IEEEPARstart{A}{ccording} to the International Telecommunication Union Radio-Communication Sector (ITU-R), the rapid growth of smart Internet-of-Things (IoT) devices and associated use cases is poised to strain the existing fifth generation (5G) networks in the near future \cite{ITUI}. This aligns seamlessly with recent statistical projections by Ericsson emphasizing that the anticipated mobile traffic load for the year 2030 and beyond is expected to skyrocket to 5016 exabytes per month, underscoring the surging demand for data-centric services and applications \cite{Eric}. This surge in data consumption is accompanied by the substantial energy consumption footprint of mobile networks, globally estimated at approximately USD 25 billion \cite{Eric2}. This estimation is further compounded by contemporary global economic challenges, marked by burgeoning inflation and a mounting energy crisis.  Consequently, a comprehensive response, involving both the evolution of communication protocols and technologies as well as a paradigm shift in the design and configuration of existing radio access networks (RANs) is needed.

Compared to 5G networks, the transition to 6G networks holds immense promise for enabling unprecedented levels of connectivity, capacity, and intelligence. With data-intensive applications such as augmented reality, virtual reality, IoT, and autonomous systems on the horizon, 6G must enable harnessing the power of artificial intelligence for efficient network management, intelligent resource allocation, and enhanced user experiences, while reducing total operational costs and energy consumption. Realizing these requirements in 6G  requires supporting stringent Key Performance Indicators (KPIs) including a connection density of $10^7$ users per square kilometer with a 1 Gbps downlink data rate, coupled with network reliability exceeding $99.99999\%$ and communication latency as low as $0.1$ msec \cite{8826541}. 

Consequently, mobile operators are currently facing increasing pressure to enhance their network performance, prompting them to explore novel solutions in order to keep up with the requirements of 6G. This typically involves either adding new network equipment/modules or adopting more efficient, flexible, and scalable alternatives. Quantitatively, the RAN represents a substantial portion (approximately 65\%-70\%) of total network capital expenditures, according to Ericsson \cite{Eric}. To effectively address the 6G requirements, a revolutionized synergistic approach is required, wherein RAN architectural innovations align harmoniously with robust energy-efficient strategies and sophisticated quality-of-service (QoS) provisions. In light of this transformative imperative for the future, next-generation mobile networks necessitate a meticulous design approach to handle limited radio resources, optimize network parameters, and manage the increasing demands of mobile data traffic. As we move towards 6G, these challenges become more complex, demanding unique and innovative solutions for future network paradigms.
These challenges have driven researchers to propose advanced RAN architectures with efficient and powerful processing techniques that can dramatically improve network performance.

 \begin{figure}[t!]
		\begin{center}
			\includegraphics[width=0.47\textwidth]{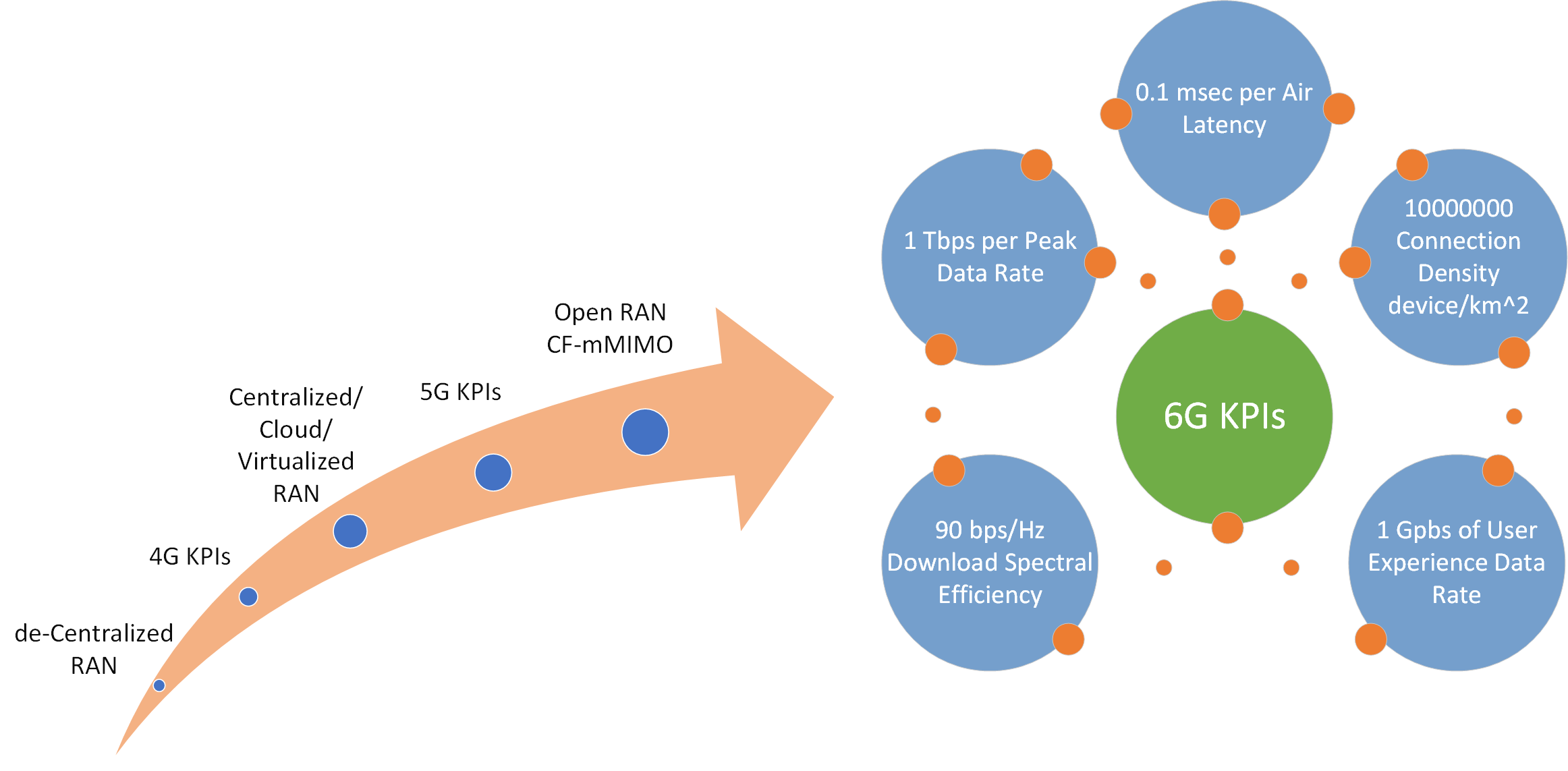}
			\caption{{Evolution of mobile generations: A comparative evaluation from 4G KPIs to 6G KPIs.}}
			\label{6GKPI}
		\end{center}
\end{figure}

	Understanding the RAN architecture could help in determining the most appropriate evolution to keep up with the requirements of 6G and beyond. The landscape of mobile communication has witnessed a remarkable evolution over the years, with each generation bringing forth new demands and technological challenges. One pivotal aspect of this evolution is the continuous transformation of the RAN architecture to meet the diverse KPIs associated with different mobile generations as shown in Fig.~\ref{6GKPI}. 

In the early 2000s, de-Centralized RAN was deployed as the nascent phase of mobile networks and was synonymous with the 2G and early 3G architectures. In de-Centralized RAN, base stations (BSs) are distributed across the service area, each serving a specific cell. This architecture sufficed for the voice-centric communication predominant during these early stages. The primary KPIs were coverage, capacity, and spectral efficiency. However, as mobile networks transitioned to 4G and later 5G, the limitations of de-Centralized RAN became evident. The exponential growth of data-driven applications demanded higher data rates, lower latency, and increased reliability. de-Centralized RAN, with its cell-based approach, struggled to provide the required capacity and spectral efficiency. This triggered the need for a more advanced architecture, leading to the emergence of distributed antenna systems (DASs) \cite{6824752}. 

The advent of 4G networks ushered in Centralized RAN, a transformative architecture that revolutionized the way RANs were structured. C-RAN centralized baseband processing in a central unit (CU), while remote radio heads (RRHs) were distributed across the coverage area. This DAS architectural shift enabled more efficient resource management, improved coordination between cells, and greater spectral efficiency. C-RAN was adept at meeting KPIs associated with 4G and early 5G, including higher data rates, lower latency, and improved reliability. The transition to C-RAN also brought about significant advantages in terms of energy efficiency and network optimization \cite{7858136}. However, as mobile networks continued to evolve, driven by emerging applications like augmented reality, virtual reality, and autonomous systems, C-RAN struggled to meet the escalating demands for ultra-high data rates, extremely low latency, and robust connectivity in high-density areas. This led to the exploration of alternative architectures, culminating in the emergence of Cell-Free Massive multiple-input multiple-output (CF-mMIMO), characterized by the deployment of a multitude of access points (APs) across the service area, effectively eliminating cell boundaries \cite{R5,R7}. Leveraging massive MIMO techniques, this architecture achieved unprecedented spatial multiplexing and interference management. These capabilities made CF-mMIMO well-suited for advanced 5G applications, providing a foundation for immersive experiences and reliable IoT connectivity. The transition from C-RAN to CF-mMIMO underscored the importance of architectural innovation in accommodating evolving mobile generation requirements \cite{R13}. 

Another viable solution to reduce RAN costs is to exploit network functions virtualization (NFV) and leverage software-based solutions and cloud-native technologies, supported by a virtualized/cloud RAN (V-RAN) architecture to optimize network resources and streamline operations. V-RAN was introduced as a pivotal step toward software-defined networking (SDN) and NFV. This architecture virtualized both baseband processing and radio functions, making network management highly agile and cost-effective \cite{8642354}. It enabled the deployment of network services as software modules, enhancing adaptability to evolving service requirements.  V-RAN excelled in addressing KPIs associated with resource utilization, service agility, and total cost of ownership. The transition from C-RAN to V-RAN marked a critical juncture in the evolution of RAN architectures. It responded to the growing demand for more flexible and scalable networks, capable of rapidly adapting to changing service needs. Virtualization technologies allowed for dynamic resource allocation, efficient scaling of network functions, and the introduction of new services without the need for extensive hardware upgrades \cite{9075181}. 

In order to address the demands of low-latency and high-reliability applications, Fog RAN (F-RAN) has gained prominence in recent years as an evolution of traditional RAN designs \cite{7555389,7558153}. Unlike the C-RAN and V-RAN, F-RAN takes a partially decentralized approach,  pushing a significant portion of network processing functions closer to the edge of the network. In F-RAN, computing resources and network functions are strategically distributed across the network, typically in proximity to BSs or APs, where processing occurs closer to where data is generated or consumed. This architecture facilitates lower-latency and high-reliability processing of data and applications, rendering it highly suitable for emerging use cases like ultra-reliable low-latency communication (URLLC). Applications benefiting from this approach include autonomous vehicles, industrial automation, and augmented reality, especially in the context of 5G and beyond. This approach also optimizes resource utilization, reduces network congestion, and enhances the overall QoS. F-RAN is also aligned with the growing trend of multi-access edge computing, which leverages edge resources for various services \cite{7537176, 9861780, 9824969}. This approach aligns with the evolving requirements of modern mobile generations, making F-RAN a significant player in the future of mobile communication.

As the landscape of mobile communication continues to evolve, the need for openness, interoperability, and innovation in RAN architecture becomes increasingly pronounced. Enter Open RAN (O-RAN), a paradigm-shifting approach that embraces the principles of virtualization, standardization, and openness to unleash the full potential of mobile networks. O-RAN redefines the traditional vendor-specific, monolithic RAN equipment by disaggregating network components into interoperable, open interfaces \cite{10024837}. It empowers network operators to mix and match components from various vendors, fostering competition and driving innovation. This approach enhances network flexibility, scalability, and cost-effectiveness while accommodating the diverse KPIs of modern mobile communication. One of the primary drivers for the adoption of O-RAN is its ability to meet the requirements of advanced 5G and upcoming 6G networks. With the proliferation of diverse applications, from URLLC for mission-critical services to massive machine-type communication (mMTC) for IoT, O-RAN provides the agility to tailor network resources according to specific needs. Furthermore, O-RAN addresses the challenges posed by the growing complexity of RAN management and the need for rapid deployment of new services. By fostering an open ecosystem of vendors and developers, it encourages the creation of innovative solutions that can be seamlessly integrated into the network \cite{10071958}.

It is crystal clear that the evolution of RAN architectures is crucial to ensure that mobile networks meet the evolving demands of different mobile generations. Each architecture represents a step forward in addressing specific challenges and optimizing performance, ultimately shaping the future of wireless communication. Researchers and network operators must continue to assess and innovate in this field to meet the diverse needs of modern mobile communication. As we delve into this survey in the subsequent sections, we will explore promising technologies and techniques contributing to the realization of advanced DAS architectures. This includes the implementation of flexible functional splits between a central processing unit (CPU) and a radio unit (RU), utilization of advanced coding strategies for both uplink and downlink transmissions, cross-layer optimization techniques, and the application of sophisticated resource allocation algorithms and hybrid architectures. Through these advancements, the groundwork is laid for fulfilling the needs of next-generation mobile networks.

 \section{Survey Scope and Organization}
In this section, we provide a clear roadmap of the survey's focus, encompassing key aspects such as survey scope and performance metrics employed to evaluate performance, a thorough review of related works, the primary contributions offered by our survey, and an overview of the survey's organizational structure.

    \subsection{Scope and Performance Metrics}
    The aim of this survey is to serve as a reference point for readers interested in DAS to help them in navigating the state-of-the-art in this area. This survey discusses the capabilities of DAS architectures, including C-RANs, CF-MIMO, V-RANs, F-RANs, and O-RANs. {It reviews} these architectures, and provides an overview of the challenges, methods, solutions, and opportunities of DAS in future mobile networks. Additionally, it discusses how DAS benefits from {complemetary} technologies such as virtualization, cloud-native networking, device-to-device (D2D) communications, reconfigurable intelligent surfaces {(RISs)}, and non-terrestrial networks. Moreover, it presents future research directions and challenges associated with the deployment of DAS architectures, {highlighting potential developments in} DAS systems and architectures.
	
		\begin{figure}[t!]
		\begin{center}
			\includegraphics[scale=0.31]{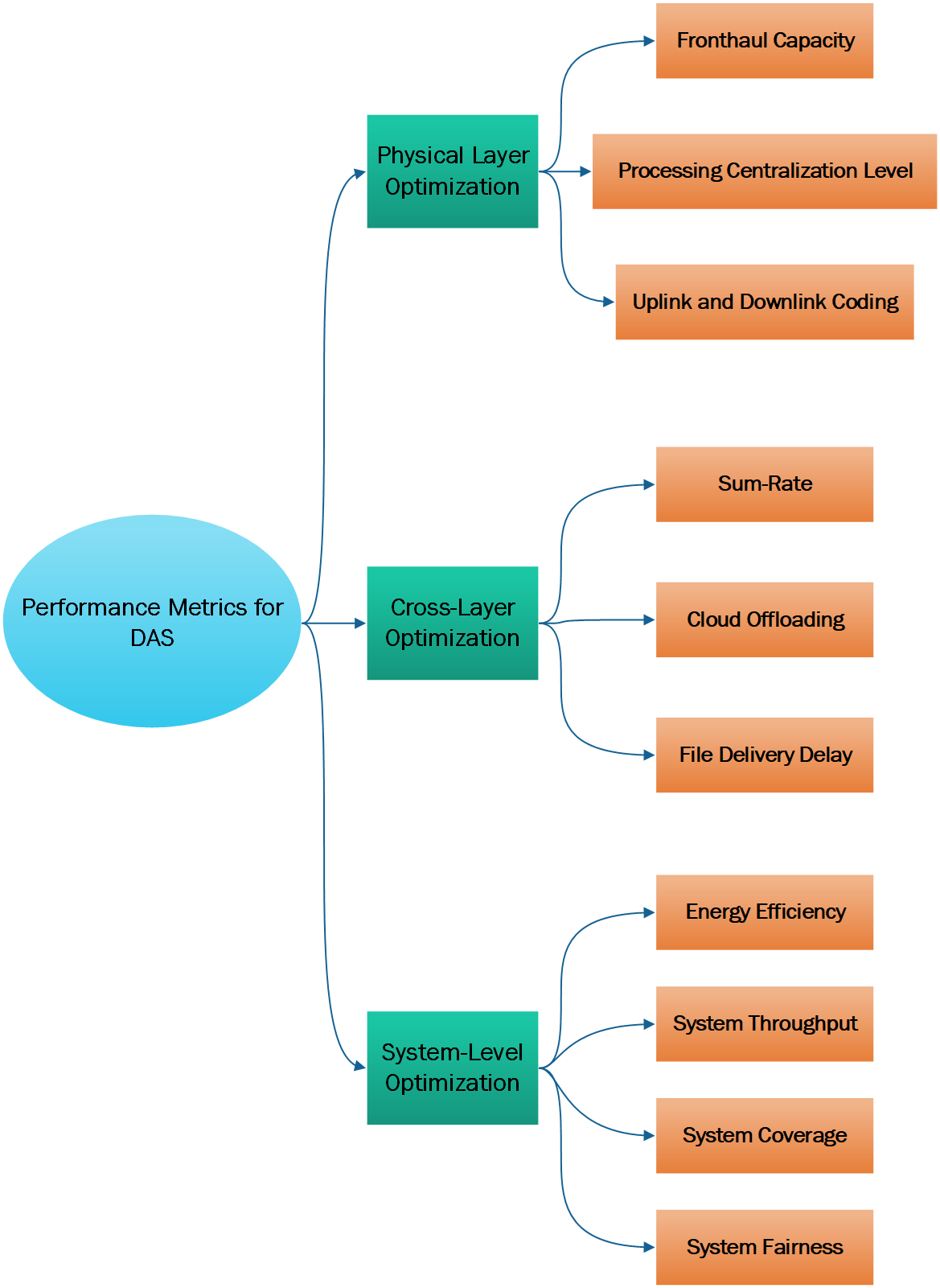}
			\caption{Performance metrics for DAS covered in this survey paper. }
			\label{PM}
		\end{center}
	\end{figure}
	Particularly, we discuss the performance metrics of DASs shown in Fig.~\ref{PM} from various perspectives. At first, we highlight {the following} three {physical layer processing aspects that are important to consider in} the design of DASs. 
	\begin{itemize}
		\item Fronthaul Capacity: We start by investigating the {capabilities/}limitations of conventional fronthaul interfaces, including the common public radio interface (CPRI) \cite{CPRI1}, open base station standard architecture initiative (OBSAI) \cite{tt}, and open radio interface (ORI) \cite{ORI}, to keep up with the requirements of next-generation RAN. Then, we {review the research efforts} that propose new techniques to lower the extremely high capacity demands of fronthaul networks, while {preserving many benefits of the centralized processing in DASs}.
		
		\item Processing Centralization Level: The selection of the appropriate functional split  (i.e., the centralization level) in the design of DAS architectures is a challenging task. To {select} an appropriate split, different criteria (e.g., inter-cell interference mitigation, traffic demand, latency requirements, etc.) need to be considered.  
		We discuss {different levels of functional split} between RUs and the CPU and {their impact on fronthaul traffic} compared to the conventional fully-centralized approach \cite{8384342}. In particular, we {elaborate} the benefits and drawbacks {of different} split options in four functional split standards, including third generation partnership project (3GPP) \cite{3GPP1}, enhanced CPRI (eCPRI) \cite{eCPRI1},  the small cell forum (SCF) \cite{SCF}, and the O-RAN Alliance \cite{ORAN-FH}. Furthermore, we {explain} how exploiting emerging cloud-based technologies (such as virtualization, SDN, and cloud-native wireless networking) {aids in} converting the DAS architecture to be RAN-as-a-service (RANaaS) through flexible functional splits that adapt to the continuously changing network requirements.

		\item Uplink and Downlink Coding:
		From an information-theoretic perspective, the DAS model is best understood as a relay network. Using different relaying strategies,  the RUs relay useful information {from the users} to the CPU. We illustrate how the design of DAS physical and data link layers can adapt to the capacity and latency limitations of the fronthaul links {through various} uplink and downlink transmission strategies. In particular, we review uplink relaying schemes that relay information {from users to the CPU through the RUs}, such as decode-and-forward (DF) \cite{quek_peng_simeone_yu_2017}, amplify-and-forward (AF), compress-and-forward (CF) \cite{6342931}, compute-and-forward (CoF) \cite{6522158}, and noisy network coding \cite{5752460}. In the downlink, we investigate data-sharing techniques \cite{6920005}, compression-based strategies, \cite{6588350}, reverse compute-and-forward (RCoF) \cite{6283033}, and reverse quantized-compute-and-forward (RQCoF) \cite{6522158}.
	\end{itemize}

	Since distributing the antenna terminals in the service area contributes significantly to improving the network's energy efficiency, achievable rates, system coverage, and fairness, we discuss how DASs play a significant role in improving them as follows.    
	\begin{itemize}
	\item Energy Efficiency: 
         In general,  energy efficiency is considered a crucial metric that balances the need for high-performance data transmission and the low power consumption. It is essential for the sustainability of wireless networks, especially with the growing demand for mobile data and the expansion of the IoTs. It is common in the literature that the energy efficiency metric is formulated as the total data transmitted over the entire network relative to the total energy consumed by the network. Considering only the transmission power, massive MIMO systems are considered energy-efficient due to the high array gain that improves the system throughput without consuming more transmit power \cite{R21}. This property can be attained in massive MIMO systems whether the antennas are co-located or distributed \cite{R20}. However, in the distributed case (DAS), the power consumed in fronthaul links in DASs reduces the energy efficiency. On the other hand, bringing the antennas closer to the users in DASs reduces the transmit power required to achieve the required QoS. Therefore, the main factor that affects the energy efficiency in DASs is the power consumption in the fronthaul links, which becomes a main concern if the number of antennas is large. In other words,  distributing the antennas close to the users improves the energy efficiency of the radio access link, but degrades the overall energy efficiency due to the fronthaul links' power consumption\cite{R32, R33}.

	\item System Throughput: The achievable rates in DASs are significantly improved due to several factors. First, since the antennas are physically distributed and connected to a CPU through  \text{fronthaul} links, the  \text{CPU} can process the received/transmitted signals  \text{of the antennas jointly} to completely eliminate or mitigate the interference, which leads to a significant increase in the achievable rates at the users. Second, the channels between the antennas and the users are more likely to be of high quality due to the small distances between them. In addition, increasing the number of distributed antennas improves the macro-diversity that helps in reducing the outage probability \cite{8972478, R3, R45}. 
	
	\item Coverage Probability: Distributing the antennas over a wide area also guarantees to improve the coverage  \text{and reduces blind spots that are caused by blockages}. This improves the coverage probability significantly since if a user has no line-of-sight to some APs, it would have line-of-sight to others \cite{8972478}. 
	
	\item System Fairness: DASs also improve system fairness by generating a uniform deliverable QoS for all the users in the considered area. In particular, the cooperation between APs removes the cell boundaries and hence avoids having users at the edge  {of a cell where they would} suffer high interference and weak channel conditions \cite{R6}.   
	\end{itemize}

{In addition to physical layer aspects, we also discuss} cross-layer optimization {aspects, focusing on the following} three important performance metrics. 
	\begin{itemize}
		\item {Sum Rate:} This measures the number of correctly received bits by the users per unit time. Maximizing the sum rate in DAS requires careful optimization {of} user scheduling and power allocation, which was shown to be a challenging task \cite{CLNC1, 4068001, 4769397, 5464889}. {Due to its combinatorial and non-convex nature, a large body of literature approached this optimization using iterative methods  \cite{4068001,  4769397,5464889,6525475,5062043}}. {From a} cross-layer optimization {perspective}, such as {optimization also includes optimizing} network coding (NC) to combine users' files for better cross-layer system performance \cite{CLNC1, 6662474, CLNC2, CLNC3, 6692154, CLNC4, 5462127, CLNC5, CLNC6}.

		\item {Cloud Offloading:} Cloud offloading is an essential cross-layer performance metric that has to be optimized to compensate for the limited capacity of C-RAN's fronthaul links \cite{CLNC3, 8288205, CLNC4}. To offload the cloud's tasks to different nodes in the DAS, popular files can be cached on the network edge (i.e., close to the end users) such as at enhanced RUs or at the users. {This enables the delivery of} such files to request users with little to no intervention from the cloud. As a result. {This caching needs to be optimized, in order to efficiently offload traffic from the fronthaul network}.

		\item {File Delivery Delay:} Delivering files to a set of users with a minimum possible delay was widely considered in the literature, e.g., \cite{NC1, NC2, NC3, NC4, 7845689, 8794557, 9512273, 9180294, 8267199, 8444481, 6882208, 1176612, 4557282, 1705002, 4675714}.  The file delivery delay metric is {most important in applications that require} immediate delivery of data for real-time applications, {such as} live video streaming. In the literature, this delay metric is defined, based on the application, as the number of transmissions \cite{7845689, 8794557, 7511238, 9512273, 9180294, 8267199, 8444481, 6882208, 1176612, 4557282, 1705002, 4567144, 4675714} or the latency in seconds \cite{RAIDNC1, 4567144, RAIDNC2, 6662474, RAIDNC3, RAIDNC4, 6692154, RAIDNC5}.  \text{Optimizing this delay requires} scheduling of the file delivery problem using cross-layer optimization. This enhances the efficiency and responsiveness of data delivery, especially in time-sensitive applications, by devising strategies and algorithms that minimize file delivery delay.

	\end{itemize}  
 
Next, we comprehensively survey and compare the state-of-the-art works on various DAS architectures and key enabling technologies for next-generation RANs, encompassing the aforementioned performance metrics.

  \begin{table*}[t!]
		\caption{Summary of Survey Papers about DAS Architectures; \cmark, \bcancel{\cmark}, or \xmark~ indicates that the topic is addressed, partially addressed, or not addressed, respectively.} 
		\centering
  	\begin{adjustbox}{width=\textwidth}
					\begin{tabular}{|c||c|c|c|c||c|c|c|c||c|c|c|c||c|c|c|c|}
			\hline
			\multirow{2}{*}{References} & \multicolumn{4}{c|}{Fronthaul Network Design} & \multicolumn{4}{c|}{Physical-Layer Coding} &  \multicolumn{4}{c|}{Cross-Layer Optimization} &  \multicolumn{4}{c|}{System-Level Optimization} \\ 
			\cline{2-17}
			& C-RAN & F-RAN & CF-MIMO & O-RAN & C-RAN & F-RAN & CF-MIMO & O-RAN & C-RAN & F-RAN & CF-MIMO & O-RAN & C-RAN & F-RAN & CF-MIMO & O-RAN \\
			\hline
			\cite{8113473} &  \bcancel{\cmark} &  \xmark & \xmark & \xmark & \xmark & \xmark & \xmark & \xmark & \xmark & \xmark & \xmark & \xmark & \xmark & \xmark & \xmark & \xmark   \\
			\hline 
			\cite{8479363} &  \cmark &  \bcancel{\cmark} & \bcancel{\cmark} & \bcancel{\cmark} & \xmark & \xmark & \xmark & \xmark & \xmark &\xmark  & \xmark & \xmark & \xmark & \xmark & \xmark & \xmark  \\
			\hline 
			\cite{9650567} &  \xmark &  \xmark & {\cmark} & {\xmark} & \xmark & \xmark & \cmark & \xmark & \xmark & \xmark & \xmark & \xmark  & \xmark & \xmark & \cmark  & \xmark   \\
			\hline 
			\cite{chen2021survey} &  \xmark &  \xmark & {\xmark} & {\xmark} & \xmark & \xmark & \cmark & \xmark & \xmark & \xmark &\xmark  & \xmark & \xmark & \xmark & \cmark & \xmark  \\
			\hline 
			\cite{elhoushy2021cell} &  \xmark &  \xmark & \cmark & {\xmark} & \xmark & \xmark & \cmark & \xmark & \xmark & \xmark & \xmark  &\xmark  & \xmark  & \xmark & \cmark & \xmark  \\
			\hline
			\cite{bassoy2017coordinated} &  \xmark &  \xmark & {\xmark} & {\xmark} & \xmark & \xmark & \cmark & \xmark & \xmark  & \xmark &\xmark  & \xmark &\xmark  &  \xmark& \cmark & \xmark  \\
			\hline 
			\cite{7845689} &  \bcancel{\cmark} &  \bcancel{\cmark} & \xmark & \xmark  &  \xmark &  \xmark & \xmark  &  \xmark &  \xmark &  \xmark &  \xmark & \xmark &  \xmark&  \xmark &  \xmark &   \xmark\\
			\hline 
			\cite{8327582} &  \bcancel{\cmark} &   \bcancel{\cmark} &  \xmark &  \xmark &  \xmark & \xmark  & \xmark &  \xmark &  \xmark &  \xmark &  \xmark &  \xmark &  \cmark &  \cmark & \xmark &   \xmark \\
			\hline 
			\cite{8367785} &  \bcancel{\cmark}  &   \bcancel{\cmark} &  \xmark &  \xmark &  \xmark &  \xmark &  \xmark &  \xmark &  \xmark &  \xmark & \xmark  &  \xmark &  \cmark &  \cmark &  \xmark &   \xmark \\
			\hline 
			\cite{Fogcomputing} &   \bcancel{\cmark} &   \cmark &  \xmark & \xmark &  \bcancel{\cmark} &  \cmark &  \xmark & \xmark & \xmark &  \xmark &  \xmark &  \xmark &   \bcancel{\cmark} &   \cmark &  \xmark &   \xmark \\
			\hline

			This survey &  \cmark &  \cmark & \cmark & \cmark & \cmark & \cmark & \cmark & \cmark & \cmark  & \cmark & \cmark  & \cmark  & \cmark & \cmark & \cmark &  \cmark \\
			\hline 
			\hline
		\end{tabular}
  \end{adjustbox}
  \label{rf}
	\end{table*}

	\subsection{Related Work and Existing Surveys}
	\label{RW}
	Several existing related survey papers investigated the design, challenges, and opportunities of different DAS architectures separately. They also reviewed different DAS architectures from a single layer perspective. We summarize these papers in Table~\ref{rf}.

 In particular, the physical layer design of fronthaul networks was explored in \cite{8113473}, which specifically addressed the challenges of optical fronthaul over the CPRI protocol in the C-RAN architecture. In \cite{8479363}, the authors concentrated on the specifications and applications of different functional split options proposed by 3GPP, offering a brief overview of other working groups and illustrating examples within a long-term evolution (LTE) network in a C-RAN context. Additionally, in \cite{9650567}, as part of its review, the authors delved into minimizing fronthaul signaling by employing source coding strategies in the context of uplink CF-mMIMO.

	In \cite{9650567, chen2021survey}, the authors focused on user-centric CF-mMIMO networks. Specifically, the authors of \cite{9650567} reviewed resource allocation schemes, channel estimation techniques, scalability issues, cell-formation methods, and fronthaul issues. The authors of \cite{chen2021survey} focused more on the technical foundations and the signal processing operations at the transmitters and receivers of CF-MIMO networks.  The authors of \cite{elhoushy2021cell} reviewed channel modeling, precoding, and detection techniques for CF-mMIMO networks, effects of practical  {limitations} on their performance, and their integration with emerging 5G technologies. 
    The authors of \cite{bassoy2017coordinated}  studied different coordinated multi-point (CoMP) clustering schemes, and how these schemes affect spectral and energy efficiencies. 
	
	From a cross-layer optimization perspective, several surveys covered different aspects of file delivery and resource allocation in DAS. For example, the authors of
	\cite{7845689} discussed network layer file delivery in C-RAN, F-RAN, and D2D communications. In \cite{8367785, 8327582}, the authors presented several file delivery and caching strategies for cellular networks, including macro-cellular networks, heterogeneous networks, D2D networks, C-RANs, and F-RANs. Both \cite{8367785, 8327582} considered  different network-layer performance metrics, including throughput, backhaul cost,  and network delay. The authors in \cite{Fogcomputing} provided an  {overview of how fog} computing strategies support delay-sensitive file requests from end users with reduced energy consumption and low traffic. Moreover, \cite{Fogcomputing} summarized physical-layer resource allocation approaches  {focusing on their} latency, bandwidth, and energy consumption in fog computing. While \cite{7845689, 8367785, 8327582} considered network layer techniques to improve system performance for file delivery and caching, \cite{Fogcomputing} considered physical layer metrics in fog computing. Compared to the aforementioned surveys,  this survey provides an extensive overview of state-of-the-art cross-layer optimizations that involve the joint implementation of physical and network layers factors for DASs.

	\subsection{Contributions}
		In contrast with the surveys introduced in
subsection \ref{RW}, this survey presents a more comprehensive literature review and comparative analysis of various DAS architectures. In particular, it  {reviews the history and evolution} of DAS systems, and provides an extensive discussion on key enabling technologies. We summarize our contributions in this survey paper as follows:
	\begin{itemize}
		\item We comprehensively review the evolution of DAS architectures from de-centralized RAN to C-RAN, V-RAN, F-RAN, and CF-MIMO to O-RAN. We also present the motivation for reconstructing and redesigning legacy DAS architectures, the state-of-the-art, the ongoing working groups, and standardization activities to satisfy the stringent requirements of next-generation RAN.
		
		\item We provide a literature review of the design of the fronthaul network demonstrating the limitations of conventional fronthaul interfaces to satisfy the staggeringly soaring capacity and latency demands. Additionally, we explore potential solutions to tackle this issue from various perspectives. We start by investigating the main four functional split standards that cover the evolution of DAS networks, including 3GPP \cite{3GPP1}, eCPRI \cite{eCPRI1},  SCF \cite{SCF}, and the O-RAN Alliance \cite{ORAN-FH}. We  {then} highlight how flexible functional split can satisfy the continuously changing network requirements. Moreover, we provide comparative analysis for various uplink and downlink coding strategies in DAS.

		\item We review  works that focused on optimizing  DAS networks via optimizing  beamforming matrices  the channel estimation operations to improve their performance in terms of achievable rates, energy efficiency, and fairness. In addition, we explore  recent works that integrated the DAS concept into emerging 6G networks such as non-terrestrial and RIS-assisted wireless networks.

        \item In addition to the physical layer perspective, we discuss DAS architectures from a cross-layer perspective through cross-layer NC (CLNC) that exploits NC decisions, rate adaptation, and power allocation.  {We also discuss} graph theory techniques  {that can be used} to tackle cross-layer optimization problems for file delivery and resource allocation in DAS. 
  
		\item Last but not least, we  {overview} existing challenges and future research directions related to different DAS architectures,  {with the aim of motivating} the academic and industry community towards realistic solutions with new technological advances.
	\end{itemize}

\subsection{Paper Structure}
\begin{figure}[!htbp] 
		\begin{center}
			\includegraphics[width=0.46\textwidth]{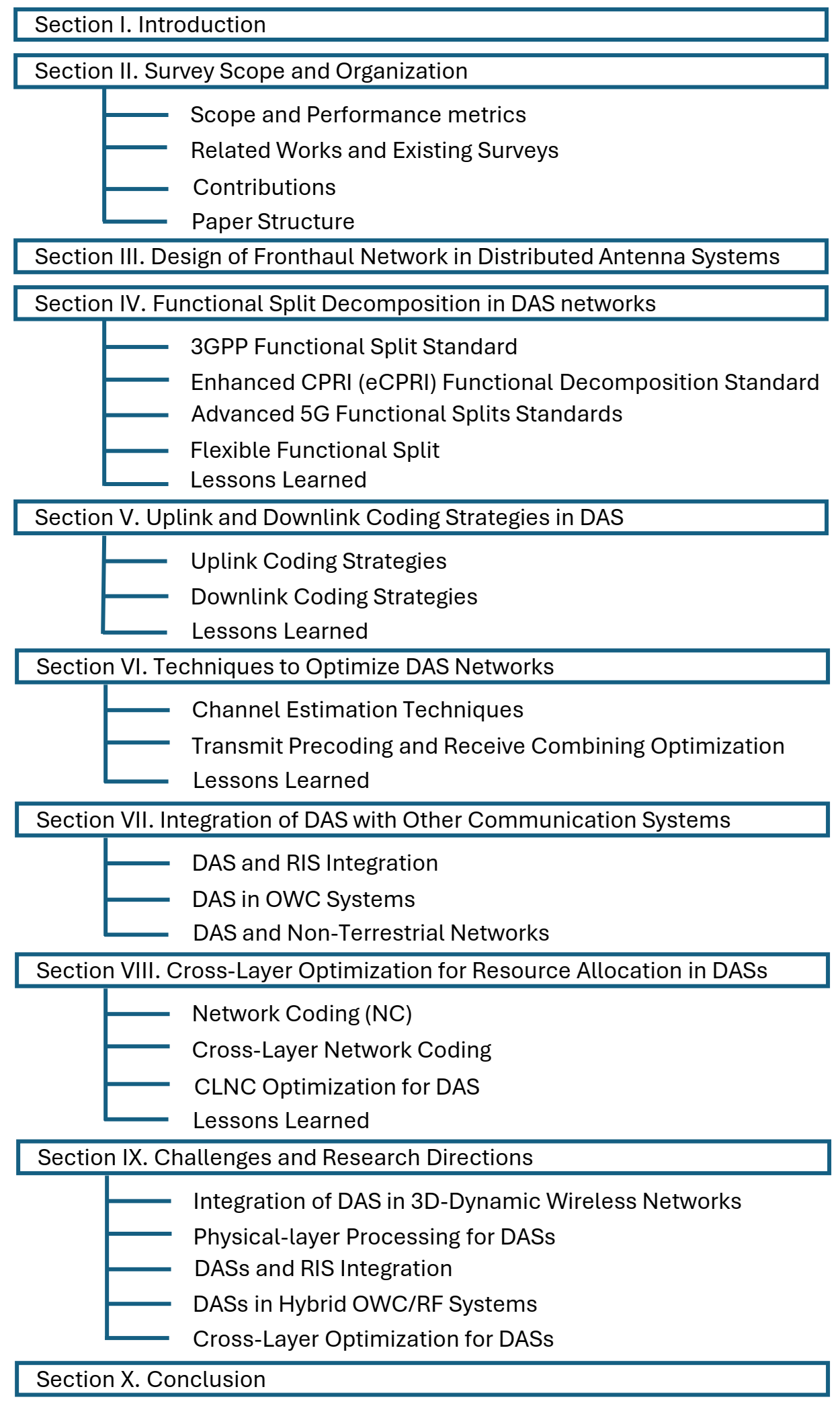}
			\caption{{The outline of this paper.}}
			\label{Outline}
		\end{center}
\end{figure}

This paper is structured as follows. In Section~\ref{Design} and Section~\ref{Split}, we review the design requirements of the fronthaul network in DAS by covering the conventional fronthaul interfaces and functional split standards.  Section~\ref{Uplink} discusses uplink and downlink coding strategies used in DAS to relay signals between users and the CPU via the distributed RUs. System-level optimization for different DAS architectures is discussed in Section~\ref{Tec} with focus on channel estimation, transmit precoding and receive combining techniques. The integration of DAS with other advanced communication technologies is discussed in Section~\ref{Integ}. Cross-layer optimization for resource allocation in different DAS architectures is surveyed in Section~\ref{Cross} with focus on graph theory techniques and CLNC.  An outlook on future research directions is provided in Section~\ref{Fut}, before closing this paper with conclusions in Section~\ref{C}. The outline of this survey paper is illustrated in Fig.~\ref{Outline}.

Note that some acronyms frequently used in this article are listed with their definitions in Table~\ref{table_AB} for ease of reference.

\begin{table*}[t!]
	\renewcommand{\arraystretch}{0.9}
	\caption{List of Abbreviations}
	\label{table_AB}
	\centering
	\begin{tabular}{|p{1.8cm}| p{5.8cm}|| p{1.6cm}| p{5.9cm}| }
	\hline
	
	Abbreviation  & Explanation & Abbreviation  & Explanation\\	
		\hline
		\hline
3D & Three-Dimensional &
3GPP & Third Generation Partnership Project\\
\hline
4G    &     Fourth Generation&
5G &     Fifth Generation \\
\hline

6G    &      Sixth Generation &
AF   & Amplify-and-Forward \\
\hline
AP & Access Point &  ARQ & Automatic Repeat Request\\
\hline

BBU & Baseband Processing Unit & BER & Bit Error Rate \\
\hline

BS   & Base Station & CAP 	& Compress-After-Precoding \\
\hline

CCoF   & Compute-Compress-and-Forward & CF   & Compress-and-Forward \\
\hline

CF-mMIMO & Cell-Free Massive MIMO & CFE   	& Compress-Forward-Estimate  \\
\hline

CLNC     &     Cross-Layer Network Coding&
CoF   & Compute-and-Forward \\
\hline

CoMP & Coordinated Multi-Point&
CP & Control Plane \\
\hline

CPRI & Common Public Radio Interface &
CPU & Central Processing Unit\\
\hline

C-RAN    &     Centralized Radio Access Network   & CSI & Channel State Information\\
\hline

CU & Centralized Unit & D2D   &     Device-to-Device \\
\hline

DAS  & Distributed Antenna System & DF   & Decode-and-Forward \\
\hline

DMPC & Distributed Multi-Pair Computation &

DU & Distributed Unit \\
\hline

ECF   & Estimate-Compress-Forward& eCPRI & Enhanced Common Public Radio Interface \\
\hline

eMBB & Enhanced Mobile Broadband &
EMCF 	& Estimate-Multiply-Compress-Forward \\
\hline

FFT & Fast Fourier Transformation&
F-RAN     &     Fog Radio Access Network\\
\hline

HARQ & Hybrid Automatic Repeat Request &
IDNC     &     Instantly Decodable Network Coding\\
\hline

iFFT & Inverse Fast Fourier Transformation&
IoT   &   Internet of Things\\
\hline

L1 	&Physical Layer in OSI Model &
L2 	& Data-link Layer in OSI Model\\
\hline

L3 	& Network Layer in OSI Model &

LTE   &   Long Term Evolution \\
\hline

MAC  	& Medium Access Control &

MAC& Media Access Control layer\\
\hline

MIMO & Multi-Input-Multi-Output &
mMIMO & massive Multiple-Input Multiple-Output  \\
\hline

MMSE & Minimum Mean-Squared Error &

mMTC & massive Machine-Type Communications \\
\hline

NC    &     Network Coding &  NFV & Network Functions Virtualization      \\
\hline

OBSAI & Open BS Standard Architecture Initiative&

O-DU & Open distributed Unit\\
\hline

O-RAN    &     Open Radio Access Network  & ORI & Open Radio Interface \\ 
\hline

O-RU & Open Radio Unit &
OSI & Open Standard Interconnection \\
\hline

OWC & Optical Wireless Communication &

PHY & Physical Layer \\
\hline

QAM & Quadrature Amplitude Modulation &

QoS   & Quality-of-Service \\
\hline

RaF   & Receive and Forward &

RA-IDNC     &     Rate-Aware IDNC \\
\hline

RAN & Radio Access Network&

RANaaS & Radio Access Network-as-a-Service\\
\hline

 RCoF   & Reverse-Compute-and-Forward&

RE& Radio Equipment Control\\
\hline
 RF & Radio-Frequency &

RIS    &     Reconfigurable Intelligent Surface \\
\hline

RLNC   & Random Linear Network Coding   &

 RB    &      Resource Block  \\
\hline
RQCoF  	 & Reverse-Quantize-Compute-and-Forward &
RRC & Radio Resource Controller\\
\hline

RRH     &    Remote Radio Head  &

RU      & Radio Unit \\
\hline

SDN & Software-Defined Networking &

SCF & Small-Cell Forum \\
\hline

 TDD & Time Division Duplex &

UAV & Unmanned Vehicle \\
\hline

UE & User Equipment &

URLLC & Ultra-Reliable Low-Latency Communications\\
\hline

VLC & Visible Light Communication &
V-RAN    &     Virtualized Radio Access Network     \\
\hline

\end{tabular}
\end{table*}

 \section{Design of Fronthaul Network in Distributed Antenna Systems}\label{Design}

	Because of the centralized nature of the processing resources in the CPU and the distributed nature of the antenna terminals in the service area, fronthaul links between the distributed RUs and the CPU are of critical importance in the design of DAS architectures. In order to build a high-reliability fronthaul network, it is necessary to take into account some interdependent technical requirements such as the RU configuration, required data rate, bit error rate (BER), latency, synchronization, and jitter, and operations administration and maintenance constraints. It is also important to consider business aspects in terms of obtaining low-cost implementation from a mobile operator point of view \cite{7009970}.  In this section,  we start by presenting traditional fronthaul interface standards and highlighting their constraints, followed by how we can overcome their limitations  to keep up with the requirements of next-generation RAN as discussed next.

	The conventional fronthaul interface has been standardized by the CPRI and OBSAI specifications \cite{7096298}. Both protocols include different RAN vendor-specific groups, thereby, full interoperability between them is not guaranteed \cite{7331128}. 
 
    The first version of the CPRI protocol was published in 2003 as a result of cooperation between leading RAN technologies vendors, e.g. Ericsson, Huawei Technologies Co. Ltd, NEC Corporation, and Nokia. Further improved versions have been released until version 7.0 in 2015 \cite{CPRI1}. CPRI specifications divide the BS into two parts including a baseband unit (BBU), or a radio equipment control (REC) unit (can be referred to as CPU), and a radio frequency unit or radio access equipment (RE). The CPRI interface separates between the radio-frequency (RF) signal and baseband signal using the physical layer (L1) and data-link layer (L2) protocols of the open standard interconnection (OSI) protocol stack \cite{7858136,9489948}. However, CPRI has a stringent BER constraint, where each fronthaul link must operate with at most a BER value of $10^{-12}$ \cite{7009970,7402275}. 
	
	In contrast, the OBSAI is another industry specification group resulting from the collaboration of joining RAN vendors, and component and module manufacturers to reduce the cost of deploying RAN architectures \cite{tt}. OBSAI was first released in 2002 and successive improved versions have been published afterwards as for CPRI. According to the OBSAI specifications, the BS is divided into four main modules including RF, transmission, processing, and control modules. The RF module transmits, receives, and amplifies RF signals and converts them from analog to digital and vice versa. The transmission module provides communicating interfaces between the different modules, as well as the conformance and interoperability test specifications. In the processing module, the baseband signals are computed and processed. The control module supports system management, configuration, and synchronization of all BS modules and generates reports to the operations administration and maintenance system \cite{8527624}. Therefore, OBSAI specifications deal with L1, L2, transport, and application layers protocols of the OSI protocol stack \cite{inbook1}. Also, the BER requirement in OBSAI is $10^{-15}$ which is more stringent than that in CPRI. 
	
	Another industry specification group called ORI was formed by the European Telecommunications Standards Institute in 2010 to introduce a fronthaul interface specification that can support multi-vendor interoperability between network elements in the RAN system \cite{ORI,7331128}. This interface is built over the CPRI protocol by removing some functions and adding other options to support the full-interoperability \cite{7009970}.  
	
	Despite the differences between the CPRI, OBSAI, and ORI interfaces, they all provide a constant bit rate in both uplink and downlink transmission. However, CPRI is the widely used protocol for the fronthaul interface by RAN vendors, especially for conventional fully-centralized 4G C-RANs, due to its efficient one-to-one in-phase/quadrature-phase samples mapping techniques between the distributed RU and the CPU \cite{7858136,7414147}. As a result, in this subsection, we confine our investigation to the CPRI specifications as a case study of the conventional fronthaul interfaces.
	
\begin{figure}[t!]
    \centering
		\includegraphics[width=0.5\textwidth]{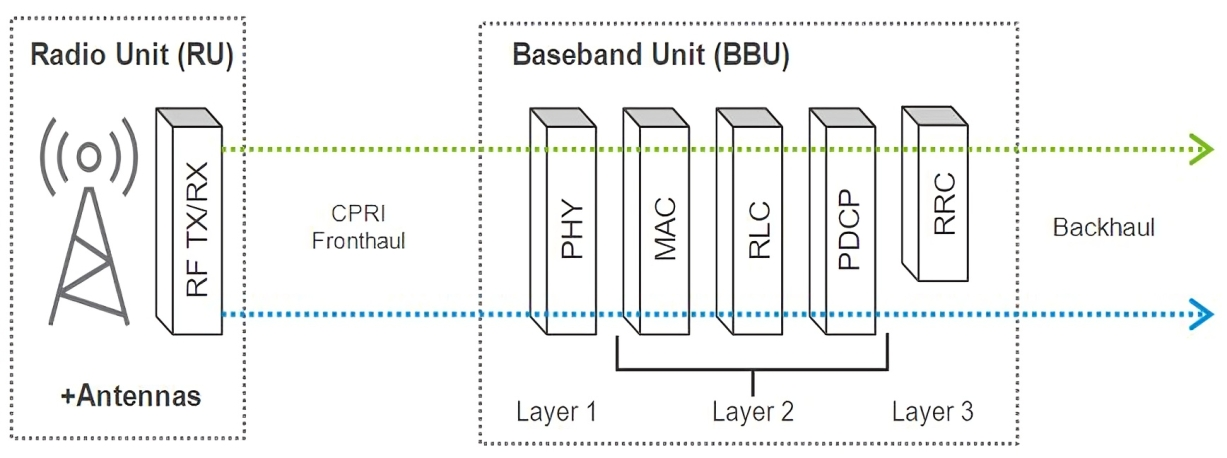}
		\caption{CPRI RAN functional split: fully separating RaF functions in distributed RU (RE) and all L1/L2/L3 processing functions of the OSI protocol stack in the CPU  (REC).}
		\label{Fig:CPRI}	
	\end{figure}

	In conventional 4G C-RAN networks, the CPRI fronthaul interface is used to fully separate the RaF functions in distributed RU and all the baseband processing functions in the CPU as shown in Fig. \ref{Fig:CPRI} \cite{7402275}. This split allows the full centralization of all L1/L2/L3 processing functions of the OSI protocol stack, at the expense of extremely stringent fronthaul capacity and latency demands \cite{ITU}. 

	CPRI can support constant data rate options from 614.4 Mbps up to 24330.24 Mbps over one RU-CPU fronthaul connection \cite{7009970,Pfeiffer:15}. Generally, information flows at the higher layers of the OSI protocol stack are less data-intensive than those at the lowest. Processing functions in the physical (PHY) layer (e.g. cyclic redundancy check, modulation, mapping, and encoding) add extra overhead to the data blocks received from the higher media access control (MAC) layer. Thus, incredibly higher data rates are required as the information flows at the PHY layer towards the RU where RaF functions are performed \cite{CPRI1}. Practically, CPRI requires an overhead of up to 30 times the 4G LTE baseband data rates, thus requiring the deployment of one or more 10 Gbps optical fiber fronthaul links to support 4G LTE data rates \cite{Chanclou2013OpticalFS}. As current 5G networks aim to support 10 times the data rates of LTE, the fronthaul rate needs an increase to over 100 Gbps, which is cost-prohibitive and operationally challenging. Particularly, the required CPRI data rate $R_{CPRI}$ depends on the employed radio-access technology, MIMO implementation, and RF channel bandwidth as shown in the following expression \cite{7009970,Oliveira2014AnalysisOT}
	\begin{equation}
		R_{CPRI} = 2 S_n A f_s b_s C_w C,
	\end{equation}
	where $S_n$ is the number of sectors, $A$ represents the number of antenna ports per sector, $f_s$	is the sampling rate for RF signal digitization (30.72 Mega samples/second per 20 MHz RF channel bandwidth) and $b_s$ is the number of bits per sample (referred as the sample width and is typically 15 bits/sample for LTE systems), $C_w$ accounts for control word overhead in the CPRI frames (typically 16/15), and $C$ denotes a line coding factor (either 10/8 for 8B/10B code or 66/64 for 64B/66B code) \cite{3GPP1,7009970}. The factor 2 is due to the in-phase and quadrature phase components. For instance, for high bandwidth demanding services such as enhanced mobile broadband (eMBB) or for massive MIMO distributed RUs, the CPRI protocol requires unreasonably high fronthaul capacity with maximum latency between RU and CPU of a few hundred microseconds. 
	
    \begin{table*}[t]
		\caption{Required CPRI fronthaul data rate in fully centralized RAN \cite{3GPP1}}
		\centering 
		\begin{tabular}{P{0.2\textwidth}P{0.1\textwidth}P{0.1\textwidth}P{0.1\textwidth}P{0.1\textwidth}P{0.1\textwidth}P{0.1\textwidth}}
			\hline
			\midrule
			\multirow{2}{*}{\textbf{Number of antennas}} &
			\multicolumn{6}{c}{\textbf{RF channel bandwidth}} 
			\\
			& \textbf{10 MHz} & \textbf{20 MHz} & \textbf{50 MHz} & \textbf{100 MHz} & \textbf{200 MHz} & \textbf{1 GHz}  \\
			\midrule
			2 & 1 Gbps & 2 Gbps & 5 Gbps & 100 Gbps & 20 Gbps & 100 Gbps \\
			8 & 4 Gbps & 8 Gbps & 20 Gbps &  40 Gbps & 80 Gbps & 400 Gbps \\
			64 & 32 Gbps & 64 Gbps & 160 Gbps &  320 Gbps & 640 Gbps & 3200 Gbps  \\
			256 & 128 Gbps & 256 Gbps & 640 Gbps &  1280 Gbps & 2560 Gbps & 12800 Gbps  \\
			\bottomrule
			\label{table:CPRI1}
		\end{tabular}
	\end{table*}
	Table~\ref{table:CPRI1} demonstrates the approximate required CPRI data rates without line coding using various RF channel bandwidths and numbers of antennas in a conventional fully-centralized C-RAN  as given by the 3GPP.  As shown in this table, with 20 MHz LTE and 2 antennas configuration, the required CPRI data rate without line coding is 2 Gbps (2.5 Gbps with line coding 10/8) for one RU-CPU fronthaul connection. Therefore, extremely high fronthaul capacity is needed in order to have a fully centralized C-RAN which is a major problem in the design of next-generation RANs \cite{LTE1,6897914}. 
	
	Moreover, the 4G LTE RAN latency requirement of about 10 ms leaves a delay budget of 100 microseconds over the fronthaul link between the CPU and RU; conversely, the maximum fronthaul distance is only around 15 km. With some 5G applications, an RAN latency of 1 ms is required. This requirement forces the distance between the RU and the CPU to be around 1 km, requiring many more locations for the CPUs, which is expensive \cite{8367785}.

	CPRI specifications have some other drawbacks that make them too far from accommodating the evolution in the RAN industry nowadays. First, CPRI has a constant data rate regardless of the change of mobile traffic per the temporal/service area dimensions \cite{CPRI1}. Even in situations when there is no active user traffic in the network, there are still CPRI streams forwarded between the CPU and distributed RUs. This underutilizes the fronthaul bandwidth thus decreasing efficiency. Second, as the number of antennas increases, the required CPRI data rate is incredibly increasing. This is a major impediment to the use of the CPRI interface in 5G networks as far as deploying CF-mMIMO networks is concerned. Moreover, advanced networking and operations administration and maintenance features of data transport standards are required which are not supported by CPRI specifications \cite{eCPRI1}. Finally, the cost of the fully-centralized DAS networks supported by the CPRI specifications is very high \cite{7414147}. 

	With the current revolution in 5G/6G networks, many mobile operators started to question the efficiency of CPRI specifications to keep up with the requirements of next-generation RAN. Currently, researchers and industry experts are working on proposing new techniques for lowering the extremely high capacity demands on the fronthaul network, while still keeping as many benefits as possible from the fully centralized processing as in conventional C-RAN.	To efficiently utilize the fronthaul network resources and better support the evolving needs of next-generation RAN, advanced functional split schemes, uplink and downlink relaying techniques, hybrid DAS and other emerging technologies, and machine learning and artificial intelligence approaches have been investigated as discussed in the following sections.

	\section{Functional Split Decomposition in DAS networks}\label{Split}
	
	In next-generation RAN, the mobile traffic is rapidly growing to volumes that exceed the capabilities of conventional fronthaul interfaces. To overcome this constraint, one solution is to include more processing functions at the distributed RUs to process their observed signals more before forwarding them to the CPU via the fronthaul network. 
	However, the critical questions here are, how many processing functions should be included locally at  RUs? and to which extent may this help in relaxing the extremely high capacity demands on the fronthaul links? To find proper answers, we must investigate all possible functional splits which determine the amount of PHY/MAC processing functions included locally at the RU and the amount of the centralized functions at the CPU. It is thus critical that the functional split between the RU and CPU should take into account some cost-effective and technical trade-offs between the required fronthaul capacity, latency, and signal processing centralization \cite{8384342}. In fact, the choice of how to split the RF and baseband processing functions in the DAS architecture depends on the QoS of the supported services (e.g. low/high throughput, low/high latency, and real/non-real-time applications) and the available transport network infrastructure (e.g. optical fiber/wireless networks including microwave, millimeter waves, or free-space optics). Moreover, it must support specific load demand or user density per given geographical service area.
	
\begin{figure}[t!]
		\includegraphics[width=0.47\textwidth]{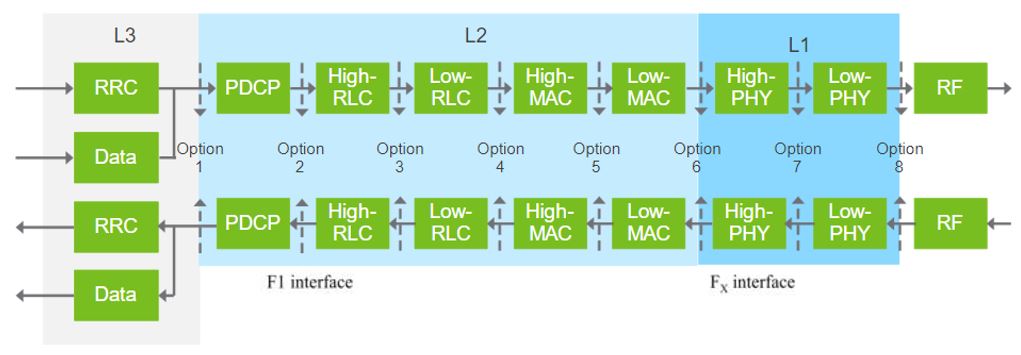}
		\caption{Signal processing functions in the LTE protocol stack and  different functional split options provided by 3GPP.}
		\label{Fig:3GFS}	
	\end{figure}

	Fig.~\ref{Fig:3GFS} illustrates the RF and baseband signal processing functions in the LTE protocol stack (still valid for the 5G new radio) and different   functional split options provided by 3GPP \cite{3GPP1}. In this figure, the signal processing chain closer to the RU is located on the right, and moving to the left means that the observed signal at the RU goes through more and more local processing before forwarding it to the CPU via the fronthaul network. The dashed arrows within Fig.~\ref{Fig:3GFS} indicate the different function split options, where the processing functions on the right of the red arrow are the functions performed locally in the RU, and the remaining functions on the left of the dashed arrow are centrally implemented in the CPU. The more processing functions left locally in the RU, the more signal processing is done before forwarding signals on the fronthaul network to the CPU, thereby lowering the  data-rate and latency demands over the fronthaul link \cite{8479363}. At the same time, this will increase the complexity of the RU configuration, limiting us from deploying cheaper, simpler, and low-power nodes. Here is where the optimization of the tradeoffs becomes paramount. 
	
	Understanding the protocols included in the CPU could help in determining the appropriate functional split. The conventional CPU support all five protocol layers of the RAN. The physical layer or PHY is mainly responsible for coding and modulation operations for the radio transmission. The actual radio transmission is performed  by the RU. For every transmit interval, the  MAC layer defines the size to be used. This transmit interval is referred to as transmit block. The MAC is responsible for the reliable transfer of data over the air. The MAC layer performs a hybrid automatic repeat request (HARQ) procedure, which uses acknowledgments and negative acknowledgments feedback from the receiver. HARQ allows the transmitter to retransmit the data packet if the packet is not correctly received. The MAC layer is also used for implementing fast control signaling, for example, activating carrier aggregation. Fast control activation is required because other signaling mechanisms through the upper radio resource controller (RRC) can be almost 10 times slower. The radio link controller (RLC) layer processes the data arriving from the upper layers into packets that fit into the transmit block defined by the MAC layer.  An incoming packet may be segmented into smaller packets or multiple packets may be concatenated into a larger packet. The packet data convergence protocol (PDCP) provides optional internet protocol packet header compression and security features. It encrypts the data that is sent over the air and protects the integrity of signaling messages to protect against malicious tempering. RRC is the top protocol layer used for major RAN functions, such as connection establishment, handover procedures, paging idle user equipment (UE), uplink power control, and so on. 
	The PHY, MAC, and RLC layers are involved in tasks that are very delay-sensitive (of the order of a few milliseconds), such as HARQ operations and fast control signaling. In contrast, the higher layer PDCP and RRC protocols undertake tasks that are more delay tolerant and can take tens of milliseconds. 
	
	Several functional split standards have been proposed for next-generation RANs   (e.g. dense small cells, CF-mMIMO, and O-RAN) that allow for a radical reduction of the required fronthaul bandwidth compared to the conventional fully-centralized approach \cite{8384342}. 
	In this section, we investigate the functional split options proposed by four groups: 3GPP, eCPRI,  the SCF, and xRAN (which became the O-RAN Alliance in June 2018). We consider the 3GPP signal processing model \cite{3GPP1}  and its functional split options introduced therein as a reference architecture for our investigation of all functional split standards. In Fig.~\ref{Fig:FS}, we map the functional split options in the 3GPP model \cite{3GPP1} to the functional split points introduced in the eCPRI specification by the CPRI group \cite{eCPRI1}, SCF standard \cite{SCF}, and the xRAN (O-RAN) group \cite{ORAN-FH}.  Fig.~\ref{Fig:FS} shows that the intra-PHY functional split point is further split into multiple sub-options at all standards. A more detailed mapping and comparison between the intra-PHY functional splits for different functional split standards is shown in Fig.~\ref{Fig:subFS}. This is to support more cooperation between all distributed RUs in the service area and provide more flexibility to meet the fronthaul network requirements as discussed next.
	
	\subsection {3GPP Functional Split Standard}
	In 4G/LTE networks, the RF and baseband processing
	functions included in the 3GPP protocol stack are divided into a distributed unit (DU) and a CU \cite{3GPP1}. The CU includes the BBU functions specified by LTE standard (still valid for 5G New Radio) like transmission of user data, mobility management, resource sharing and allocation, session control, etc., except those functions allocated exclusively to the DU. The CU also manages the operation of DUs over the fronthaul transport network. 
	The CU may also be referred to as BBU, CPU, or REC. 
	In contrast, the DU  includes part of the baseband processing functions besides the RF functions, depending on the proposed functional split option. 
	This split allows for the coordination of performance features such as required throughput, latency, and cost. In 3GPP, DU may be referred to as RRH, RE, RU, or AP.

	\begin{figure}[t]
		\includegraphics[width=0.47\textwidth]{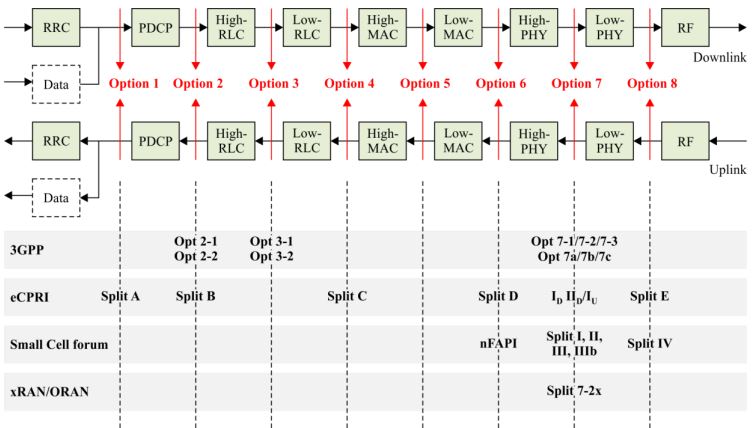}
		\caption{Mapping the functional split options in the 3GPP model \cite{3GPP1} to the functional split points  in the eCPRI specification \cite{eCPRI1}, SCF standard \cite{SCF}, and the O-RAN standard \cite{ORAN-FH}.}
		\label{Fig:FS}	
	\end{figure}

	\begin{figure}[t]
		\includegraphics[width=0.47\textwidth]{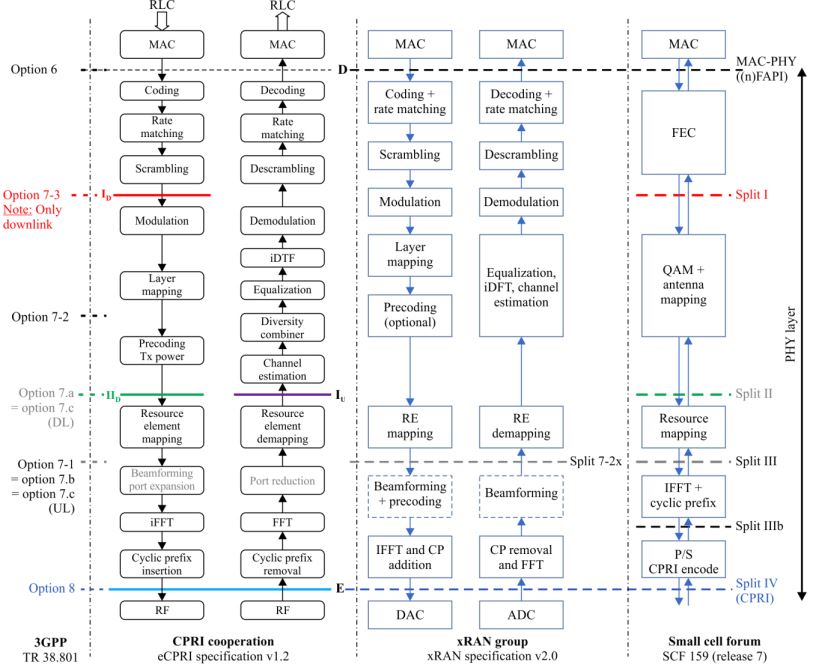}
		\caption{A more detailed mapping and comparison between the intra-PHY functional splits for 3GPP model \cite{3GPP1},
			eCPRI specification \cite{eCPRI1}, SCF \cite{SCF} and O-RAN standards\cite{ORAN-FH}.}
		\label{Fig:subFS}	
	\end{figure}

	\begin{table*}[!t]
		\tiny
		\caption{Comparison between different 3GPP high-level functional split options} 
		\centering
		\begin{tabular}{P{0.01\linewidth}P{0.03\linewidth}P{0.03\linewidth}P{0.03\linewidth}P{0.03\linewidth}P{0.25\linewidth}P{0.25\linewidth}P{0.08\linewidth}}
			\hline
			\midrule		   
			\textbf{Option} & \textbf{Split} & \textbf{Uplink data rate } & \textbf{Downlink data rate} & \textbf{Max. one-way latency}  & \textbf{Advantages} & \textbf{Drawbacks} & \textbf{References}   \\ [0.5ex] 
			\hline 
			1 & PDCP/RRC & 3 Gbps & 4 Gbps & 10 ms & \begin{itemize}
				\item Lower and load-dependent data rate is required  on the fronthaul link. 
				\item Support user-plane separation.
				\item Support faster mobility, measurement configuration, and reporting control.
				\item Suitable for edge-computing and cashing applications.
			\end{itemize} & \begin{itemize}
			\item Highest complex RU configuration.
			\item Lowest centralization.
			\item Don't support inter-cell coordination.
		\end{itemize} & \cite{7550569,7980777,8113473,9107209,9433522,9120828,9684900}   \\\hline 
		2 & RLC/PDCP & 3024 Mbps & 4016 Mbps & 1.5-10 ms & \begin{itemize}
			\item Lower and load-dependent data rate is required  on the fronthaul link. 
			\item Lower fronthaul latency constraints.
			\item Enable mobility coordination. 
			\item Suitable for wireless fronthaul networks. 
			\item Supports deploying networks composed of an aggregation of different radio access technologies.
		\end{itemize} & \begin{itemize}
		\item Have some limitations in coordinated scheduling between multiple RUs.
		\item Don't support CoMP functionalities.
	\end{itemize} & \cite{6771075,7331128,6897914,7996493,7550569,long,10.5555/3181071,8320765,8113473,9433522,8736299,9594852,9300210,8985486,8419201,9322543,9107209,9013336,8845147,9627736,9373366,BORROMEO2022108931,9120828,inbook2,YOUNIS2022107,9717286,9684900} \\\hline
	3 & intra-RLC & lower than option 2 & lower than option 2 & 1.5-10 ms  & \begin{itemize}
		\item Lower and load-dependent data rate is required  on the fronthaul link. 
		\item Lower fronthaul latency constraints. 
		\item Higher reliability can be obtained.
		\item Suitable for wireless fronthaul networks. 
		\item Suitable for non-ideal transmission scenarios.
	\end{itemize} & \begin{itemize}
	\item Latency-sensitive in some scenarios.
\end{itemize} & \cite{8320765,9433522,9107209,9627736,inbook2,9684900} \\\hline
4 & RLC/MAC & 4.5 Gbps & 5.2 Gbps & approximately 100 $\mu$s & \begin{itemize}
	\item Lower and load-dependent data rate is required  on the fronthaul link. 
	\item Suitable for non-ideal transmission scenarios.
	\item Used in scenarios where a local scheduler is required.
\end{itemize} & \begin{itemize}
\item The close relation between RLC and MAC functions disappears in this split.
\item Doesn't fit shorter subframe sizes applications in 5G networks.
\item No benefits for LTE systems.
\end{itemize} & \cite{10.5555/3181071,8320765,9433522,8761941,9107209,9627736,inbook2,9120828,9684900} \\\hline
5 & intra-MAC & 7.1 Gbps & 5.6 Gbps & hundreds of $\mu$s & \begin{itemize}
	\item Lower and load-dependent data rate is required  on the fronthaul link. 
	\item Real-time processing functions are included in the RU.
	\item Lower fronthaul latency requirements depending on the interaction between scheduling functions in CPU and RU.
	\item Supports long-distance fronthaul link between the CPU and RU.
	\item Suitable for non-ideal transmission scenarios.
\end{itemize} & \begin{itemize}
\item Complex interface between CPU and RU.
\item Hard to split the scheduling operations over the CPU and RU.
\item Have some limitations in minimizing the inter-cell interference and deploying CoMP functionalities.
\end{itemize} & \cite{7306542,7414147,7444125,7504140,6666,10.5555/3181071,inbook1,4444,8113473,9433522,8736299,8761941,9107209,9627736,9120828,9684900} \\ \hline
6 & MAC/PHY & 7.1 Gbps & 5.6 Gbps & 250 $\mu$s & \begin{itemize}
	\item Lower and load-dependent data rate is required  on the fronthaul link.
	\item Suitable for non-ideal transmission scenarios.
	\item Centralized scheduling is possible.
\end{itemize} & \begin{itemize}
\item Only 20\% (L2/L3 functions) of the total baseband processing functions is centralized.
\item Limit the deployment of some CoMP functionalities and 5G schedulers.
\item Need an inband protocol for resource block (PB) allocation, MIMO
processing, and modulation.
\item Very high fronthaul latency requirements
\end{itemize} & \cite{7414147,10.5555/3181071,8346013,8254712,8025034,8320765,9594852,9433522,8378032,9300210,9200734,Marotta:19,9322543,9107209,9013336,8845147,9373366,inbook2,9120828,9717286,9684900,9678321,GARIMA2022100674} \\\hline
\end{tabular}
\label{table:3GPP1}
\end{table*}

	\begin{table*}[!t]
		\tiny
		\caption{Comparison between different 3GPP low-level functional split options} 
		\centering
		\begin{tabular}{P{0.01\linewidth}P{0.03\linewidth}P{0.03\linewidth}P{0.03\linewidth}P{0.03\linewidth}P{0.28\linewidth}P{0.23\linewidth}P{0.08\linewidth}}
			\hline
			\midrule		   
			\textbf{Option} & \textbf{Split} & \textbf{Uplink data rate } & \textbf{Downlink data rate} & \textbf{Max. one-way latency}  & \textbf{Advantages} & \textbf{Drawbacks} & \textbf{References}   \\ [0.5ex] 
			\hline 
7.3 & High PHY & 15.2 Gbps & 9.8 Gbps & 250 $\mu$s & \begin{itemize}
	\item Lower required data rate on the fronthaul link due to signal modulation.
	\item The required fronthaul data rate depends on the cell load. 
	\item Suitable for non-ideal transmission scenarios.
	\item Centralized scheduling is possible.
\end{itemize} & \begin{itemize}
\item  Option 7.3 uplink is not defined by the 3GPP.
\item Including modulation in the RU increases its complexity. 
\item limits the deployment of some CoMP functionalities and 5G schedulers.
\item Need an inband protocol for RB allocation, MIMO processing, and modulation. 
\item Very high fronthaul latency requirements
\end{itemize} & \cite{7414147,7980777,8644108,10.5555/3181071,9433522,9249096,8985486,Marotta:19,iet1,9120828,9148093,9684900,9678321}  \\\hline
7.2 = 7.a = 7.c(DL) & low PHY/high PHY & 15.2 Gbps & 9.8 Gbps & 250 $\mu$s & \begin{itemize}
	\item Variable and moderate required  data rate on the fronthaul link.
	\item The required fronthaul bandwidth scales with the used spectrum instead of the number of antennas.
	\item Used for time-sensitive networking technologies and packet-based fronthaul networks
	\item Enables CoMP techniques, joint transmission, and reception between all RUs.
	\item Suitable for non-ideal transmission scenarios.
	\item Centralized scheduling is possible.
\end{itemize} & \begin{itemize}
\item Sub-frame level timing interactions between the low PHY processing in RU and high PHY processing in CPU may occur.
\item Need an inband protocol for RB allocation. 
\item Very high fronthaul latency requirements
\end{itemize} & \cite{9433522,8959317,8644108,8891191,9249096,8985486,MEI2020368,Marotta:19,9492647,9373366,9148093,3GPP1,3GPP2,6934902,6923535,7331128,7306544,6771075,7145716,7504140,6666,7841885,7503792,Lund,Obaidi,10.5555/3181071,long,10.1007/978-3-319-52171-8_3,4444,8250956,8113473,8384342,9120828,iet1,8801953,BORROMEO2022108931,9684900,9678321} \\\hline
7.1 = 7.b = 7.c(UL) & low PHY & 60.4 Gbps  & 9.2 Gbps & 250 $\mu$s & \begin{itemize}
	\item Lower required fronthaul data rate compared to option 8.
	\item Low RU complexity.
	\item Enables CoMP techniques, joint transmission and reception between all RUs.
	\item Suitable for non-ideal transmission scenarios.
	\item Centralized scheduling is possible. 
\end{itemize} & \begin{itemize}
\item  High constant data rate required on the fronthaul link, especially on the uplink.
\item The required fronthaul data rate scales with the number of antennas.
\item Very high fronthaul latency requirements.
\end{itemize} & \cite{9439518,9433522,8959317,8891191,9249096,8642354,8985486,Das:20,Marotta:19,9492647,6771075,9148093,6886880,7096298,6934902,7145716,7306542,7331128,7414147,6923535,7511579,7504410,7503792,7841885,10.5555/3181071,inbook1,8250956,8025007,7996632,8025037,8169875,8252720,iet1,8252881,9120828,BORROMEO2022108931,9684900,9690182,9678321}  \\\hline
8 & RF/PHY & 157.3 Gbps & 157.3 Gbps & 250 $\mu$s &  \begin{itemize}
	\item Supports all the fully-centralized signal processing capabilities.
	\item Lowest RU complexity.
\end{itemize} & \begin{itemize}
\item Extremely high constant data rate is required on the fronthaul link.
\item The required fronthaul data rate scales with the number of antennas.
\item Very high stringent fronthaul latency requirements.
\end{itemize} & \cite{9433522,8642354,Das:20,9200734,Marotta:19,9492647,9373366,9148093,6771075,3GPP2,6882182,7511579,7950267,7504410,6897914,7247156,7414147,6934902,7444125,10.1007/978-3-319-52171-8_3,7504140,KANI201542,7096298,7331128,7306544,7306542,7145716,4444,8045624,6923535,GOMES201550,7958544,7402275,7894280,7925770,10.5555/3181071,555,7980777,8113473,8250956,inbook1,iet1,8025007,8584159,9120828,9724184,BORROMEO2022108931,9684900,9690182,9678321} \\\hline 
\end{tabular}
\label{table:3GPP}
\end{table*}

3GPP standard consists of eight main functional split options including several sub-options as shown in Fig.~\ref{Fig:FS} \cite{3GPP1}. Each option in the 3GPP standard offers a different trade-off between the fronthaul capacity demand, latency requirements, and the benefits of signal processing centralization as discussed next. Note that Tables~\ref{table:3GPP1} and \ref{table:3GPP} present the following options.

\begin{itemize}
	\item In option 8 (RF/PHY split), all the baseband (L1/L2/L3) signal processing functions are centralized at the CPU, whereas only the RF functions (e.g. RF sampler, quantizer, and  up/down converter) are left in the RU site. This option is called CPRI-like split because it corresponds to the conventional fully-centralized 4G C-RAN approach where
	the CPRI is used as a baseline interface \cite{6897914}. Option 8 exploits the benefits of centralized baseband signal processing that enables coordination between the distributed RUs, thus allowing them to perform joint transmission and reception. Consequently, this option supports CoMP, efficient resource allocation, and interference mitigation techniques \cite{7444125}. This also allows for the virtualization of many network services over software hosted on commercial/consumer off-the-shelf servers. Further, this simplifies more the RU configuration which can handle multi-radio access technology and reduce the RAN CAPEX/OPEX \cite{3GPP1, 7504410}. Because of including the ARQ function in the CPU, this option can be used for the non-ideal transmission scenarios \cite{3GPP1}. However, centralizing all the baseband processing functionalities in option 8 places extremely high and constant capacity demands on the fronthaul network with stringent latency constraints, whether user traffic is present or not \cite{7996632}. As illustrated in Table~\ref{table:3GPP}, using 100 MHz RF channel bandwidth, 8 MIMO layers, 32 antennas, 256-quadrature amplitude modulation (QAM), and $2 \times 16$ in-phase/quadrature-phase bit width\footnote{Note that this configuration is deployed for both uplink and downlink.}, the required constant fronthaul data rate for both uplink and downlink transmissions is 157.3 Gbps \cite{3GPP2}. Reference \cite{7487973} showed that although the maximum resource sharing in the CPU can be obtained using  option 8, the required fronthaul data rate is the highest compared to other functional split options.  Also, in \cite{7343513,7504410}, numerical results revealed that further reduction in the required fronthaul data rates can be obtained by moving to other functional split options than option 8. This limits the use of option 8 to scenarios where high-quality optical fiber infrastructure is available.
	
	\item In option 7.1 (low PHY split), as shown in Fig.~\ref{Fig:subFS}, the cyclic prefix insertion/removal,  inverse/fast Fourier transformation (iFFT/FFT), and beamforming port expansion/reduction functions are left in the RU \cite{3GPP1}.  Using the Fourier transformation, the transmitted signals are represented over the fronthaul link as subcarriers in the frequency domain. In the uplink, through using the cyclic prefix removal and FFT that transform the observed signals at the RU into the frequency domain, the RU can remove the guard subcarriers which is 40$\%$ of the LTE traffic \cite{6923535}. This further decreases the required data-rate over the fronthaul link as compared to option 8. However, a constant data rate over the fronthaul link is still required because the resource element mapping that can detect unused subcarriers and achieve variable data-rates, is still centrally implemented at the CPU. In \cite{3GPP1,3GPP2}, split 7.b in both uplink and downlink and 7.c in uplink only are equal to option 7-1. As illustrated in Table~\ref{table:3GPP}, using 100 MHz RF channel bandwidth, 8 MIMO layers, 32 antennas, 256-QAM modulation (same configuration for both uplink and downlink), the required constant uplink and downlink fronthaul data rates are 60.4 Gbps and 9.2 Gbps, respectively \cite{3GPP2}.  In \cite{7996632}, authors illustrate how split option 7-1 reduces the actual throughput of the fronthaul link with 43.8\% as compared with option 8. 
	In \cite{7504410}, numerical results reveal that the required fronthaul data rate using the proposed split option 7-1 achieves a reduction of 30\% to 40\% than that in option 8. This option also enables CoMP techniques, joint transmission, and reception between all RUs, and is suitable for non-ideal transmission scenarios as the ARQ is left in the CPU \cite{3GPP1,NGMN}.  3GPP considers this option for the uplink transmission only \cite{3GPP1,3GPP3}.
	
	\item In option 7.2 (low PHY/high PHY split), besides including the cyclic prefix insertion/removal and iFFT/FFT functions,  the RU includes the resource element demapping, channel estimation, and diversity combiner functions in the uplink and precoding and the resource element mapping functions in the downlink as shown in Fig. \ref{Fig:subFS} \cite{3GPP1}. Using this option, the transmitted signals over the fronthaul link are represented as subframe symbols in the frequency domain.  This further decreases the required fronthaul data rate but at the expense of increasing the complexity of the RU configuration and reducing the centralized  processing capabilities in the CPU.  Starting from this option, all remaining 3GPP functional splits have a variable required fronthaul data rate because of including  the FFT and the resource element mapping functions locally in the RU \cite{Obaidi,Jin:21}. As a result, the fronthaul interface needs to satisfy a certain QoS constraint to ensure priority for time-sensitive traffic. Therefore, this option can be used for time-sensitive networking technologies and packet-based fronthaul networks \cite{Lund}. Also, this option enables CoMP techniques, joint transmission, and reception between all RUs, and is suitable for non-ideal transmission scenarios as the ARQ is left in the CPU \cite{3GPP1,NGMN}.
	In this option, the required fronthaul data rate depends on the used spectrum,  number of transmitted symbols over the fronthaul link, number of quantized bits per symbol, and control information used for further high PHY and MAC processing \cite{10.5555/3181071}. In \cite{3GPP1,3GPP2}, split 7.a in both uplink and downlink and 7.c in downlink only are equal to option 7-2.  As illustrated in table \ref{table:3GPP}, using 100 MHz RF channel bandwidth, 8 MIMO layers, 32 antennas, and 256-QAM modulation (same configuration for both uplink and downlink), the required  uplink and downlink fronthaul data rates are 15.2 Gbps and 9.8 Gbps, respectively, reducing the needed fronthaul capacity by a factor of 10 compared to option 8 \cite{3GPP2}.  3GPP considers this option for the downlink transmission only \cite{3GPP1,3GPP3}. 
	
	\item Option 7.3 (High PHY split) further includes more processing functions at the RU.   As shown in Fig.~\ref{Fig:subFS}, modulation/demodulation, equalization, and inverse discrete Fourier transform functions in the uplink, and layer mapping in the downlink are included locally in the RU \cite{3GPP1}. Since signal modulation is left in the RU site, a lower required data-rate is expected over the fronthaul link compared to other option 7 splits. Particularly, based on the used modulation order, several bits are assigned to each symbol which decreases the required fronthaul data-rate. Using this option, the transmitted signals are represented over the fronthaul link as codeword bits. In this option, because of the increased latency over the fronthaul link, this may reduce the benefits of having wider channel bandwidth and shorter subframes, thereby limiting the deployment of some CoMP functionalities and 5G schedulers \cite{NGMN,10.5555/3181071}. However, it still can be used for non-ideal transmission scenarios since the ARQ is left in the CPU \cite{3GPP1,NGMN}. The 3GPP considers this option for the downlink transmission only \cite{3GPP1,3GPP3}. However, the required downlink data-rate over the fronthaul link for this option was not defined in both \cite{3GPP1} and \cite{3GPP2}. In \cite{SCF}, the required fronthaul data-rate was expressed by the SCF. As illustrated in Table~\ref{table:3GPP}, using 100 MHz RF channel bandwidth, 8 MIMO layers, 32 antennas, and 256-QAM modulation (same configuration for both uplink and downlink), the required  uplink and downlink fronthaul data rates are 15.2 Gbps and 9.8 Gbps, respectively \cite{SCF}. In \cite{7121511}, numerical results revealed that using split option 7.3 can reduce the required mobile fronthaul transmission bandwidth by 90\% as compared to option 8 (traditional C-RAN). Numerical results in \cite{7343513} illustrated that the proposed option 7.3 can reduce the required fronthaul optical bandwidth by 92\% as compared with option 8 while improving the throughput of cell-edge users by 116\% compared with option 6.  Reference \cite{Miyamoto:16} showed that the proposed option 7.3 can further reduce the required fronthaul bandwidth up to 97\% compared to option 8. Experiments in \cite{7830261} showed that option 7.3 can reduce the required fronthaul optical bandwidth for both uplink and downlink transmissions by 90\% as compared to option 8 with an signal-to-noise ratio penalty of less than 2 dB for CoMP in the uplink. 
	
	In general, all three sub-splits in option 7 offer a good trade-off relation between the RU configuration complexity, fronthaul capacity demands, and inter-cell cooperation. As a result, these sub-splits become promising candidates for high-capacity 5G networks in dense urban areas.
	
	\item In option 6 (MAC/PHY split), by adding scrambling/descrambling, rate matching, coding/decoding functions, all the PHY, and RF processing functions are left locally in the RU site as shown in Fig. \ref{Fig:subFS} \cite{3GPP1}. Therefore, the CPU handles only the MAC and network layers (L2/L3) processing functions which are approximately 20\% of the overall implemented baseband processing functions \cite{NGMN}. Using this option, the transmitted signals over the fronthaul link are represented as transport blocks (information bits). This leads to a further reduction in the required fronthaul bandwidth \cite{3GPP1}. As illustrated in Table~\ref{table:3GPP1}, using 100 MHz RF channel bandwidth, 8 MIMO layers, 256-QAM modulation (same configuration for both uplink and downlink), the required uplink and downlink fronthaul data rates are 7.1 Gbps and 5.6 Gbps, respectively \cite{3GPP2}. 
	As a result, this option has been chosen by the SCF for their proposed lower layer split and standardized in the network functional application platform interface initiative  \cite{SCF,8845147}. In contrast, because of the increased latency over the fronthaul link, this split may reduce the benefits of having wider channel bandwidth and shorter subframes, thereby limiting the deployment of some CoMP functionalities and 5G schedulers \cite{NGMN,10.5555/3181071}. However, it still can be used for non-ideal transmission scenarios as the ARQ is left in the CPU \cite{3GPP1,NGMN}. In \cite{7343513}, although option 6 has lower cell-edge user throughput, it has the lowest required fronthaul data-rate compared to options 7.3 and 8. Reference \cite{7503792} showed the perceived user throughput and the total cost of ownership for different functional splits, concluding that intermediate splits like option 6 are the most promising ones.

	\item In option 5 (intra-MAC split), the RF, physical, and some parts of the MAC functions (e.g. HARQ) are included locally in the RU site while keeping the overall scheduler in the CPU \cite{3GPP1}. This further limits the benefits of centralized processing at the CPU because most of the baseband processing functions are performed locally at the RU \cite{4444}. Reference \cite{8255992} showed that option 5 has some limitations in minimizing the inter-cell interference, but it further decreases the required data-rate over the fronthaul link compared to options 7 and 8. As illustrated in Table \ref{table:3GPP1}, using 100 MHz RF channel bandwidth, 8 MIMO layers, 256-QAM modulation (same configuration for both uplink and downlink), the required uplink and downlink fronthaul data rates are 7.1 Gbps and 5.6 Gbps, respectively \cite{3GPP2}. Furthermore,  from this option split and in all other remaining options, the time-critical processing functions in the HARQ are deployed at the RU \cite{4444}. This means that the time-critical processing is no longer dependent on the transmission of the needed data over
	the fronthaul, which dramatically reduces the delay constraints over the fronthaul link. This also allows for longer distance fronthaul links between the RU and CPU \cite{long}. As a result, using this option may increase the latency over the fronthaul link, thereby limiting the deployment of some CoMP functionalities \cite{NGMN,10.5555/3181071}. However, this option is still suitable for non-ideal transmission scenarios as the ARQ is left at the CPU \cite{3GPP1,NGMN}.  At the same
	time, the centralization of the scheduling functions at the CPU allows for coordinated transmissions among different small cells \cite{9627736}.
	
	\item In option 4 (RLC/MAC split), the RF, PHY, and MAC processing functions are left locally in the RU while keeping the PDCP and RLC functions in the CPU \cite{3GPP1}. Therefore, the close relation between RLC and MAC functions disappears in this split. This option receives RLC protocol data units   in the downlink while sending MAC service data units  in the uplink. In 5G networks, it is expected to have shorter subframe sizes that require more frequent actions taken by the scheduler at the CPU to meet the varying channel conditions and traffic demands requirements. Therefore, more frequent control signals  between MAC and RLC are required to determine the required size of the next RLC protocol data units. As a result, this puts some constraints on using this option in 5G networks \cite{10.5555/3181071}. Reference \cite{3GPP1} illustrated no benefits obtained for LTE systems using this split. However, this option is still suitable for non-ideal transmission scenarios as the ARQ is left at the CPU \cite{3GPP1,NGMN}. As illustrated in Table \ref{table:3GPP1}, using 100 MHz RF channel bandwidth, 8 MIMO layers, 256-QAM modulation (same configuration for both uplink and downlink), the required uplink and downlink fronthaul data rates are 4.5 Gbps and 5.2 Gbps, respectively \cite{3GPP2}.
	
	\item In option 3 (intra-RLC split),  the RF, PHY, MAC, and low RLC are left in the RU site while keeping the high RLC and PDCP  in the CPU \cite{3GPP1}. 
	Two sub-options are available based on the real-time/non-real-time functions split. In option 3.1, the split is based on the ARQ. The low RLC is composed of synchronous RLC functions including segmentation and concatenation operations, whereas the high RLC contains the ARQ and other minor asynchronous RLC processing functions. In contrast, the split in option 3.2 is based on the transmitting and receiving RLC entities. The low RLC is composed of a transmitting transparent mode  RLC entity, a transmitting unacknowledged mode RLC entity, a transmitting side of acknowledged mode, and the routing function of a receiving side of AM, which are related to downlink transmission.
	On the other side, the high RLC includes a receiving transparent mode RLC entity, receiving unacknowledged mode RLC entity, and a receiving side of acknowledged mode except for the routing function, and reception of RLC status reports, which are related to uplink transmission.
	Generally, option 3 reduces the required latency constraints over the fronthaul link because the real-time scheduling is performed in the RU \cite{8320765}. Also, higher reliability can be achieved using this split, thereby it is suitable for wireless fronthaul networks \cite{3GPP1}. Furthermore, it is still suitable for non-ideal transmission scenarios as the ARQ is left at the CPU \cite{3GPP1,NGMN}. 
	As illustrated in Table~\ref{table:3GPP1}, using 100 MHz RF channel bandwidth, 8 MIMO layers, 256-QAM modulation (same configuration for both uplink and downlink), the required uplink and downlink fronthaul data rates were defined in \cite{3GPP2} as lower than option 2.
	
	\item In option 2 (RLC/PDCP split), the RRC and PDCP functions are kept in the CPU while including the RF, PHY, MAC, and RLC functions in the RU. 
	The main benefit of this split is the possibility to have a network deployment composed of an aggregation of
	different radio access technologies \cite{inbook2}. 
	Another interesting feature of this split is its ability to generate copies of the traffic
	directed to a certain user and forward it to several RUs \cite{8113473}. This opens up the possibility to deploy efficient mobility
	algorithms.
	In addition, this split further reduces the required latency constraints over the fronthaul link to the order of
	milliseconds because all real-time processing functions are performed locally in the RU site \cite{7331128}. Furthermore, since the added PDCP header to each internet protocol packet is very small,  the transmitted fronthaul data is approximately equivalent to the users' data \cite{8845147}. 
	As illustrated in Table~\ref{table:3GPP1}, using 100 MHz RF channel bandwidth, 8 MIMO layers, 256-QAM modulation (same configuration for both uplink and downlink), the required uplink and downlink fronthaul data rates are 3024 Mbps and 4016 Mbps, respectively \cite{3GPP2}. The 3GPP has recommended this option for fixed wireless access  applications, where coordinated scheduling is not required and bandwidth and latency requirements over the fronthaul network are relatively relaxed \cite{3GPP1,3GPP3}. This split also has a standardized interface by the SCF \cite{SCF}.
	Moreover, option 2 is considered a base for an X2-like architecture because of the similarity on the user plane. However,  some functionalities may be different as some new procedures are required in the control plane (CP). Particularly, there are two possible sub-options available in this split.
	Option 2.1 is called the 3C-like split because it splits the user plane only, similar to the 3C architecture in LTE dual connectivity  \cite{8113473}. In contrast, option 2.2  separates the RRC and PDCP for the CP and the PDCP for the user plane into different central entities \cite{7980777}. As a result, option 2.2 requires coordination of security configurations between different PDCP instances \cite{3GPP1}. 
	
	\item In option 1 (PDCP/RRC split), only the RRC is kept in the CPU while including all other RF and baseband processing functions locally in the RU site \cite{3GPP1}.
	This allows for performing the user data processing at the RU which is beneficial for edge-computing and caching applications, especially in F-RAN architectures \cite{3GPP1,9678343}. Also, the centralization of the RRC in this option enables the customization of the deployments and having faster mobility algorithms, measurement configuration, and reporting control \cite{SCF}. This option is called the 1A-like split because it is similar to the 1A architecture in LTE dual connectivity \cite{8113473}. This option is also called the CP/user plane split because the CP functions are included in the RRC, whereas the entire user plane functions are performed in the other baseband functions included in the RU \cite{7980777}. Reference \cite{7980777} concludes that the full separation between user plane/CP in combination with the full-centralization of the control
	functions in option 1 would follow the principles of SDN.
	However, this split makes the RU more complex, large and consumes more power than the simplified RU of option 8. Also, it doesn't support inter-cell coordination, thereby it might not be recommended for scenarios where we have many cells connected to the CPU. 
	The required data-rate of option 1 over the fronthaul network is much simpler because the entire protocol stack resides at the RU site.  As illustrated in Table \ref{table:3GPP1}, using 100 MHz RF channel bandwidth, 8 MIMO layers, 256-QAM modulation (same configuration for both uplink and downlink), the required uplink and downlink fronthaul data rates are 3 Gbps and 4 Gbps, respectively \cite{3GPP2}. Also, reference \cite{7550569} shows how the latency constraints can be relaxed to tens of
	seconds if only the RRC functions are centralized. 
\end{itemize}

In summary, 3GPP proposed eight functional split options, depending on which baseband processing functions are included at the DU or RU and
which functions are moved to the CU or CPU. In practice, the most popular options under investigation are (1) option 8 (CPRI-like split) in 4G fully-centralized RAN \cite{7402275};
(2) option 7  (intra-PHY split) in 5G RAN networks \cite{8644108}, where some
PHY layer processing functions are performed at the RU before its transmission; (3) option 6 (MAC/PHY split) in dense small-cell deployments \cite{SCF}, where RF and PHY layer functions are included
in the RU; (4) option 2 (PDCP/RLC split) in wireless fronthaul transport networks \cite{3GPP3}, where the PDCP and RRC functionalities are kept
in the CPU, while RF, PHY, MAC, and RLC functions are included in the RU; and (5) option 1 (PDCP/RRC split) in edge-computing services and F-RAN architectures \cite{9678343}, where user plane functions are performed in the RU and the CP operations are kept in the CPU. Options 7 and 8 are user load independent and require high-capacity and low-latency demands over the fronthaul links between RU and CPU, whereas these requirements are relaxed by moving to higher splits \cite{7402275,6923535}.

\subsection{{Enhanced CPRI (eCPRI) Functional Decomposition Standard}}

The conventional CPRI interface (option 8 in the 3GPP functional split) requires extremely high-capacity and ultra-low latency demands over the fronthaul links of the 4G fully-centralized C-RAN.
With the aim of cost reduction, backwards compatibility, and hardware reuse, the industry cooperation (i.e., NEC, Nokia, Huawei, and Ericsson)
involved in the specification of CPRI have  released
an evolved CPRI (eCPRI) specification in 2018 \cite{eCPRI1}.
In the eCPRI specification, the RU is called  the eCPRI Radio
Equipment (eRE), whereas the CPU is known as the eCPRI Radio Equipment Control (eREC) \cite{eCPRI1}. In between these two units lies the fronthaul network. The eCPRI protocol is a promising technology for 5G RAN. 
It supports flexible and efficient radio data transmission via using a packet-based fronthaul network (e.g. Ethernet or internet protocol). This makes the fronthaul capacity demands depend on
the actual resource block (RB) utilization at any given time \cite{KANI201542}. This also allows for the implementation of
other benefits such as point to multi-point
architecture, statistical multiplexing gain, traffic aggregation, and low quantization resolution \cite{9221322}. As a result, several CF-mMIMO architectures in the literature employ eCPRI-based functional splits \cite{9838846,arxiv1,arxiv2,ALBREEM202279,8960442,9319703}.
\begin{figure}[t]
	\centering
  \includegraphics[width=0.45\textwidth]{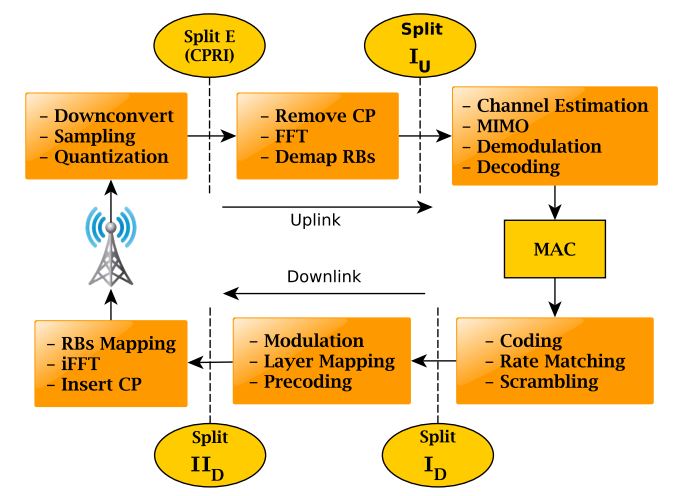}
	\caption{eCPRI vision of 5G intra-PHY processing chain\cite{eCPRI1}.}
	\label{Fig:eCPRI}	
\end{figure}

Unlike conventional CPRI protocol, the eCPRI specification enables more flexibility in splitting the 5G new radio processing functions between the CPU and RU which reduces the high capacity requirements over the fronthaul network while limiting the complexity of RU \cite{8384342,9748751,9221322,9492417,8736787, 8250956,9013123}.
The signal processing decomposition in the eCPRI specification takes the 3GPP functional split standard as the starting point. From there,
eCPRI specification supports five inter-layer
functional splits named A, B, C, D, and E which are equivalent to options 1, 2, 4, 6, and 8 (CPRI-like split), respectively in the 3GPP functional split standard as shown in Fig. \ref{Fig:FS} \cite{3GPP1, ITU}. One additional set of intra-PHY functional splits, two for the downlink and one for the uplink, named $\text{\rom{1}}_\text{D}$, $\text{\rom{2}}_\text{D}$, and $\text{\rom{1}}_\text{U}$ is included between split E and D.  Any combination of the different intra-PHY downlink/uplink splits is possible.
As shown in Fig.~\ref{Fig:subFS}, split $\text{\rom{1}}_\text{U}$ in eCPRI is equivalent to split 7.a/b, aggregated split $\text{\rom{2}}_\text{D}/\text{\rom{1}}_\text{U}$ is equivalent to option 7.a/b and option 7.c in downlink, and split $\text{\rom{1}}_\text{D}$ is equivalent to option 7.3 \cite{3GPP1, eCPRI1}.  The main difference between split $\text{\rom{1}}_\text{D}$ and $\text{\rom{2}}_\text{D}$ is that split $\text{\rom{1}}_\text{D}$ works with bits, whereas splits $\text{\rom{2}}_\text{D}$ and $\text{\rom{1}}_\text{U}$ are oriented to IQ data \cite{eCPRI1}. Due to similarities in inter-layer functional splits between the eCPRI specifications and the 3GPP model, we confine our eCPRI functional decomposition study to the intra-PHY splits to demonstrate how they can support MIMO networks, carrier aggregation, and CoMP functionalities efficiently while taking into their account the fronthaul bandwidth and latency requirements.
The eCPRI vision of the 5G intra-PHY processing chain is indicated in Fig.~\ref{Fig:eCPRI}.
\begin{figure}[t]
	\includegraphics[width=0.47\textwidth]{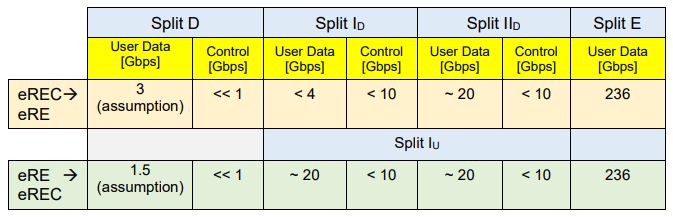}
	\caption{PHY layer splits bit rate estimations \cite{eCPRI1}.}
	\label{Fig:eCPRIrate}	
\end{figure} 

Generally, splits E, $\text{\rom{1}}_\text{D}$, $\text{\rom{2}}_\text{D}$, $\text{\rom{1}}_\text{U}$, and D have extremely strict latency requirements, while
splits A,B, and C have relaxed latency requirements over the fronthaul links \cite{9045398}. Regarding the needed fronthaul data rate, there are three PHY layer processes that 
mostly increase the required fronthaul capacity. These three processes are modulation, the port-expansion combined with the beamforming process, and the IFFT being in combination with the cyclic prefix process. As a result, only split E and, under some circumstances, splits $\text{\rom{1}}_\text{D}$, $\text{\rom{2}}_\text{D}$, $\text{\rom{1}}_\text{U}$ have high fronthaul bandwidth demands. By moving the split towards the left in Fig.~\ref{Fig:FS} the required fronthaul bandwidth will be lowered and vice versa. 
In addition, the needed data rate over the fronthaul link between the eREC and eRE depends on the number of MIMO layers, support MU-MIMO or not, and the number of antennas \cite{eCPRI1}.
In \cite{eCPRI1}, a tabular representation 
of fronthaul capacity demands is presented as indicated in Fig. \ref{Fig:eCPRIrate}. Using 100 MHz RF channel bandwidth, 8 downlink MIMO layers, 2 downlink MIMO layers, 256-QAM modulation (same configuration for both uplink and downlink), the required uplink and downlink fronthaul data-rates for the eCPRI PHY layer splits are calculated \cite{eCPRI1}. According to this study, 
$\text{\rom{1}}_\text{U}$ and  $\text{\rom{1}}_\text{D}$  have the best bit rate optimizations among the eCPRI intra-PHY splits.
In \cite{8384342,9013123}, numerical results  demonstrate a fronthaul capacity reduction by approximately
1.7 times in the case of split $\text{\rom{1}}_\text{U}$ compared to split E.  Split $\text{\rom{2}}_\text{D}/\text{\rom{1}}_\text{U}$ can also offer  more than 5-fold bandwidth saving in the fronthaul bit-rate compared to the CPRI-like split \cite{8385866,8526294}. 

In summary, In order to keep up with the requirements of next-generation RAN, the fronthaul traffic should be packetized and transmitted over  packet-based transport
networks. Additionally, advanced functional splits than CPRI need to
be considered for the design of the fronthaul network, given the extremely high capacity and ultra-low latency requirements of CPRI transport. As a result, the CPRI interface has been evolved to eCPRI \cite{eCPRI1} by the CPRI group  with advanced  functional split decomposition to be used for packet-switching networks. The most popular eCPRI splits are splits E, $\text{\rom{1}}_\text{U}$, $\text{\rom{1}}_\text{D}$, and
$\text{\rom{2}}_\text{D}$ which are used in CF-mMIMO networks \cite{9838846,arxiv1,arxiv2,ALBREEM202279,8960442,9319703}.

\subsection{{Advanced 5G Functional Splits Standards}}
\begin{figure}[t]
	\centering
 \includegraphics[width=0.45\textwidth]{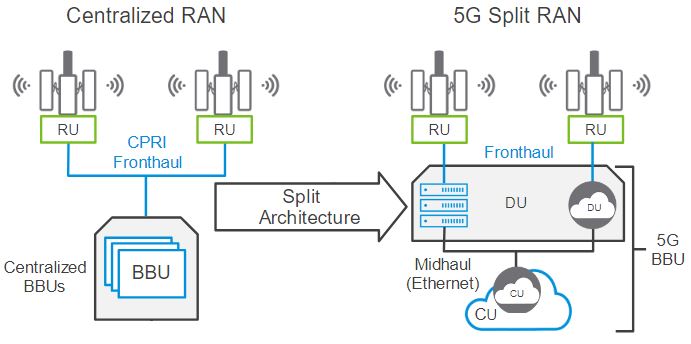}
	\caption{RAN evolution to 5G.}
	\label{Fig:RANsplit}	
\end{figure}

In the leap forward from 4G LTE to the 5G new radio architecture, the key evolution in the DAS systems is that all the processing functions that were usually centralized at the CPU, are now disaggregated into three different units: the CU, the DU, and the RU as shown in Fig. \ref{Fig:RANsplit} \cite{8723481}. This allows for increasing the flexibility to finely allocate resources to each one of the three units and improving their resource utilization and management. Moreover, splitting the processing functions throughout the three units in the RAN can decrease the transport bandwidth and latency demands at each unit and minimize their computational complexity as well as reduce the RAN CAPEX/OPEX costs \cite{9798822}.

CUs are software-based units hosted in a cloud-based platform that maintain BBU-like functionalities, especially the non-real-time processing functions. In contrast, DUs can be a mixture of software-based or physical technologies that contain some near-real-time processing functions related to the RU. The split between CU and DU is hardly impacted by the type of physical infrastructure. The DU is placed close to the RU than the CPU sites in 4G C-RAN architecture to serve delay-sensitive 5G applications, such as autonomous vehicles and factory automation, whereas the CU can be placed in a centralized location further away from the DU. The central location of the CU allows larger deployments and makes implementations very conducive to virtualization.  The 5G split architecture requires an additional transport network solution for the CU to DU interface. This new transport network is called midhaul or F1 interface based on 3GPP \cite{TS}. Different deployment scenarios can be also considered including independent RU, DU, and CU locations, co-located DU and CU, RU and DU integration as well as RU, DU, and CU integration. Generally, a centralized CU can control DUs in an 80 km radius. Expected distance between RU
and DU is in the range of 1-20 km, DU-CU is 20-40 km, backhaul connection between CU and the core network is up to 300 km \cite{8479363}. 
Furthermore, 5G split architectures are designed to be “inherently” intelligent. However, the key considerations of future RAN design are still its size, processing complexity, and power consumption.  
\begin{figure}[t]
	\centering
 \includegraphics[width=0.45\textwidth]{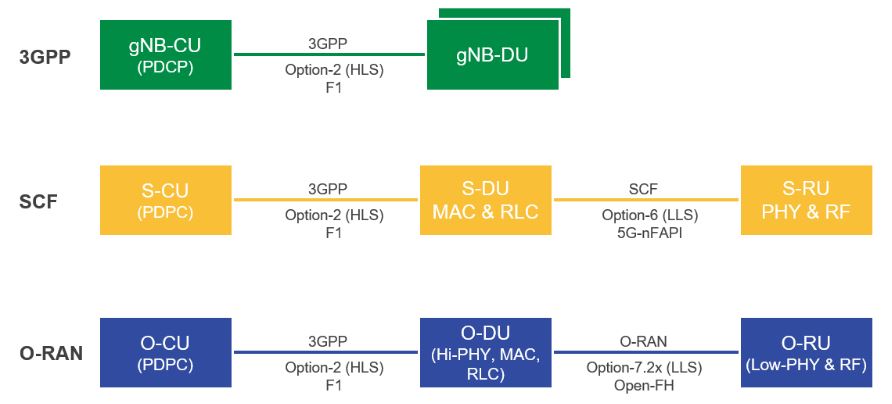}
	\caption{5G functional split \cite{SCF}.}
	\label{Fig:CUDU}	
\end{figure}

The functional split architecture between these three units allows for utilizing different distributions of protocol stacks between CU, DUs, and RUs depending on midhaul and fronthaul transport network availability and network design. The choice of optimal 5G functional split option depends on the deployment scenarios and standards as shown in Fig. \ref{Fig:CUDU}. In 2017, 3GPP selected option 2 (PDCP/RRC) as the higher layer functional split point (referred to as F1 interface or midhaul) \cite{TS}, whereas they are open with either option 6 for MAC/PHY (L2/L1) split or option 7 for intra-PHY (intra-L1) split as the lower layer functional split point (referred as $\text{F}_\text{X}$ interface) \cite{TR,77777}. Cascaded functional split architecture can also be considered to add additional scalability and flexibility on the design of the fronthaul network \cite{ITU}. 
Further evolution was needed to reduce the RU configuration complexity, increase cost resiliency and competitiveness, bring more intelligence to the network, and support efficient transport utilization. The SCF \cite{SCF} and O-RAN Alliance \cite{ORAN-FH} evolved their specifications to support 5G functional splits with two additional architectures.

In \cite{SCF}, the SCF studies the requirements of different functional split options, especially for densification scenarios, where many small/femtocells need to be deployed as shown in Fig. \ref{Fig:FS}.
Particularity, The SCF standardizes option 6 (MAC/PHY) as a lower layer split between the small cell RU   and small cell DU  units in the network functional application platform interface initiative \cite{SCF}. 
This interface models the traffic generated by the MAC/PHY split by translating the information exchanged by these functions into UDP  packets to be transmitted over a packet-based network \cite{9014264,TakahiroKubo20172017XBL0041}. It also enables the O-RAN ecosystem by allowing any small cell RU or small cell DUs from different vendors to connect to any small cell RU. Moreover, it encourages innovation and competition among suppliers of small cell platform software and  hardware by providing a common application platform interface \cite{8845147}.

In contrast, the O-RAN alliance defined option 7.2x (intra-PHY) for the lower layer split between the open RU (O-RU) and open DU (O-DU), deployed over where an e-CPRI fronthaul interface \cite{ORAN-FH,Yajima2019OverviewOO}. 
The intra-PHY split is the favorable approach by the O-RAN alliance for its less RU configuration complexity and it supports various fronthaul bandwidth requirements as well as it has high virtualization capabilities \cite{9685992,9887515}. In addition, split 7.2x supports the block floating point IQ compression and de-compression technique to further reduce transport bandwidth \cite{Silva_Ramalho_Almeida_Medeiros_Berg_Klautau_2022}. With option 7.2x, two variants are available: 7.2a and 7.2b based on where precoding occurs. In split 7.2a, the precoding and resource element mapper are included in the O-DU, whereas the O-RU handles beamforming, iFFT/FFT, and cyclic prefix functions. If precoding is included in the O-RU, then this is 7.2b split \cite{9439518}. As the amount of baseband processing functions performed in the O-RU or included in the O-DU is a critical parameter in the O-RU configuration, split 7.2a-based O-RUs are much lower in cost and simpler compared to split 7.2b-based ones. In contrast, split 7.2b requires less fronthaul bandwidth as compared to split 7.2a because including the precoding functions in the O-RU  further decreases the fronthaul bandwidth demands, especially when the number of MIMO streams is greater than MIMO layers. However, it requires more processing and memory requirements in the O-RU.

On the other side, while there can be different splits for the higher-layer split, the only one being considered in both the SCF and O-RAN alliance de-facto between the DU and CU is option 2, where the latency on the midhaul link should be around 1 msec \cite{SCF,ORAN-FH}.

In summary, several standard authorizations are engaged in
addressing the challenging aspects regarding the design
of feasible transport networks for 5G RAN scenarios. The most
noticeable change in 5G new radio architecture is that all the baseband functions that were usually present at the CPU, now
are now split into the CU, DU, and RU. This  RAN functions disaggregation allows for relaxing the stringent requirements over the transport network and maximizing performance. 3GPP selected option 2 as the higher layer functional split point between the CU and DU, whereas they are open with either option 6 or option 7 as the lower layer functional split point between the RU and DU. The SCF standardized option 6  as a lower layer split between small cell RU and small cell DU, while selecting option 2 for the higher layer split between small cell RU and small cell DU. Finally, the O-RAN alliance defined option 7.2x as a lower layer split between O-RU and O-DU, while keeping option 2 for the midhaul interface between O-CU and O-DU.

\subsection{{Flexible Functional Split}}
The selection of the appropriate functional split  (i.e., the centralization level) still remains a challenging task in the design of DAS architectures. Different criteria (e.g., inter-cell interference mitigation, traffic demand, latency requirements, ..., etc.) have to be considered in order to make such a decision. Following the daily variations in network requirements and service demands, implementing a fixed functional split (as proposed by the aforementioned functional split standards) is not a viable solution in the long run \cite{8255992}. Therefore, the flexibility of dynamically selecting the optimal functional split is essential to efficiently utilize the baseband processing resources and fronthaul bandwidth \cite{8408566}.

By applying the principles of virtualization to the RAN functionalities, the baseband processing functions no longer require special proprietary hardware to perform, and can instead be run on a cloud-based platform \cite{8113473}. Because of hardware/software decoupling flexibility, a dramatic decrease in hardware costs and application agility is achieved as applications can be upgraded easily or swapped altogether, which is not easier with traditional hardware \cite{7534741}. This converts conventional DAS architectures to V-RAN or RANaaS, where decomposing RAN operations into physical and virtual infrastructures can be done \cite{Frauendorf2023}. V-RAN uses virtualization technologies such as NFV, cloud-native applications, Kubernetes, and containers to deploy the DU and CU functionalities as software-based units hosted over commercial off-the-shelf or x86 servers \cite{9500721,7958544}.  
By exploiting the capabilities of V-RAN technologies, we can easily achieve network scalability with flexibly centralized processing functions. 
Additionally, this allows for reaping the benefits of different functional splits by enabling mobile operators to dynamically change the split option according to the available network resources and their needs. Since different functional splits are characterized by significantly different fronthaul capacity and latency requirements, mobile operators can support specific QoS configurations for each provided service (e.g., high data rate, low latency).  Furthermore, the flexible functional split allows mobile operators to support the varying traffic demand and user density in each geographical area.

Several recent research works in academia and industry investigated the benefits of acquiring a flexible functional split DAS architecture \cite{3GPP1,SCF,8432268,NGMN}. 
In \cite{6898939}, a novel RANaaS concept was introduced in which centralization of processing and management is flexible and can be dynamically adapted to the network requirements. The paper also provides a trade-off between full centralization, as in 4G C-RAN, and partial centralization, as in today’s 5G networks taking into account the network characteristics as well as actual service needs. 
In \cite{7744804}, a high-level overview of a scalable and flexible 5G RAN was presented, which supports multiple functional splits to adapt to user traffic variations. Reference \cite{7929675} proposed a C-RAN architecture that simultaneously supports different functional splits for each DU in the network to improve energy efficiency and user experience. In \cite{8845147,9013336}, an adaptive RAN design was introduced that can  dynamically adapt its centralization level  from/to option 6 to/from option 2 at runtime without service interruption based on  user traffic demands.  Simulation results show that the proposed dynamic model significantly improves the achievable data rates compared to statically centralized solutions. The work in \cite{9844117} 
proposed a flexible functional split selection scheme for the DUs in 5G C-RAN by adopting to traffic heterogeneity under limited midhaul capacity constraints. Results demonstrate that the flexible selection of functional split options among each CU-DU pair achieves 90\% centralization over the fixed functional split for a given midhaul capacity.
In \cite{8255992,8408566}, virtual network embedding algorithms were proposed to jointly utilize the fronthaul bandwidth and minimize the inter-cell interference by flexibly selecting the appropriate functional split option for each cell. The proposed algorithms are operating between functional splits 6, 7, and 8. The simulation results indicate that baseband processing requirements and fronthaul capacity constraints are drastically changed, depending on the flexible functional split used. The authors also suggest considering different functional split options for daylight and night based on the users' distribution and traffic variation. Reference \cite{9399108} showed that a dynamically centralized C-RAN architecture can achieve substantial inter-cell interference reduction even in limited fronthaul capacity scenarios.  
In \cite{8283832}, two schemes were proposed for selecting the optimal functional split option in a C-RAN architecture. The first one is a nearly-optimal backtracking scheme that yields a performance upper bound. The second one is a low-complex greedy approach which provides the best trade-off between the distance to the optimal solution and computational load reduction. 

Migration from fixed functional split architectures to flexible functional split ones can be a rewarding process for mobile operators in terms of reducing total network costs over the long run. 
The work in \cite{9075181} evaluated the migration cost to C-RAN with both full-centralization of network functions and partial-centralization by using flexible function splitting. Numerical results demonstrate that partial-centralization configurations have an optimal total cost of ownership with lower crossover time compared to C-RAN with full-centralization. In \cite{7248612}, a graph-based framework was proposed to flexibly split and place the baseband processing functions in the network that can efficiently reduce the fronthaul cost at the expense of increased computational cost. The numerical results showed that the possibility of centralized placement increases as the delay requirements become more stringent. Reference  \cite{9316303} investigated how dynamically selecting the functional split serves C-RAN in limited-capacity and limited-budget scenarios. Simulation results show that intelligently selected functional splits can reduce the required fronthaul bandwidth by 40\% and the total network cost by  35\% compared to conventional C-RAN architectures.  Reference \cite{9781611} investigated the profitability and feasibility of the dynamically-adapted 5G RAN architectures. Results showed that the dynamically-adapted configurations can achieve drastically  cost reductions compared to static configurations. In \cite{9724184}, an integer linear programming-based BBU placement problem with a flexible functional split selection approach was investigated to simultaneously minimize the required number of fronthaul optical fiber connections and BBU hotels, thereby reducing the total network cost. Numerical results demonstrated that the proposed scheme can achieve substantial network cost reduction with  25\% and 50\% compared to the  fixed functional split scheme for small and large network scenarios, respectively. In \cite{9946423}, a deep reinforcement learning approach was proposed that jointly optimizes the flexible placement of the DU and CU network functions in the regional and edge open-cloud nodes and associates the users to RUs to minimize the end-to-end delay of users and the O-RAN deployment cost. Numerical results showed that the proposed scheme can reduce the average end-to-end user delay by up to 40\% and the O-RAN deployment cost by up to 20\% compared to the conventional O-RAN architecture.
Reference \cite{8486243} investigated the design challenges of V-RAN based on topological data from three different mobile operators. The paper proposed a novel analytical model that significantly minimizes the V-RAN cost by jointly choosing the optimal functional split and the routing paths between the RUs and CUs. The proposed scheme is operating between functional splits 2, 6, and 7. The simulation results revealed that multi-access edge computing and F-RAN architectures can increase substantially the RAN cost as it pushes the majority of the V-RAN processing functions placement back to RUs. They also indicate that  pure C-RAN architecture (unvirtualized) is rarely a feasible solution for upgrading existing networks.

Another advantage of the flexible functional split in DAS architectures  is its potential to improve resource allocation and energy efficiency. By distributing the workload among different entities, DAS can effectively utilize the network resources and reduce the power consumption of individual network components. This can lead to significant savings in terms of both energy costs and environmental impact.
In \cite{8280524},  a hybrid C-RAN/V-RAN architecture with a data center on the edge was proposed, where the DU’s functionalities can be virtualized and split at several appropriate options. The numerical results showed that the proposed architecture can reduce the required midhaul bandwidth by 42\% compared to the fully centralized C-RAN, and decrease the power consumption by 35\% compared to the case where all the processing functions are distributed at the edge. Reference  \cite{7997127} investigated the interplay of the midhaul bandwidth and energy efficiency, with the flexible placement of the baseband processing functions at the edge or central cloud in a hybrid C-RAN/V-RAN architecture. In \cite{8386280}, a two-step recovery scheme was proposed to orchestrate the lightpath transmission adaptation and functional split reconfiguration that can preserve the V-RAN fronthaul connectivity even when network capacity is scarce. The work in \cite{8718107} jointly combined the task offloading with the flexible functional split in F-RAN architecture. The proposed model aimed to split the task according to the appropriate functional split between DU and CU for cooperative execution when the task is offloaded. Numerical results showed that the proposed scheme can significantly reduce the execution delay and energy consumption. In \cite{9912981}, a meta-heuristic optimization scheme was proposed to optimize the functional split problem in a hybrid Cloud Fog RAN architecture. Numerical results demonstrated that the meta-heuristic model can achieve statistical optimality equal to the integer linear programming in terms of energy-efficiency and network coverage. Reference \cite{8612914} proposed a novel flexible functional split scheme for F-RAN architectures that enables the non-orthogonal coexistence of eMBB and URLLC  services by processing the eMBB communications centrally at a cloud, whereas handling URLLC traffic at the edge nodes. This solution achieves high spectral efficiency for eMBB traffic via centralized baseband processing while satisfying the low latency requirements of the URLLC services using edge processing. The work in \cite{8790771} proposed an online reinforcement learning approach to flexibly select the appropriate functional split option that can effectively deliver on-demand contents with minimal average latency in F-RAN with multiple cache-enabled enhanced RRHs (eRRHs). 

In \cite{8761941}, a new orchestration framework for jointly optimizing the functional split and end-to-end resource allocation on a user basis was proposed for a multi-sited C-RAN infrastructure. Simulation results showed that the jointly and dynamically selected functional split and resource allocation scheme outperforms cell-centric functional split approaches. The work in \cite{8422754} proposed a user-centric approach that optimizes the split of the CU while taking into account the requirements of its processing network functions and the availability of cloud infrastructure. Simulation results demonstrated that the proposed scheme optimizes both bandwidth and processing usage as well as minimizing energy consumption compared to cell-centric, distributed, and centralized C-RAN architectures.
In \cite{8403571,8917915}, two offline and online heuristic algorithms were proposed to jointly optimize the selected optimal functional split option with the corresponding user data transmission duration and transmitted power in a C-RAN architecture with renewable powered RUs. Numerical results showed the proposed scheme can maximize the achievable throughput while satisfying the average fronthaul rate constraint. 
In \cite{8877874}, an Apt-RAN model was proposed to optimize the energy consumption of the CU pool and the number of handovers, considering different flexible functional split options. Numerical results showed that lower-layer functional split options have high energy consumption at CU as compared to higher-layer splits.
The work in \cite{8693874} proposed a flexible functional split model for a downlink C-RAN that can dynamically configure each active RU to use either data-sharing (DS) or compression-after-precoding (CAP) in order to minimize the total power consumption while considering the limited fronthaul capacity, fronthaul power consumption, and QoS requirements. Numerical results  showed that the proposed model can significantly reduce the total power consumption compared to the pure DS  and CAP schemes. Reference \cite{9751689} proposed a time-averaged stochastic model to jointly optimize the selection of the appropriate functional split and minimize the average network power consumption in elastic optical fronthaul networks. Numerical results showed that the proposed scheme can reduce the average power consumption by up to 70\% compared to a traditional C-RAN. In \cite{9690182}, a flexible centralization scheme was proposed to maximize the overall network spectral efficiency for disaggregated RANs. The proposed scheme jointly considered the user association, RU clustering, fronthaul network routing, and baseband processing unit placement. Numerical results demonstrated that the proposed scheme can achieve a 1.33-times spectral efficiency gain compared to conventional methods, but also provides 1.47 and 1.27 multiplexing benefits for networking and computing resources, respectively.


Recently, the network slicing concept has gained much attention with the recent advances in  SDN and NFV technologies, enabling logically isolated subnetworks for different purposes in the same physical network infrastructure. Through leveraging the benefits of flexible functional split configuration, mobile operators can address the issues deriving from the various requirements of a multi-service environment, especially dynamically adopting the network resources and relaxing the high fronthaul bandwidth constraints.
In \cite{8761081}, a joint RAN slicing and functional split scheme was proposed to optimize the centralization level and achievable throughput. Numerical results showed that even though in terms of centralization level the proposed scheme has more costs, it can better satisfy the network requirements and also achieve higher throughput in the network. The work in \cite{8762089} analyzed the relations between the optimal functional split, computational resource allocation requirements, and network slicing deployment constraints in a C-RAN. Simulation results showed that deploying VNFs in cloud-based 5G networks can achieve high resource utilization efficiency and increase the supported number of network slice chains.
Reference \cite{9615541} proposed a mixed-integer quadratically constrained programming scheme to dynamically select the most appropriate functional split for each network slice separately in a multi-tier 5G O-RAN architecture. Simulation results demonstrated that through leveraging the dynamic functional split benefits, the proposed scheme can efficiently utilize the physical network resources and better satisfy different network slice constraints compared to fixed functional split configurations. In \cite{9617786}, a mathematical framework was proposed to dynamically select the optimal functional split option per each network slice and the optimal slice size in terms of achievable data-rate, that maximizes the mobile operator's total revenue profits. Reference \cite{9627736} proposed an open-access framework is proposed that includes flexible functional splitting for different slices and slice subnets while considering different network-sharing policies from 5G specifications. The paper highlightd the importance of flexible functional splitting as a key enabler to deal with the heterogeneous and varying requirements of communication services, leading to a considerable network slice cost reduction. Numerical results also showed that flexible functional splitting can reduce the average load on physical fronthaul links by a factor of 3.

Fronthaul network is identified as a major element of SDN-based 5G network architectures \cite{8361844}. Reference \cite{7010528} investigated the potential of deploying SDN principles in optimizing C-RAN performance that can be dynamically configured
to enable certain customized services. The paper also highlighted a number of challenges to be overcome, such as heterogeneity, latency, and communication protocol over the fronthaul network. The work in \cite{8386082}  proposed an end-to-end 5G X-haul architecture over the converged SDN fronthaul network, where the fronthaul interface is based on two different NGFI split options. Reference \cite{8428882} proposed a hierarchical layered SDN framework over NGFI to support dynamically different functional splitting options in C-RANs.
In \cite{Marotta:19}, accurate estimation techniques of available fronthaul bandwidth and the associated real-time selection of the appropriate functional split  were investigated for converged SDN access networks. In \cite{8422555}, a new cooperation scheme was proposed to manage the adaptive flexible functional split in 5G networks, while satisfying the resource availability constraints in 5G SDN converged access networks. Simulation results showed that the proposed SDN converged scheme enables the mobile network to take advantage of the highest possible centralization of the baseband processing functions and simultaneously adapt to current traffic demand by exploiting flexible functional split adaptively aligned with the varying traffic demand. The work in \cite{9061000} proposed an SDN-based orchestration model to dynamically select the most appropriate functional split while taking into account the availability of transport resources and the time-varying radio performance constraints. Numerical results showed that the proposed flexible RAN model can achieve better utilization of transport resources compared to the conventional C-RAN architecture. 

Reference \cite{8386298} experimentally demonstrated a reconfigurable and flexible fronthaul interface with analog radio-over-fiber   integration that can optimally serve different 5G services. The work considerd implementing a flexible functional split for options 8 and 7-1 over point-to-point optical fiber connections. The numerical results showed that this integration can achieve more than 15\% lower latency compared to option 7-1, which is an efficient solution to support the stringent latency requirements of URLLC. In \cite{7996632},  an implementation of NGFI was presented that is able to perform functional split options 8 and 7-1, both transported over Ethernet fronthaul in a C-RAN architecture.

In terms of challenges, one of the main issues with implementing a flexible functional split in DAS is the need for advanced coordination and communication among the different network entities. This can be complex and requires careful planning and coordination to ensure that the network functions are distributed in an optimal manner. Further research and development in this area can help to unlock these benefits and realize the full potential of flexibly-configured DAS networks.

In summary, following a fixed functional split between RU and CPU proved by practice that it is not a viable solution over the long run, and may lead to stringent bandwidth and latency constraints being imposed on the fronthaul network, thus making its design and deployment more costly and challenging \cite{Dalgitsis2022}. 
As we are going beyond the 5G milestone, next-generation RAN aims to provide integrated network processing functions as a service hosted over a cloud-based platform, to handle varied traffic loads and different types of devices. For this, mobile operators are moving to leverage the benefits of the recent advances in NFV and SDN technologies. These technologies help virtualize the network architecture to create enhanced communication capabilities and resource optimization techniques, such as flexible functional split and network slicing that significantly relax stringent fronthaul bandwidth requirements. In addition, network virtualization and flexible functional split help to reduce the huge investments implied by 5G, due to their capabilities to meet the varying capacity and latency requirements. Hence, flexible functional split approaches are key enablers for 5G fronthaul network design.

\subsection{Lessons Learned}
The evolution of wireless communication networks, particularly in the context of 5G and beyond, has highlighted the importance of flexible functional splits in fronthaul network design. Fixed functional splits between the RU and the CPU have proven to be impractical in the long run due to stringent bandwidth and latency constraints imposed on the fronthaul network. To address these challenges, it is crucial to consider advanced functional splits beyond traditional interfaces like CPRI. The introduction of the eCPRI interface with its functional split decomposition has paved the way for packet-switching networks and enabled the use of splits E, $\text{\rom{1}}\text{U}$, $\text{\rom{1}}\text{D}$, and $\text{\rom{2}}_\text{D}$ in CF-mMIMO networks.

Standardization bodies, such as 3GPP, SCF, and the O-RAN alliance, have played significant roles in addressing the design complexities of transport networks for 5G RAN scenarios. These bodies have defined various functional split options at different layers, allowing for the disaggregation of baseband functions across the CU, DU, and RU. By selecting appropriate split points, the stringent requirements over the transport network can be relaxed, leading to improved performance and reduced deployment costs.

As we move beyond 5G, next-generation RANs aim to provide integrated network processing functions as services hosted on cloud-based platforms. NFV and SDN technologies are instrumental in achieving this goal. They enable network architecture virtualization, offering enhanced communication capabilities and resource optimization techniques such as flexible functional splits and network slicing. These approaches significantly relax fronthaul bandwidth requirements and accommodate varying capacity and latency demands, thereby reducing investment costs. In practice, flexible functional splits are crucial enablers for the design of 5G fronthaul networks. They allow for the efficient utilization of resources, address stringent requirements, and provide the necessary flexibility to accommodate future network evolution. By embracing advances in NFV, SDN, and virtualization technologies, operators can unlock the full potential of next-generation RANs while optimizing performance and cost-effectiveness.

Based on the processing functionalities included locally in the RU, various relaying strategies and signal processing techniques can be performed for both uplink and downlink in order to make efficient use of the fronthaul links. This has been the topic of many studies recently as discussed next.

\section{Uplink and Downlink Coding Strategies in DAS}\label{Uplink}

One way to limit the fronthaul signaling requirements is through using efficient uplink and downlink source and channel coding schemes. From an information-theoretic perspective, the DAS model is best understood as a relay network. The RUs can be considered relaying nodes that facilitate communication between the CPU and the user terminals. A major drawback of DAS architectures is that their performance is constrained by the limited capacity of fronthaul links that connect the distributed RUs with the CPU. This puts a constraint on the amount of exchanged traffic between the RUs and the CPU. 
In order to tackle this problem, various uplink relaying schemes were proposed to relay user data traffic from distributed RUs to the CPU, such as DF \cite{quek_peng_simeone_yu_2017}, AF, CF \cite{6342931}, CoF \cite{6522158}, and noisy network coding \cite{5752460}. 
Additionally, in the downlink, the CPU transmits data to users via the distributed RUs using cooperative relaying schemes such as data-sharing techniques \cite{6920005}, compression-based strategies, \cite{6588350}, RCoF \cite{6283033}, and RQCoF \cite{6522158}. These techniques have gained significant attention in recent years due to their potential to enhance the overall network performance. However, despite the ongoing effort, the optimal relaying strategy is still to be found \cite{8432465}. 

The interesting but essential questions here are: What is the appropriate way for the RUs to relay as much useful information as possible to the CPU while satisfying the limited fronthaul capacity constraints of the fronthaul links? Is it using local detection, compression, or computation? How can the physical and data link layer design of a DAS architecture adapt to the capacity limitations of the fronthaul links? Also, what should the CPU do to recover the original messages? Should it use successive or joint decoding? 

To properly answer these questions, in this section, we discuss various uplink and downlink relaying strategies that utilize the limited fronthaul capacity constraints in different ways for different DAS architectures, along with their corresponding
optimization methods and frameworks for finding appropriate solutions. We conclude by providing design insights learned from such optimization.

\subsection{{Uplink Coding Strategies}}
In the uplink, multiple users communicate their codewords with the CPU through the distributed RUs. In an ideal case with infinite-capacity fronthaul links, each RU can forward its exact observed signal to the CPU, and the full joint-decoding gain can be achieved. Therefore, the resulting uplink channel is modeled as an instance of a single-input-multiple-output (SIMO) or MIMO multiple-access channel. In practice, however, the fronthaul links have limited capacity constraints, thus each RU should relay as much useful information as possible from its observation to the CPU in order to mitigate the inter-cell interference and improve the overall network performance. As a result, from an information-theoretic point of view, the uplink channel is modeled in this case as a two-hop multiple-access relay channel \cite{8114545}.

Different relaying strategies can be used to relay traffic from RUs to the CPU.
The first typical approach is to make each RU \textit{decodes} the codewords of its scheduled users, and then re-encodes and \textit{forwards} them for collaborative transmission to the CPU; this is the so-called DF strategy \cite{1499041}. Although this technique offers significant advantages, the RU is ultimately interference-limited as the number of transmitted codewords increases. This also scarifies the potential benefits that the user terminals can gain  
from the jointly centralized processing capability at the CPU \cite{quek_peng_simeone_yu_2017}. In addition, it introduces additional delay and requires high signal processing capability at the distributed RUs which significantly increases the processing power as well as the network complexity and cost, which can be a disadvantage in systems with limited computational resources  \cite{7573000}. Also, DF is sensitive to errors in the received signal. If the RU node is unable to correctly decode the signal, it will not be able to forward it to the CPU, which can result in transmission errors and loss of data \cite{4305394}.

Higher data rates can be attained by decoding integer linear combinations of simultaneously transmitted users' codewords that ``mimic'' the channel realization at each RU, as long as they are all drawn from the same linear codebook; this is the so-called CoF strategy. CoF has recently been
employed in DAS systems \cite{6522158} including C-RAN \cite{8644380,9148793} and CF-mMIMO systems \cite{7962724,9755878,JIANG2023} to significantly reduce the fronthaul load, increase the system throughput as well as offer protection against noise and inter-cell interference.
The key idea of the CoF scheme is that  user terminals encode their messages using nested lattice linear codes, and an RU suppresses interference by not decoding the users' codewords but rather \textit{computes} a weighted sum function (integer linear combination) of them and \textit{forwards} it to the CPU \cite{6034734}. 
The advantage of using nested lattice codes is that they are closed under modulo-addition, {and hence} the (modulo-) integer linear combination of lattice codewords is still a valid lattice codeword \cite{WPNC}.  A nested lattice codebook is constructed by nesting
set of lattice points of a fine lattice confined within the fundamental Voronoi region of a coarse shaping lattice \cite{8186899}.
At the CPU, after receiving all the linear combinations {forwarded by the RUs}, it {uses the inverse of integer coefficient matrix to} recover the {codewords} \cite{6522158}. However, this strategy is sensitive to channel estimation errors and requires the integer coefficients matrix to be invertible \cite{5205840}. 
In the original CF \cite{6034734}, all constructed nested lattice codes at user terminals share a common coarse lattice, thereby all users are constrained by the same power constraint. In \cite{7028519,6736658,7017572,7517355}, asymmetric CoF schemes were proposed that allow asymmetric construction of coarse lattices and unequal power allocation across users, as a result, this improves the achievable sum rate and computation rate as compared to \cite{6034734}. In \cite{8756704}, this asymmetric approach was employed in CF-mMIMO  networks whose results show a significant reduction in the required backhaul capacity compared to \cite{6034734}. 
To solve the possibility of having a full-rank failure of the integer coefficient matrix in CoF, references \cite{6736536,7349245} utilized the successive interference cancellation (SIC) strategy, where a successive computation of multiple codeword combinations is required to enlarge the achievable rate and enhance system performance. In \cite{9622183}, power control algorithms were proposed for both parallel computation and successive computation in CF-mMIMO networks. Simulation results showed that the successive computation approach outperforms the parallel computation in terms of achievable sum-rate. 

Since the decoded integer linear combinations of users' codewords at  RUs are in general correlated, the performance of the CoF can be improved if the computed integer linear combination of users' codewords at each RU is further compressed using a lattice-based quantizer before forwarding it to the CPU to reduce the information redundancy \cite{7275188}; this is the so-called \textit{Compute-Compress-and-Forward} (CCoF) strategy.
Further improvements on CCoF were proposed in \cite{7572049}, where the computed integer linear combinations at RUs are remapped to equations of the users' input symbols, forwarded compressed in a lossy manner to the CPU, and are not required to be linearly independent.

As an alternative to the DF, CoF, and CCoF schemes, the RUs can offload the decoding process to the CPU using  AF or CF strategies. As a result, the uplink channel is converted into a virtual MIMO multiple-access channel \cite{7465790}. 
In the AF, the RU simply acts as a repeater that amplifies the observed signal it receives and forwards a scaled version of it to the CPU, where beamforming gains can be exploited. However, the amplification process can introduce noise and distortion into the transmitted signal as well as amplify the interference from other wireless devices or signals, which can degrade the performance of the system \cite{4305387}. Additionally, AF requires the RU node to have a stable power supply in order to amplify the received signal. This can be challenging in some scenarios, such as in remote or inaccessible locations, where the power supply may be unreliable or unavailable \cite{6034234}.

In contrast, in the CF scheme, RUs first downconvert their observed RF signals to baseband signals, which are analog in nature, then compress the baseband signals and forward the corresponding compression indices, which are represented by digital codewords, via their fronthaul links to the CPU. Once the CPU receives all the indices, the CP decompresses these compressed signals in order to obtain a distorted version of the received signals across all the distributed RUs, then recovers the original users' messages based on the entire set of decompressed signals \cite{6342931}.
Intuitively, the required resolution of the compression level at each RU is determined by the capacity of the fronthaul link. A limited fronthaul link's capacity would imply “coarser compression”, which in the rate-distortion theory is reflected by a  larger compression distortion and as a result, decreases the users' achievable rates, and vise versa \cite{6824778}. 

In \cite{5730555}, the use of CF scheme for C-RAN was justified from an information-theoretic prospective. Results showed that CF can achieve the information-theoretic capacity of the network within a constant gap for a Gaussian multi-message multicast network which depends only on the network topology and is independent of other channel parameters.
At the CPU, the original users' messages can be recovered by joint decompression (of the quantized signals forwarded by RUs) and decoding (of the users' codewords) using joint processing and decoding for both the decompression and message decoding simultaneously. Although this technique is information-theoretically better justified, its computational complexity is extremely high and increases exponentially with the number of user terminals \cite{5594708}. Thus, several low-complexity distributed lossy compression techniques with comparable performance have been proposed using sequential (successive) decompression and decoding strategies, which differ based on the amount of utilized side information in the decompression process at the CPU. One of these strategies is independent compression or single-user compression in which RUs' observed signals are compressed and decompressed independently without capitalizing on the presence of any side information at the CPU. \cite{quek_peng_simeone_yu_2017}. This may lead to high compression distortion as it ignores any correlations among the received signals at the RUs \cite{6924850}.
Thus, by taking advantage of the correlated received signals at the CPU, once one of the compressed signals is decompressed, it can serve as side information in decompressing the remaining compressed signals. As a result, RUs can capitalize on this side information, and the needed amount of fronthaul capacity for the compression process can be reduced \cite{7465790,6824778,7590010}. This compression strategy is called Wyner-Ziv compression \cite{1055508}, and can achieve users' rates equal to that achieved by joint decompression and decoding through using SIC for message decoding at the CPU \cite{6601765,7590010}. In \cite{6342931}, a robust compression method was proposed to cope with uncertainties on the correlation of the received signals in C-RAN.
Moreover, for further improvement in the performance of CF in a Gaussian channel setting, lattice-based Wyner-Ziv compression schemes have been proposed in \cite{706450,6089357,5766129}. 
Inspired by the CoF scheme, a distributed lossy compression strategy called integer-forcing source coding was proposed in \cite{7745894}, where the receiver estimates integer linear combinations of the compressed signals, and then recovers the compressed signals themselves. Particularly, a full-rank integer coefficient matrix can be optimized by exploiting the correlations between the compressed signals. This technique 
was studied in a C-RAN scenario and  showed to perform similarly to Wyner-Ziv compression with SIC \cite{8262720}. 
References \cite{quek_peng_simeone_yu_2017,8262720} showed that in an uplink C-RAN, CoF schemes are better than CF under small fronthaul capacity values. However, distributed lossy CF schemes outperform CoF at moderate and large fronthaul capacities. 

CF strategies are also favorable relaying schemes for CF-mMIMO. Due to the large number of antennas at the APs in CF-mMIMO systems, employing low-resolution analog-to-digital converters to quantize the observed signals at the AP is a feasible and practical solution to reduce the fronthaul load, power consumption, system complexity as well as the hardware cost.
Based on what information is quantized and forwarded to the CPU and where channel estimation is performed, several CF scenarios have been studied in CF-mMIMO systems. In
the first approach, each AP compresses its observed pilot and data signals separately and sends their compressed versions over its limited-capacity fronthaul link to the CPU, and then, the CPU performs the channel estimation, the design of combining vectors, and the data detection; this is the so-called \textit{Compress-Forward-Estimate} (CFE) strategy \cite{8422865,8645433,9110901,8891922,Bashar20201,9123382}. Following this way, centralized combining methods can be implemented over a limited fronthaul network by compression at the APs.
The second approach is to first estimate the channel at each AP, and then each AP compresses the estimated channels and data signals separately and forwards their compressed versions to the CPU that recovers the channel state information (CSI) and performs data recovery using centralized combining; this is the so-called \textit{Estimate-Compress-Forward} (ECF) \cite{8891922}. Since the compression is implemented at the APs, ECF reduces the fronthaul load in a distributed fashion. 
In the third approach, each AP first performs the channel estimation, multiplies the observed data signals by a local combining vector computed based on the local channel estimate, compresses and forwards the results to the CPU that only performs data recovery; this is the so-called \textit{Estimate-Multiply-Compress-Forward} (EMCF) \cite{8761134}. Therefore, by implementing the channel estimation and EMCF can further reduce the fronthaul load compared to other compression approaches. 
The four approach is similar to the EMCF. However, the received signal at the CPU is further multiplied by the receiver filter coefficients to minimize the interference and improve the performance; this is the so-called \textit{Estimate-Multiply-Compress-Forward-Weight} (EMCF-Weight) \cite{8761134,9123382,8422865}. The channel  estimation and design of the combining vectors are also implemented at the APs in a distributed way. As a result, EMCF-weight has a significant performance improvement in terms of the fronthaul load and the achievable uplink rates. Reference \cite{8756286} studied the effect of different quantization approaches on the fronthaul link. In the first approach, the AP quantizes the CSI and data signals and forwards their digital representations to the CPU, whereas the second scheme forwards only a quantized weighted signal version to the CPU.
Simulation results demonstrated that the first approach has a slightly higher achievable uplink rate than the second one. As the number of antennas per AP increases, this difference in performance between the two schemes decreases. Moreover, under the network setup proposed in \cite{8756286}, results showed that a performance near to that of a perfect fronthaul can be obtained if the number of quantization bits is set to be greater than 7.

Despite the efficiency and simplicity of CF schemes in reducing the fronthaul load in DASs, the uniform quantization approach can result in quantization error, leading to significant performance degradation. 
As a result, in an uplink DAS, CoF schemes are better than CF under small fronthaul capacity values. However, distributed lossy CF schemes outperform CoF at moderate and large fronthaul capacities \cite{quek_peng_simeone_yu_2017,8262720}. A Comparison of the uplink coding strategies is illustrated in Table~\ref{uplink_coding}.

Next, we investigate how the cooperative downlink  coding strategies play a vital role in ensuring reliable and efficient data transmission from the CPU to the end users in DAS. 
	\begin{table*}[t]
		\caption{Comparison of the uplink coding strategies} 
		\centering
		\begin{tabular}{P{0.1\linewidth}P{0.25\linewidth}P{0.25\linewidth}P{0.2\linewidth}}
			\midrule		
			\hline
			\textbf{Scheme} & \textbf{Processing at RU} & \textbf{Processing at CPU}  & \textbf{References}  \\ 
			\hline
			DF & Decode the codewords & Recover the original messages & \cite{quek_peng_simeone_yu_2017,7573000}\\
			\hline
			CoF & Compute integer linear combinations of the codewords & Recover the codewords& \cite{6522158,8644380,9148793,7962724,6034734,WPNC,5205840,7028519,6736658,7017572,7517355,6736536,7349245,9622183}\\
			\hline
			CCoF & Compute integer linear combinations of the codewords, then compress these combinations & Decompress the compressed equations and recover the codewords  & \cite{7275188,7572049} \\\hline
			AF & Amplify the received signal & Recover the original messages& \cite{7465790,4305387,6034234} \\
			\hline
			Single user Compression & Compress the observed signal  & Decompress the compressed signals independently & \cite{quek_peng_simeone_yu_2017,6924850} \\
			\hline
			Wyner-Ziv Compression & Compress the observed signal  & Utilize side information to decompress the compressed signals & \cite{7465790,6824778,7590010,1055508,6601765,6342931,706450,6089357,5766129} \\ 
			\hline
			integer-forcing source coding & Compress the observed signal & Estimate integer linear combinations of the compressed signals, then solve these equations & \cite{7745894,8262720} \\
			\hline
			CFE & Compress observed signal and CSI & Estimate channel and design combining & \cite{8422865,8645433,9110901,8891922,Bashar20201,9123382} \\
			\hline
			ECF & Estimate the channel and compress observed signal and CSI &  Design combining & \cite{8891922} \\
			\hline
			EMCF & Estimate the channel, design combining, and compress observed signal & Recover the original messages & \cite{8761134} \\\hline
			EMCF-weight & Estimate the channel and design combining, and compress observed signal & Design receiver filter and recover the original messages & \cite{8761134,9123382,8422865} \\\hline

	   \end{tabular}
       \label{uplink_coding}
	\end{table*}

\subsection{{Downlink Coding Strategies}}
In de-centralized RAN, each scheduled user is served by a single BS and experiences interference from all neighboring BSs. The benefit of centralized signal processing in DAS architectures arises from the ability of multiple RUs to cooperatively serve users, which minimizes the effect of unwanted interference. Since the messages intended for different users in DAS, all originate from the CPU, the CPU can relay useful information about the user messages to the distributed RUs via the fronthaul links, thus allowing the RUs to perform network-wide beamforming in order to achieve interference mitigation and cooperative transmission.
Similar to the uplink, if the fronthaul links have infinite capacities, all the information of users' messages can be forwarded from the CPU to all RUs, thus achieving full cooperation. In this ideal case, the downlink channel can be modeled as a SIMO or MIMO broadcast channel. However, practically, due to the finite fronthaul capacity, the amount of information that can be forwarded to RUs is limited. Here, the main question for the CPU is to determine the most useful information about users' messages to be conveyed to the RUs in order to minimize interference as possible. As a result, the channel can be expressed as an instance of a two-hop broadcast-relay channel \cite{8359129}. 

Different cooperative transmission schemes that enable collaboration among  RUs can be used to forward traffic from the CPU to RUs. A straightway method for the CPU to enable cooperation is by \textit{sharing} each user's message with multiple RUs to form a cooperative cluster, which then forms locally beamformed signals  to serve their scheduled users. This technique is commonly referred to as DS \cite{5997324}. Ideally, in order to achieve full cooperation among RUs, the CPU should share the users' messages with all serving RUs. The benefit of the data-sharing approach is that the RUs receive clean messages that can be utilized for joint transmission. However, such full cooperation maybe not be feasible to be achieved because of the limited fronthaul capacity constraint, which imposes restrictions on the size of the RU cluster that can collaborate in serving a user \cite{6920005}.  A significant improvement can be obtained using common messages and rate-splitting optimization techniques \cite{5997324,8359129,8732995}. 

Alternatively, since all the users' data are available at
the CPU, the encoded and beamformed signals can be centrally  constructed at the CPU. Subsequently, these signals can be compressed and transmitted to the RUs for further processing  \cite{6588350}. Particularly, the signals beamformed at the CPU have the potential to emulate the benefits of full cooperation. However, the beamformed signals become analog instead of discrete like the raw data in the data-sharing strategy. Therefore, it is necessary to compress these signals before transmitting them over digital fronthaul links with limited capacity. The primary benefit of the compression-based approach is the efficient utilization of fronthaul capacity when transmitting beamformed signals containing multiple users' messages over the fronthaul link. However, the downside of this approach is the introduced compression distortion, which leads to a decrease in the achievable downlink rates. In general, the reduction in the achievable rates is influenced by the compression rate, which is determined by the capacity of the fronthaul links and the compression strategy implemented at the CPU \cite{quek_peng_simeone_yu_2017}. Further improvement can be achieved  through the implementation of the  \textit{multivariate compression}, where the baseband signals are jointly compressed at the CPU \cite{6588350}. This approach enables correlation among the compression distortions in the signals of different RUs. By designing the correlation to cancel out compression distortions at the user side, the achievable rates for users can be improved as the added distortions {(due to interference)} are mitigated.

In the context of downlink transmission in CF-mMIMO networks, compression strategies remain favorable. Two compression approaches have been explored in CF-mMIMO. The first approach is CAP \cite{8690797,8902784,8678745}, where the centralized signal is first precoded and then compressed at the CPU before being transmitted to the APs. This makes CAP suitable for centralized precoding. The second approach is \textit{precoding-after-compress} \cite{10.1007/s11277-018-6038-1}, where simple compression is performed at the CPU by separately quantizing the symbols for each user. Each AP then receives the symbols and designs precoding vectors for each user, making precoding-after-compress suitable for distributed precoding.

As another option inspired by CoF in the uplink transmission, the CPU can send linear combinations of the users' codewords to the RUs using structured lattice codes. This strategy is called RCoF \cite{6283033}.
Furthermore, a variant called RQCoF \cite{6522158} can be utilized, where the linear combinations of codewords are quantized at the CPU before being transmitted to the RUs. 

In general, when fronthaul capacities are moderate to high, the compression-based strategy consistently outperforms data-sharing, RCoF, and RQCoF strategies, as it achieves low compression distortion. However, in scenarios with small fronthaul capacities, the data-sharing, RCoF, and RQCoF schemes are more practical and offer better performance compared to compression-based strategies, which suffer from high compression distortion \cite{7362826, 6283033}.

To achieve maximum end-to-end performance, it is necessary to optimize both the uplink and downlink schemes. While the exact characterization of the downlink DAS capacity is still an open problem, most research studies have focused on separately or jointly optimizing the schemes using uplink-downlink duality to achieve downlink rates that are greater than or equal to the uplink rates \cite{7541570, 8635883, 9893884, 9687122,miretti2023uldl}. A Comparison of the downlink coding strategies is illustrated in Table \ref{downlink_coding}.
\begin{table*}[t]

	\caption{Comparison of the downlink coding strategies} 
	\centering
	\begin{tabular}{P{0.1\linewidth}P{0.25\linewidth}P{0.25\linewidth}P{0.2\linewidth}}
		\midrule		
		\hline
		\textbf{Scheme} & \textbf{Processing at CPU} & \textbf{Processing at RU}  & \textbf{References}  \\ 
		\hline
		Data Sharing  & Sharing users' messages with RUs & Broadcast the codewords & \cite{5997324,6920005,8359129,8732995}\\
		\hline
		Independent Compression & Encode, beamform, and compress users' signals & Decompress the compressed signals independently & \cite{quek_peng_simeone_yu_2017,6588350}\\
		\hline
		Multivariate Compression  & Jointly compress users' signals & Jointly decompress the compressed signals & \cite{6588350} \\\hline
		CAP  & Precoding design and compression & Decompress the compressed signals & \cite{8690797,8902784,8678745} \\
		\hline
		Precoding-After-Compress  &  Simple compression of users' signals & Precoding design and Decompression & \cite{10.1007/s11277-018-6038-1}  \\
		\hline
		RCoF  & Compute integer linear combinations of the codewords & Recover the codewords & \cite{6283033} \\ 
		\hline
		RQCoF  & Compute integer linear combinations of the codewords, then compress these combinations & Decompress the compressed equations and recover the codewords &  \cite{6522158} \\
		\hline	
	\end{tabular}
    \label{downlink_coding}
\end{table*}

\subsection{Lessons Learned}
This section illustrates  different uplink and downlink relaying strategies in DAS that aim to optimize network performance while considering the limited fronthaul capacity constraints. The fronthaul links connecting the distributed RUs with the CPU pose a limitation on the amount of traffic exchanged between them.
For the uplink transmission from users to the CPU, various relaying schemes have been proposed including  DF, AF, CoF, CF, and CCoF. Each strategy has its advantages and disadvantages in terms of performance, computational complexity, and sensitivity to channel conditions.
In the downlink transmission from the CPU to users, cooperative relaying schemes are used. Techniques like data-sharing, compression-based strategies, RCoF, and RQCoF have been studied. These techniques leverage the distributed RUs to improve network performance by relaying data from the CPU to users. The optimal relaying strategy is still an active area of research.  The optimal relaying strategy for DAS architectures is still an open research question. Various optimization methods and frameworks have been proposed to find appropriate solutions based on the specific requirements and constraints of the network. These methods involve the optimization of coding schemes, power allocation, compression levels, and decoding strategies.

Generally, the choice of relaying strategy depends on the specific DAS architecture, fronthaul capacity constraints, and desired network performance. By studying and optimizing these strategies, valuable insights have been gained on how to effectively utilize limited fronthaul capacity in DAS and enhance network performance.

\section{Techniques to Optimize DAS Networks}\label{Tec}
In this section, we review some optimization techniques that aim to improve the performance of DAS networks in terms of data rate, energy efficiency, coverage probability, and systems' fairness. Some of these optimization techniques include channel estimation, power control, resource allocation, and downlink/uplink beamforming. Then we investigate the physical layer operations and how these operations should be optimized.  In addition, we discuss the integration of DAS networks in emerging technologies such as RISs, optical wireless communications (OWCs), and non-terrestrial networks. 
\subsection{Channel Estimation Techniques}
One of the main operations that should be implemented at transmitters and receivers is channel estimation. Channel estimation is an important procedure to improve the uplink and downlink transmission performance. Acquiring accurate CSI helps in mitigating interference, reducing the bit error rate, and designing robust beamforming techniques.  There are channel estimation schemes that are designed for time division duplex (TDD) systems and others that are designed for frequency division duplex systems. Since the channels in the frequency division duplex system are not reciprocal, it needs a much larger overhead than the TDD scheme \cite{6449246}. Thus, we focus more on TDD scheme since the frequency division duplex system is not common in DAS networks. In TDD, each coherence block is divided into three different phases: A channel estimation phase, a downlink data transmission phase, and an uplink data transmission phase \cite{R21}. In the channel estimation phase, the users send uplink pilot signals to the APs, which estimate the channel of each user from these pilots. Then the CSI parameters are transmitted to the cloud for detecting the received signals and beamforming the transmitted signals exploiting the channels reciprocity \cite{R6}. In particular, the cloud exploits the fact that the uplink channel is the same as the downlink channel within each coherence block. The TDD channel estimation overhead scales linearly with increasing the number of users and is independent of the number of APs \cite{8387197}. 

At the user end, the users may rely on the channel hardening when decoding their received signals, or estimate the channel from the received signals to avoid downlink pilot overhead \cite{R6}.   However, in DAS networks, the authors of \cite{8379438, 8799031} showed that  CF-mMIMO provides a low degree of channel hardening compared to the co-located massive MIMO systems. They showed that the downlink channel is better to be estimated either using pilot transmission at the expense of extra overhead or using the received signal.  The authors of \cite{8799031} proposed that instead of estimating the individual channel of each AP at the users, the APs can send the same pilots to the users (i.e., the user $i$ will receive the same pilot signals from all the APs). Therefore, the users can estimate the accumulated channel from all APs. 

One main issue in DASs with massive MIMO systems is that the channel estimation operations should be selected to allow system scalability. Scaling the system to a larger system can be implemented by proposing techniques that reduce the fronthaul overhead as the number of users increases, proposing simple beamforming and power allocation techniques, and reducing the exchanged data among APs. One way to reduce the exchanged CSI among APs is to propose beamforming approaches based on local CSI estimation\footnote{Local CSI of AP $i$ refers to the channels between the APs $i$ and all users (i.e. each AP does not know the channels between the other APs and users).}. 

To accurately estimate the local CSI, orthogonal pilots should be assigned for all users. However, if the number of users is high or the channel coherence interval is small, it is roughly impossible to assign a unique orthogonal pilot to each user. Therefore, pilot reuse or non-orthogonal pilots are used for channel estimation, which causes a phenomenon called pilot contamination. Users that use the same pilot or non-orthogonal pilots experience mutual interference. In particular, pilot contamination leads to increasing the channel estimation error and the correlation of the channel estimates of the users that use non-orthogonal pilots \cite{R20}. To mitigate the effect of pilot contamination, the pilots should be allocated to the users efficiently. 

The authors of \cite{6812159} characterized the low-rankness of covariance matrices of the users' channels for improving the channel estimation using pilot decontamination and mitigating the interference using spatial filtering. The authors of \cite{R3}, \cite{R5} compared the greedy pilot assignment with the random assignment approach. In the random pilot assignment approach, each user is assigned a random pilot, but there is a probability that two nearby users are assigned to the same pilot, which leads to high mutual interference. In the greedy approach, the pilots are first distributed to the users randomly, and then the user with the lowest achievable rate updates its pilot sequence in a way to minimize the effect of the pilot contamination. The authors of \cite{8780756} proposed a location-based greedy pilot assignment approach exploiting the location information in assigning pilots rather than using the random assignment as an initial step for the greedy approach.
The authors of \cite{R25, R26} proposed a structured pilot assignment, where each two adjacent users are guaranteed to use different pilot sequences to ensure marginal pilot contamination. In \cite{8914726}, the authors proposed a pilot assignment approach based on a tabu-search method to reduce the high pilot contamination effect.  Under the user-centric user association approach, the authors of \cite{7080877} proposed to assign the pilots to the users such that the users that are served by the same BS in their cluster must be assigned with orthogonal pilots. The pilots were allocated based on a graph coloring algorithm such as the
Dsatur algorithm. 

To minimize the channel estimation overhead in ultra-dense C-RANs, the authors of \cite{8247283} proposed to estimate only the intra-cluster CSIs and depend only on the large-scale channel information of the inter-cluster CSI. The authors also applied the Dsatur algorithm to find the minimum number of required pilots that can be assigned to all the users in the system.
Another way to reduce the channel estimation overhead is to consider non-orthogonal pilot allocation. The study in \cite{9154310} formulated a biconvex optimization problem to jointly estimate the channels and detect the data symbols. The problem was solved using the forward-backward splitting approach and their proposed scheme reduced the channel estimation overhead by more than 2 times compared to  the orthogonal pilot sequence-based methods. In \cite{9201112}, the authors proposed to estimate the channels via a superimposed technique, where data symbols and pilot signals are transmitted simultaneously within the coherence intervals and both are correlated to a certain level. The authors showed that the proposed superimposed pilot transmission outperforms the traditional pilot transmission in terms of normalized mean-squared error.

To decrease the computational complexity of the CSI acquisition,  deep-learning-based approaches have been introduced to estimate and/or predict the required CSI, and they proved an observable improvement compared to traditional approaches. The authors of \cite{8815888} employed denoising convolutional neural networks to provide fast and accurate channel estimates in cell-free mmWave massive MIMO systems. 
 To exploit the temporal correlation over the coherence intervals, the authors of \cite{obeed2022alternating} developed an alternating channel estimation and prediction scheme in order to reduce the channel estimation overhead and improve the system throughput. The considered scheme in \cite{obeed2022alternating} proposed that each user sends a pilot signal in a given number of coherence intervals, where minimum mean-squared error (MMSE) was used to estimate the channels of this coherence interval and a deep neural network was used to predict the channels of the remaining coherence intervals. In this way, the overhead is reduced by at least $50\%$, which leads to improving the system throughput.
 
\subsection{Transmit Precoding and Receive Combining Optimization}
In DASs, precoding design at transmitters and combining design at receivers are required to maximize achievable rates, energy efficiency, minimum rates, or secrecy rates. For downlink transmission, the distributed APs first estimate the users' channels, then use them to simultaneously beamform the signals to the users using the same time and frequency resources. Whereas, in uplink transmission, all the users transmit their signals to the distributed APs, which in turn use combining techniques to achieve the desired QoS (e.g., sum-rate). The beamforming and combining techniques can be classified in the literature into three techniques: 1) fully centralized, 2) fully distributed, and 3) partially distributed. 

In fully centralized techniques, the detection and beamforming techniques are implemented at the CPU, which means that the channel estimates of all users must be collected at the CPU/cloud. The main advantage of centralized implementation is that the best beamforming/combining techniques can be employed due to the availability of global CSI at a central entity. 

For uplink transmission, several papers considered centralized detection schemes, e.g.,  \cite{R45, 7869024, 7465790}. The authors of \cite{R45} proposed to use the  MMSE-combining as a scheme that utilizes the global CSI to improve detection performance\footnote{The MMSE combiner is a vector that is designed to minimize the mean squared error between the actual and the estimated data signals given the estimated channels \cite{R21}.}.  They also showed that using the optimal MMSE processing in a centralized implementation improves the spectral efficiency and reduces the fronthaul overhead. 
The authors of \cite{7869024} proposed to use MMSE-combining to maximize the received signal-to-interference and noise ratio of each user. 
The authors in \cite{7465790} proposed to use a weighted-MMSE-combining with successive convex approximation to design the transmit beamformers and the quantization noise covariance matrices to maximize the weighted sum-rate of multi-antennas users at multi-antennas APs in a C-RAN system. 
Note that these centralized precoding or detection schemes require global CSI availability at the CPU, which limits their scalability, especially in CF-mMIMO systems.   For scalable systems, low-complexity precoding approaches that can be implemented by knowing local CSI only are more practical \cite{R20}.

To reduce the MMSE combiner complexity, the authors of \cite{7869024}, \cite{R20} proposed the partial-MMSE combiner that considers only the channel estimates of the users that are served partially by the same APs. Partial-MMSE reduces the complexity and number of the required channel estimates to calculate the beamforming vector at the expense of a slight increase in the received interference. To guarantee scalable cell-free systems, the authors proposed a simpler precoding approach, called local-PMMSE, where each AP only uses the channel estimates of the connected users to design the beamformer for each user. In local-PMMSE, there is no need to forward the channel estimates to the CPU since each AP can design the beamformer (combiner) locally, and this alleviates a burden on the fronthaul links \cite{R20}.      
 Based on a given serving APs for each user, the authors of \cite{R20} showed that the partial-MMSE approach outperforms the local-PMMSE, which in turn outperforms the conjugate beamforming in terms of achievable rate performance. Using only the large-scale fading  information to design beamformers and combiners is desirable in real-life systems due to its low computational complexity and low fronthaul signaling \cite{ 7869024, 9739189}.   Based on large-scale fading decoding, the authors of \cite{9739189} proposed to split the uplink decoding process between the CPU and the APs to trade-off between the fronthaul signaling and the spectral efficiency.
 
 For downlink transmission, the authors of \cite{1683918} showed that the optimal precoding approach in multiuser MIMO systems is dirty paper coding. However, due to its high complexity, dirty paper coding is not a practical solution in CF-mMIMO systems. Hence, linear centralized precoding approaches are proposed and investigated in CF-mMIMO systems \cite{R12,R34}. The authors of \cite{R34} used maximum ratio transmission and zero-forcing precoding approaches to mitigate the interference at the users. They showed that the zero-forcing beamforming approach significantly outperforms the maximum ratio transmission approach. In the zero-forcing scheme, the precoding vector of each user is designed to be orthogonal to the channels of all other users. Different from the distributed precoding, the authors of \cite{R12} showed that the average rate of users increases unboundedly as the numbers of antennas and users increase if a centralized precoding scheme is used (e.g., zero-forcing). A more advanced precoding scheme was proposed in \cite{R20}, which is the MMSE precoding. The MMSE approach optimizes the combining weights at each AP to minimize the mean square error between the combined signal and the desired signal. It takes into account the channel conditions, interference, and noise statistics to improve the overall received signal quality. In \cite{8579566}, the authors designed beamforming vectors to minimize the total consumed power subject to an achievable rate constraint when the CSI of the users is imperfectly known. The authors used an iterative algorithm and Lagrange dual decomposition method to find an optimal solution for the downlink beamforming vectors. In \cite{8660693}, the authors jointly optimized the downlink transmission for hybrid analog-digital antenna arrays in C-RAN systems. In particular, the authors used a weighted-MMSE approach and block coordinate descent to jointly design the RF beamforming, the digital beamforming, and the fronthaul compression to maximize the weighted sum-rate.

Due to its simplicity and distributed nature, the authors of \cite{R3}, \cite{R5}, \cite{R6} proposed to use the conjugate beamforming approach. The authors of \cite{R34} compared the performance of the conjugate beamforming approach with a low-complexity zero-forcing beamforming approach. At the expense of fronthaul load, authors of  \cite{R34} showed that the proposed low-complexity zero-forcing approach outperforms the conjugate beamforming significantly.  The authors of \cite{8599043} proposed a modified conjugate beamforming scheme that eliminates self-interference completely, while preserving the simplicity of the conjugate beamforming approach, at the expense of exchanging  limited channel information among APs. To avoid the need for information exchange among APs, the authors of \cite{9340389} proposed an enhanced conjugate beamforming technique, where the beamforming vector is normalized with respect to the channel estimate. The aim of using this technique is to boost the channel hardening at the users which improves the downlink spectral efficiency.  The authors of \cite{9108198} designed an optimal beamformer using a bisection algorithm to maximize the signal-to-interference and noise ratio among all users.
The authors of \cite{atzeni2020distributed} proposed a distributed cooperative precoding that overcomes the need for fronthaul transmission of CSI exchange. They achieve that by engaging in an over-the-air signaling approach.  This approach guarantees that the total signaling amount does not scale up with the number of users or the APs. 
Four coordination levels among APs were investigated in \cite{R45} to provide a thorough analysis of uplink CF-mMIMO systems.  The authors concluded that CF-mMIMO can significantly outperform small cells and cellular massive MIMO by using local or global MMSE combining.

\subsection{Lessons Learned}
Although spreading antennas over areas reduces latency and improves data rates, coverage, and fairness, it presents several challenges such as deployment cost, power consumption, fronthaul issues, signal processing complexity, etc. To mitigate the effect of these challenges, optimizing DAS operations is required. 

Scaling DAS over wide areas increases the number of served users. This increases the channel estimation overhead or makes the pilot contamination more severe. Therefore, intelligent channel estimation methods are required to manage pilot contamination and make DAS more scalable. Most of the recent works argued that deep learning-based channel estimation methods are less complex and provide more accurate estimates compared to traditional methods. This is due to the fact that deep learning-based methods fuse spectral, temporal, and spatial information to provide a deeper understanding of the channel change over spectrum, time, and space. However, due to the explosive increase in the number of users and the 6G promises to provide a decent QoS for very high-mobility users, the existing approaches to reduce pilot contamination or channel estimation overhead may not be satisfactory to meet the expected requirements.

Since DASs promise to spread a massive number of distributed APs, interference must be managed efficiently to avoid degrading the rates of the end users. The common way to manage the interference is to propose robust transmit precoding and receive combining approaches. However, this requires the DAS to be more centralized so the interference among APs can be mitigated. For instance, adopting the conjugate beamforming approach for transmission would produce significant interference at the users, while the zero-forcing or the MMSE precoding approaches (which require CSI sharing among APs) would significantly reduce the interference.   The existing literature indicated that the user-centric design is the solution to balance between managing the interference and making DASs more scalable.  In such a design, the range of the serving APs for each user must be addressed carefully to manage this trade-off efficiently.

\section{Integration of DAS with Other Communication Systems}\label{Integ}
In this subsection, we review existing works that considered the DAS in different communication systems. We focus on the integration of DAS with  RIS-assisted communication systems, OWC systems, and satellite networks. 

\subsection{DAS and RIS Integration} 
\label{Sec:RIS-DAS}
An RIS is a two-dimensional surface consisting of several reflecting elements whose phases and amplitudes can be digitally-controlled to reflect the incident signals in desirable directions (i.e., change the phase of the incident signals) \cite{alouiniR26}. The main advantages of these surfaces are: i) ease of deployment, ii) low-cost, iii) passive nature, and iv) low-energy consumption \cite{9756313}. 
In contrast, each antenna in DASs must be associated with an RF chain and digital circuits to implement signal processing operations leading to a higher cost and energy consumption compared to a reflecting element. Despite this advantage of RIS elements compared to antennas, their contribution to the system performance is low compared to an antenna due to the double fading effect. Nonetheless, the deployment of multiple RISs can support DASs through reducing the number of APs, thus improving the energy efficiency of the system.

Several works studied the roles that RISs can play in DASs  \cite{10431714, 9473521, 9298890, 9352948, 9448858, 9665300, 9911685}. The authors of \cite{10431714, 9473521} showed that deploying RISs in cell-free networks helps in extending the system coverage and maximizing the minimum users achievable rates. The authors showed that the RIS provides nearly a 2-fold increase in the minimum
rate and a 1.5-fold increase in the per-user rate. The authors of \cite{9298890} formulated and solved the problem of maximizing the weighted sum-rate via optimizing the beamforming matrices at the APs and the RISs of an RIS-aided cell-free system.  The study in \cite{9352948} proposed to aid the CF-mMIMO systems with multiple RISs to improve the energy efficiency of the system. The problem in \cite{9352948} was formulated to maximize energy efficiency by optimizing the beamforming matrices at the BSs and RISs. It is concluded that the energy efficiency increases with increasing the number of RISs and their sizes to some point where it starts to decrease \cite{9298890,9352948}. This decrease is due to the increase of the hardware static
power consumption of each RIS with the number of
elements in each RIS.  The authors of \cite{9448858} showed that the energy efficiency of the worst user in a wideband cell-free system can be improved by employing multiple RISs. One of the practical challenges in RIS and DASs is the channel estimation overhead and the spatial correlation among RIS's reflecting elements. Hence, the authors of \cite{9665300} investigated the effect of the channels' spatial correlation and proposed a simple channel estimation scheme that takes into account the overhead. The authors also optimized the phase shifts to minimize the channel estimation error and derived a closed form for the ergodic downlink and uplink net throughputs. They concluded that the contribution of RIS to cell-free systems is high when the AP-user links are weak with a high probability. The authors of \cite{10640072} proposed a reflection pattern modulation-aided RIS-assisted CF-mMIMO system for energy-efficient uplink transmission. They introduced a system that optimizes the energy versus spectral efficiency trade-off by activating only part of the RIS blocks.

RISs can also be used to improve the secrecy rate of different communication systems \cite{8723525, 9206080}. The authors of \cite{9911685} investigated the role of RISs in improving the secrecy rate of a cell-free system. The authors utilized semi-definite relaxation and successive convex approximation to jointly optimize the beamforming of the APs and RISs in order to maximize the weighted secrecy rate of the system. Recently, RIS was integrated with terrestrial distributed networks to improve different metrics, such as network connectivity \cite{alabiad2023effectivenessreconfigurableintelligentsurfaces, 10437661} and energy efficiency \cite{10437618}. 

\subsection{DAS in OWC Systems}
RF communication and OWC  use different spectrums to transmit messages. Each has its own advantages and disadvantages.
On the one hand, OWC provides high data rates, but has some limitations such as the blockage effects and the significant channel attenuation with distance. On the other hand, RF can provide ubiquitous connectivity but with lower data rates. Thus OWC and RF communication systems can complement each other by mitigating each other's drawbacks.  Integrating these systems can lead to reducing the received interference (RF spectrum doesn't interfere with the optical spectrum), improving the achievable data rates, and improving the system’s fairness. The authors of \cite{8113550, 8779676} investigated a visible-light communication (VLC)  CF-mMIMO system and studied the problem of user association. The authors in \cite{8574984} studied a VLC DAS, where the APs and the users are assigned to each other based on a user-centric approach aiming at maximizing energy efficiency. 

The authors of \cite{liu2022joint} suggested employing the cell-free concept to reduce the potentially excessive handover in VLC systems which is caused by their small cell size. The authors proposed a joint AP grouping and user clustering to manage the interference. The authors of \cite{almehdhar2022hybrid} and \cite{almehdhar2022user} introduced the cell-free concept into indoor hybrid VLC/RF networks. However, the studies in \cite{almehdhar2022hybrid, almehdhar2022user} only focused on associating the users to OWC and RF networks without considering the problems of power allocation, beamforming design, resource allocation, channel estimation, and clustering formation.
\subsection{DAS and Non-Terrestrial Networks}
One limitation of terrestrial networks is their limited ability to provide ubiquitous connectivity to rural areas. Although terrestrial DAS have been proposed to improve wireless coverage, their deployment cost to cover wide areas is prohibitive. In addition, terrestrial networks in general fail to provide services when natural disasters happen. Therefore, researchers increased their focus on studying non-terrestrial networks and how they integrate with terrestrial networks. In the integrated terrestrial and non-terrestrial networks,   a 3-dimensional (3D)  network architecture (consisting of terrestrial APs and aerial APs) is dedicated to enhancing the coverage and providing connectivity to unconnected areas  \cite{giordani2020non}. In particular, satellites, high-altitude platform systems, and aerial unmanned vehicles (UAVs) are envisioned to integrate with terrestrial networks to meet the 6G stringent requirement of ubiquitous connectivity \cite{giordani2020non}.

The authors of \cite{riera2022scalable} proposed to use low-earth orbit satellites to support terrestrial DAS in order to maximize the minimum per-user rate and the sum-rate of the integrated system. The approach in  \cite{riera2022scalable} transfers the users that limit the terrestrial DAS performance to be served by the satellite segment. Due to the long distance between the satellites and the handheld devices, the link attenuation is high and direct-to-satellite (i.e., direct from a handheld device to a satellite without a ground station assistance)  communication is challenging. Therefore, the authors of \cite{abdelsadek2023broadband} proposed employing the DAS concept in satellite networks to improve the data rates at handheld devices. This can be enabled by the expected ultra-dense low-earth orbit satellite deployment and the interconnection between the satellites through optical links. The authors of \cite{abdelsadek2021future} studied the problems of optimizing pilot assignment, duplexing mode, power allocation, beamforming, and handover management in satellite DAS. The authors of \cite{10622597} investigated a user-centric multi-satellite communication system aimed at maximizing the weighted sum rates of users through joint beamforming. A centralized algorithm was proposed, assuming perfect CSI, alongside a low-complexity distributed algorithm using multi-agent deep reinforcement learning for practical implementation. Aiming at maximizing the uplink spectral efficiency for the ground users, the authors of \cite{10634184} proposed a new CF-mMIMO architecture combining low Earth orbit satellites with UAVs as flying APs. The authors derived expressions for direct and cascaded channels, proposed local combining strategies, and introduced a dynamic clustering framework to reduce interference and enhance user service. 

Although a few papers studied the integration of satellite and terrestrial networks using the DAS concept, many problems are untapped and need to be tackled in future works. Examples of these problems are studying the Doppler effect,  latency,  synchronization,  user association, backhaul issues, and CPU selection. In Section \ref{Sec:Integration}, we discuss these issues in detail.  In the next section, we discuss cross-layer optimization techniques that consider both the physical layer and network layer viewpoints of DASs.

\section{Cross-Layer Optimization for Resource Allocation in DASs}\label{Cross}
Resource allocation via cross-layer optimization is crucial to achieve high performance in DASs, e.g., C-RANs, F-RANs, and D2D-aided F-RANs. Cross-layer optimization (i.e., physical and network layers joint optimization) can target different QoS metrics, including  throughput/sum-rate maximization, cloud offloading maximization, and delay minimization under a QoS constraint. This is usually presented through maximizing a wide-network utility or minimizing a cost function that considers user scheduling using NC and power allocation with some constraints.

The cross-layer performance of the aforementioned systems  is constrained by the scarcity of radio resources and the limited capacity of fronthaul links. To efficiently utilize the available  RBs in such systems, traditional schemes consider assigning a single user to each RB on each RRH, thus transmitting only one content from that RB/RRH. However, this single user assignment becomes highly challenging  with the tremendous increase in the number of users nowadays. Consequently, NC \cite{NC1} is employed to mix users' data (packets)  and
simultaneously allocate different users to the same RB. To assess the merits of NC, consider a network composed of one transmitter and two users each requesting one packet and already has obtained the other packet. Without NC (i.e., uncoded), the transmitter needs two transmissions to complete the reception of all packets at both users. However, NC reduces the number to a single transmission by encoding both packets. The   requested and received packets of users are known in the literature as side information \cite{NC1}. By combining this side information  using NC, multiple users can be served simultaneously on the same RB/RRH. Thus, significant improvements can be achieved in transmission efficiency, content delivery, throughput, and delay in DAS. This section discusses the optimization of  DASs  from a cross-layer perspective using NC.

\subsection{Network Coding (NC)}
Since 2000, NC, initiated by Ahlswede \textit{et al.} \cite{NC1}, has drawn researchers' attention as a method for  improving content delivery and achieving maximum information flow in different communication networks, e.g., \cite{NC2, NC3, NC4}. NC maximizes information flow by combining users' data at the source and intermediate nodes in a network \cite{NC1} (in the network layer) and sending the combination instead of sending the raw data as in the simple RF (i.e., uncoded). For a system of many users and packets, the key potential of NC is to ensure partial or  full reception of the packets at multiple users, while reducing data traffic by sending combined packets. This enables maximizing the throughput and minimizing the communication delay. 
Based on the application, two NC design methods are distinguished in the literature, random linear NC (RLNC) and instantly decodable NC (IDNC). 

In RLNC, at each transmission, the sender broadcasts a coded packet obtained by combining all source packets using random and independent coefficients generated from a given Galois Field   \cite{RLNC_1,RLNC_2, RLNC_3, RLNC_4,  4557282}. RLNC has drawn much interest of the research community due to its (i) ability to recover packets without feedback and  (ii) optimality in reducing the number of packet transmissions in broadcast scenarios. It was initially used in point-to-multipoint models and then expanded in different systems, including  multipoint-to-multipoint \cite{RLNC_1} and D2D communications \cite{9036050}. However, it is only suitable for applications that tolerate high delay as it does not support progressive and instant packet decoding. Moreover, it is inefficient in multicast scenarios that different groups of users are interested in different subsets of packets. From a complexity point of view, the  computational complexity of decoding in RLNC scales cubically with the number of packets \cite{RLNC_2}, which is prohibitive for  battery-powered devices.

On the other hand, IDNC is a good candidate to overcome the aforementioned issues of RLNC, especially for applications that require real-time optimization, such as video streaming. It relies on progressive and instantaneous decoding, which means that the received coded packets are decoded instantly, and the retrieved packets are used immediately at the application layer. IDNC exploits users' side information in the network to enhance performance. In IDNC, the instantaneous decoding of the requested packets is achieved by simple encoding and decoding operations. The transmitter combines packets using binary XOR, which can efficiently be implemented on a device. At the user side, packets are decoded at their reception instant using binary XOR. Due to its simple encoding/decoding operation and instant decodability, IDNC has shown excellent abilities to substantially improve transmission efficiency, packet recovery, throughput and delay over broadcast erasure channels \cite{NC1, NC2, NC3, NC4, 7845689, 8794557, 9512273, 9180294, 8267199, 8444481, 6882208, 1176612, 4557282, 1705002, 4675714, 6874566, 9982443, 10229924}. Such simple features led to tremendous progress in the analysis and the development of simple online algorithm design for IDNC that can be represented as a graph model, known as IDNC graph. To represent all packet combinations that are instantly decodable by a subset or all users, the authors in \cite{5683677}  originally proposed the
IDNC graph.

\begin{figure}[t!]
	\begin{center}
		\includegraphics[width=0.35\textwidth]{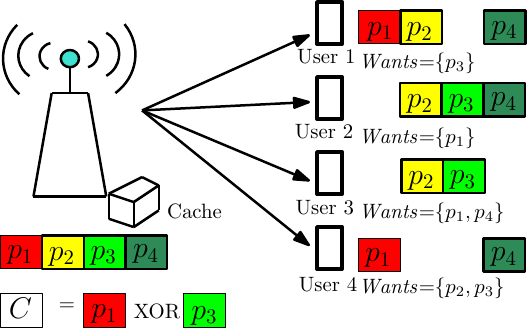}
		\caption{A simple C-RAN model containing $4$ users, $4$ packets, and $1$ RRH. The side information of users is also shown, where each user has received some packets and wants some other packets. For example, user $1$ already received packets $p_1$, $p_2$, $p_4$ and	requests $p_3$.}
		\label{fig1_NC} 
	\end{center}
\end{figure}

\begin{figure}[t!]
	\begin{center}
		\includegraphics[width=0.45\textwidth]{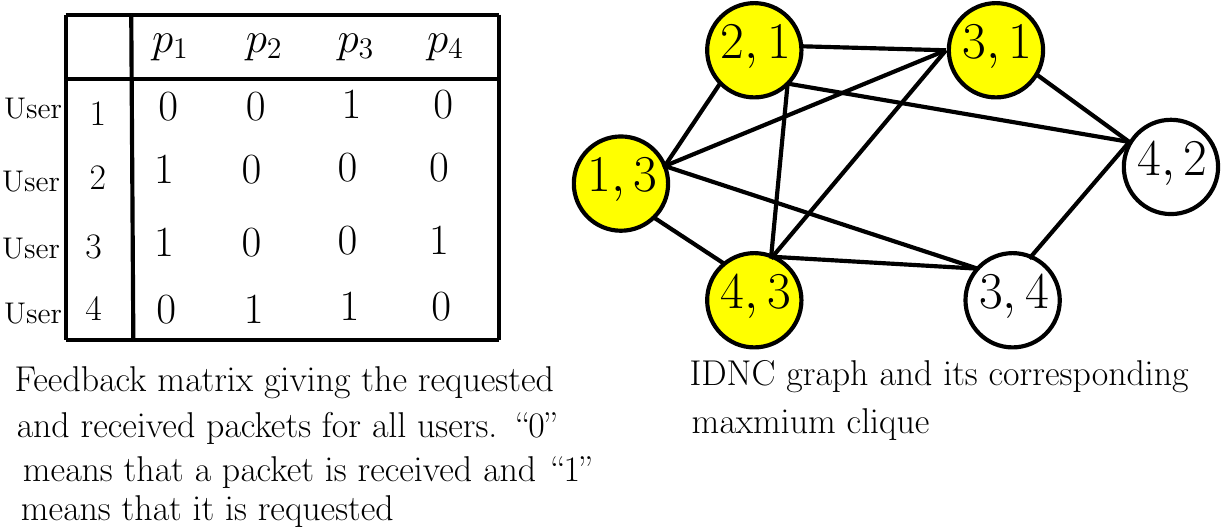}
		\caption{IDNC graph representing the coding possibilities for the
system depicted in Fig. \ref{fig1_NC}. Each vertex represents a requested packet by one user, i.e., vertex $(2, 1)$ represents the indices of user	 $2$ and packet $1$. Each clique in the IDNC graph represents a feasible IDNC combination. For example, one possible clique is: $\{(1, 3), (2, 1), (3, 1), (4, 3)\}$			that corresponds to transmitting the combination $1\oplus 3$ to the targeted users $1$, $2$, $3$, and $4$.}
		\label{fig2_NC} 
	\end{center}
\end{figure}

Fig. \ref{fig2_NC} illustrates the construction of the IDNC graph in a simple C-RAN model with one RRH that is given in Fig. \ref{fig1_NC}. The arbitrary users' side information is shown on the right side of Fig. \ref{fig1_NC}. Based on users' side information, the corresponding IDNC graph is constructed by first generating a vertex for each user and each packet, e.g., packet $p_3$ that is requested by user $1$ is represented by a vertex $(1, 3)$. The fundamental idea of constructing the IDNC graph is to find a feasible packet combination with this feature: two packets are wanted by two users are combinable if these two users can both benefit from the combined packet, making it instantly decodable for both users. This situation happens if and only if the two users are requesting the same packet or if the packet wanted by each user is already received at the other user.
The goal of IDNC is to satisfy many users with one combined packet. For example, the RRH transmits the combination $p_1\oplus p_3$ to  users $1, 2, 3$, $4$. In general, finding the optimal packet combination on such graphs that provides the highest objective function to a given network's metric requires an exhaustive search over all possible packet combinations, which is clearly non-feasible for large network sizes \cite{7845689}. Without loss of generality, solving the IDNC-related problems (e.g., completion time and decoding delay that are extensively investigated in \cite{6882208, 8444481, 9180294, 8794557, 9512273, 9391701}) is equivalent to solving the maximum clique problem \cite{5683677}, which is nondeterministic polynomial-hard \cite{10.5555/574848}. A detailed analysis of the developed algorithms, complexity, and applications of maximum clique problems can be found in \cite{9044331}.

\subsection{Cross-Layer Network Coding}

IDNC optimizes different performance metrics of DAS (e.g., completion time and decoding delay) by combining users' data in the network layer and sending the combination to multiple users. However, it abstracts the physical layer conditions to packet erasures \cite{8794557, 9512273, 9391701}. Thus, IDNC ignores the channel heterogeneous capacities of different users in the selection of a combined packet. Instead, the transmitter of an IDNC combination selects the minimum achievable capacity of all the associated users in the combination. This results in serving a large number of users but unfortunately leads to physical layer throughput degradation, which degrades the physical-layer performance of DAS. Studying the trade-off between the performances of the network layer and the physical layer from a content delivery perspective in  different system models of DAS has been a topic of research interest for years \cite{7511238, 7864466, 6662474, 5072351, 5424026,5462127, 4567144, 6692154, 9519548}.  On the one hand, transmit a combined IDNC packet with a high rate will take a shorter time but will be only successfully received by users with high channel capacities. On the other hand, transmit a combined IDNC packet with a lower rate will be received by more users but will take a longer delivery time. To improve the physical layer performance and ensure the QoS of users of DAS, the authors of \cite{7511238, 7864466, 6662474, 5072351, 5424026,5462127, 4567144, 6692154, 9982443} included the channel capacity as another factor in the selection of IDNC packets. As a result, the IDNC packet selection depends on the side information of the users and their channel capacities.

This improved version of IDNC is called rate-aware IDNC (RA-IDNC), which was first studied in \cite{RAIDNC1, 7511238, 6662474} and applied in C-RAN \cite{RAIDNC5} and point-to-multipoint system \cite{6692154}. RA-IDNC has the potential to serve a significant set of users with relatively good transmission rates. Due to this feature, RA-IDNC has been considered in a number of works, e.g., \cite{RAIDNC1, 5072351,5462127, 4567144, 6692154, RAIDNC2, RAIDNC3, 5424026, RAIDNC4} to enhance DAS. For example, the authors of \cite{RAIDNC2}, \cite{RAIDNC5}  proposed RA-IDNC scheme to minimize the completion time in C-RAN systems. However, the authors of \cite{RAIDNC5}   assumed that each RRH maintains a fixed power and that the transmission rate of all RRHs is determined by the weakest RRH, i.e., the RRH that supports the lowest transmission rate. Since this reduces the QoS for some users, the authors of \cite{CLNC1} introduced CLNC to optimize the employed rates in RA-IDNC using power control for each RRH in
a C-RAN. In CLNC, combining users' requests depends on NC, users’ rates, and the
transmit power of each RRH, which can be jointly optimized to improve the  sum-rate \cite{CLNC1, CLNC2,  CLNC5a} in C-RANs, cloud offloading \cite{CLNC3, CLNC4,  CLNC5} in F-RANs, and delay \cite{CLNC6, 9791405} in a F-RAN aided D2D system.

The  key potential of CLNC is that it balances between cross-layer throughput and network delay with negligible decoding complexity, which ensures a better QoS for real-time streaming applications in DAS. In such applications, users need to stream data from RRHs in DAS (e.g., C-RAN, F-RAN) with the minimum possible delay. Consider that a popular video representing a frame of packets is requested by a set of users located in a playground, where many users want to stream this popular video. At any given time, assume that users have already received some packets and missed some other packets due to channel impairments. To stream that video without any interruption, users should receive their missed packets with minimum delay. One way is to pre-load that frame which can be done at low rates or at off-peak times, both of which are inefficient. CLNC performs a joint optimization of NC and transmit rate and power for efficient and fast file delivery, which meets the delay requirements and streaming quality of DAS. This makes CLNC a promising technique for significantly improving users' QoS, thanks to the following properties.

\begin{itemize}
	\item The immediate decodability of the received packets makes them useful at the reception time to meet the real-time streaming application requirements.
	
	\item Encoding at the transmitter is performed using a simple binary XOR operation. Likewise, decoding at the users is performed using a simple binary XOR operation, which overcomes the high computational complexity of RLNC.
	
	\item Since CLNC depends on instant decodability, any received packets that are non-useful for the users are discarded. Thus, CLNC overcomes the need for buffers to store these non-useful packets, which enables the design of cost and energy-efficient devices.
	\item CLNC balances between the physical layer and network layer system's performance by adapting a transmission rate  and power allocation that supports the smallest channel
capacity of all network-coded scheduled users.
\end{itemize}

Table~\ref{table_1NC}
summarizes the performance of RLNC, IDNC, and CLNC according to various criteria.

\begin{table*}[t!]
	\renewcommand{\arraystretch}{0.9}
	\caption{Performance of RLNC, IDNC, CLNC According To Various
		Criteria.}
	\label{table_1NC}
	\centering
	\begin{tabular}{|p{2.1cm}|| p{3.5cm}| p{3.7cm}| p{3.7cm}| p{2.5cm}|}
		
		\hline
		
		Criteria $\backslash$ Scheme & RLNC & IDNC & CLNC & References\\
		\hline
		\hline
		Throughput & Optimal & Sub-optimal & Sub-optimal/optimal& \cite{4068001, 4769397, 4675714, 6882208}\\
		\hline
		Delay  & Huge delay & Moderate & Moderate & \cite{4068001, 4675714, CLNC5}\\
		\hline
		Complexity  & Scales cubically & Polynomial & Exponential & \cite{4068001, 4675714, 6882208}\\
		\hline
		Encoding & Mix using random coefficients from GF & XOR operation & XOR  operation & \cite{4068001, 1705002, 8444481}\\
		\hline
		Decoding & After receiving the whole frame & Instantaneous decoding per file &  Instantaneous decoding per file & \cite{4557282, 9180294, 	CLNC3, CLNC5}\\
		\hline
		Buffer Size  & Equal to the frame's size & No buffer needed & No buffer needed & \cite{4557282, 9180294, 	CLNC3, CLNC5}\\
		\hline
		Overhead & Moderate & Minimal & Minimal & \cite{4068001, 9180294, CLNC6}\\
		\hline
		Layering & Network layer & Network layer & Physical and network layers & \cite{4068001,  9180294 , CLNC6}\\
		\hline
		Feedback Signals & Minimal & Performance heavily depends on feedback & Performance heavily depends on feedback & \cite{4068001, 4675714, CLNC4}\\
		\hline
		Methodology & File combinations & File combinations using users' side information & File combinations and rate/power adaptation & \cite{4068001, 9180294, CLNC6}\\
		\hline
	\end{tabular}
\end{table*}

\subsection{CLNC Optimization for DAS}
Cross-layer optimization of user scheduling and power allocation in DAS can be solved jointly or iteratively, see for example \cite{6588350, 8027131, 7342981, 6811617}. While the joint solution is expensive in terms of computational complexity, the classical iterative one is sub-optimal. Besides the joint and classic techniques to solve cross-layer optimization problems, graph theory methods have been widely considered in the recent literature  of C-RANs, F-RANs, D2D, and F-RAN-aided D2D, e.g., see \cite{RAIDNC1, RAIDNC2, RAIDNC3, RAIDNC4, RAIDNC5, CLNC1, CLNC2,CLNC3,CLNC4,CLNC6, CLNC5a, 9791405}. Graph theory
techniques establish a framework with an exact graph representation of scheduling decisions and objective transformation, and numerous theorems that facilitates the analysis and
optimization of a problem. The framework also offers  combinatorial
techniques that allow the quantitative solving of a
problem. In general, researchers utilize graph theory to solve a typical cross-layer optimization problem in DAS, and the solution relies on designing a graph, where its vertices represent the possible scheduling and power levels. The  vertices of the graph are connected via edges that satisfy problem's constraints while vertices' weights reflect the objective function. In addition to its application in DAS, graph theory has garnered significant interest in various settings, such as aerial-terrestrial integrated networks for federated learning \cite{10550002, 9844152, 9932020}, RISs \cite{10437661}, and D2D communications \cite{10061474, 10199618}. 

Unlike the previous works that used either IDNC graph or RA-IDNC graph, the CLNC graph also considers transmission rate and power allocation for multiplexing users. The CLNC graph is introduced in \cite{CLNC1} which takes user multiplexing using NC,  rate adaptation, and power allocation into consideration. 

An instantly decodable combination that satisfies the CLNC conditions is used to retrieve a new wanted packet by a user if and only if 
\begin{enumerate}
	\item The user can properly receive the combination with a rate that is below the capacity of the channel between a user and an RRH.
	\item The user can re-XOR the combination with its previously received packets to retrieve a new packet. 
\end{enumerate}

\begin{figure}[t!]
	\begin{center}
		\includegraphics[width=0.47\textwidth]{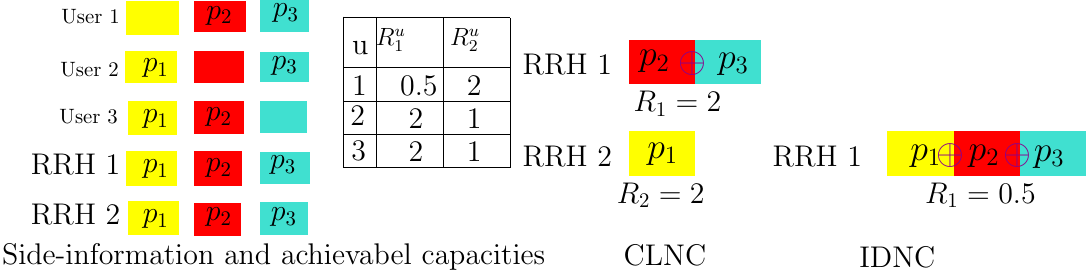}
		\caption{\label{fig3_NC} A simple C-RAN system composed of $3$ users, $3$ packets, and $2$ RRHs. Each user possesses $2$ packets and wants $1$ packet. For example, user $1$ possesses packets $2$, $3$, and wants packet $1$. The whole set of packets is stored by each RRH. Given a certain power level to each of the RRHs, assume that users' rates in bits/s are provided in a table on the middle of the
			figure. }
	\end{center}
\end{figure}

The main concept of CLNC for cross-layer optimization can be illustrated in the following example. Fig. \ref{fig3_NC} considers a simple C-RAN model that consists of $2$ RRHs, $3$ users, and $3$ packets. In order to maximize cross-layer throughput for this example, many possible solutions can be found as summarized in Table~\ref{table_2NC}. For example, the RRH $1$ XORes packets $2$ and $3$ into $2\oplus 3$ and which it transmits with a rate of $2$ bits/s to the set users $2$ and $3$. RRH $2$ transmits packet $1$ with a rate of $2$ bits/s to user $1$.  Given this, we have the following decoding process
at the users. User $2$ gets packet $2$ by XORing $2\oplus 3$ with packet $3$ and user $3$ gets packet $3$ by XORing packet $2\oplus 3$ with packet $2$. Therefore, the achievable overall throughput in this scenario is $6$ bits/s as each served user will simultaneously receive $2$ bits/s. Table~\ref{table_2NC} highlights that the uncoded scheme can only serve one user from each RRH with the maximum transmission rate and IDNC can schedule all the users in the network with the lowest transmission rate of all the scheduled users. CLNC strikes a balance between these two, thus it is more efficient in maximizing the cross-layer throughput and minimizing the completion time. Table~\ref{table_2NC} also depicts the average sum throughput maximization for the considered C-RAN and highlights that the CLNC significantly outperforms the uncoded and IDNC schemes.

\begin{table*}[t] 
	\renewcommand{\arraystretch}{0.9}
	\caption{Performance of Uncoded, IDNC, CLNC According To Various
		Criteria.}
	\label{table_2NC}
	\centering
	\begin{tabular}{|p{3.2cm}|| p{3.99cm}| p{3.99cm}| p{4.99cm}|}
		
		\hline
		
		Criteria $\backslash$ Scheme & Uncoded & IDNC & CLNC \\
		\hline
		\hline
		Solution & RRHs 1 and 2 transmit packets $2$ and $1$, respectively. & RRH $1$  XORes packets $1$, $2$, and $3$ into $1\oplus 2\oplus 3$. & RRH $1$ XORes packets $2$ and $3$ into $2\oplus 3$. RRH $2$ transmits  packet $1$ to user  $1$. 		\\
		\hline
		Number of scheduled users  & $2$ & $3$ & $3$\\
		\hline
		Adopted transmission rate  &  $2$ bits/s &  $0.5$ bits/s &  $2$ bits/s \\
		\hline
		Cross-layer throughput  & $4$ bits & $1.5$ bits & $6$ bits\\
		\hline
		Delay  & $1.5$ seconds & $2$ seconds & $1$ second \\
		\hline
	\end{tabular}
\end{table*}

With such potential features of CLNC graph optimization, researchers utilized graph theory to solve resource allocation problems in DAS, e.g., C-RAN, F-RAN, and D2D communications. For example, the authors of \cite{6811617, 7248768} utilized graph theory to solve a vanilla version of cross-layer optimization problem by considering only user scheduling and one packet to be transmitted on each RRB in C-RAN. Wang \textit{et al.} \cite{7864466} explored graph theory techniques to solve the completion time problem using cross-layer optimization with NC in point-to-multipoint systems.
The authors of \cite{7342981} proposed joint and iterative solutions for the optimization problem of user scheduling and power allocation in C-RAN using graph theory.   The works \cite{6692154, 5072351} addressed general and simple NC with a cross-layer design in the traditional network configuration with one BS. In particular, the authors of  \cite{6692154}  used cross-layer design to solve an optimization problem with the new formulation that the coding decisions are based on both the distribution of files at each user in the network and the instantaneous rate each user is experiencing. The authors of \cite{8288205, 6874566, 7511238} investigated the use of cross-layer graph design in different settings, e.g., multi-cached BS systems and cross-layer D2D communications.  For example, D2D systems are considered in \cite{6874566, 7511238}, where the authors proposed cross-layer design to speed up the recovery for a geographically close and fully and partially connected group of users. While all the previous works assumed perfect channel estimation for cross-layer optimization in DAS, reference \cite{6809217} extended the setting to imperfect channel and rate estimation. Recently, CLNC was used with machine learning to tackle the problem of packet placement and delivery in the F-RAN-aided D2D system \cite{CLNC5a}.

\subsection{Lessons Learned}
NC has shown excellent abilities in achieving maximum information flow in broadcast networks. Early research focused on developing NC strategies that attain maximum throughput by mixing all data packets for encoding and operations over Galois field for decoding. However, the decoding computational complexity of  these NC strategies is expensive. Thanks to its low complexity and manageability feature, instantaneous NC is a promising paradigm for cross-layer optimization in DAS.

There is a significant impact of scheduling many users to the RRHs   using NC on the physical layer performance. Scheduling many users to the same RRH improves the system's performance from a network layer perspective, but it degrades the system's performance from a physical layer perspective by selecting the minimum rates of all scheduled users. Therefore, this Section reveals that it is crucial to consider both NC at the network layer and the physical layer factors (i.e., transmission rate and power), to improve cross-layer system performances. In this regard, cross-layer optimization balances the system's performance from the network layer and physical layer perspectives. However, this requires (i) high overhead signaling since users need to update transmitting node with their side information status and (ii) computational complexity for algorithm optimization every time network changes, such as side information and network topology.

Although cooperation of transmitting nodes is a key enabler for delivering users' data in DAS, it requires careful optimization for user scheduling to different transmitting nodes, given that the performance of DAS is constrained by the scarcity of radio resources. CLNC, as a technique employed to effectively multiplex users to the same RB of the transmitters, supports such cooperation while overcoming the limitation of radio resources. This satisfies the tremendous increase in network densification. However, the main issue is the way of solving CLNC-related problems and the required computational complexity. While graph theory solutions show promising improved performances for DAS, designing the CLNC graph and finding the feasible CLNC solution is expensive in terms of algorithm execution time.

\section{Challenges and Research Directions}\label{Fut}

This subsection provides some future research directions that the authors believe are worth investigating. 

\subsection{Integration of DAS in 3D-Dynamic Wireless Networks}
\label{Sec:Integration}
Anticipated advancements in 6G wireless network technology are expected to result in a flexible and scalable network architecture that expands both horizontally and vertically, incorporating mobile BSs such as UAVs and satellites (which we refer to here as  3D-dynamic wireless networks). Furthermore, it is expected that the number of users will increase significantly, necessitating greater bandwidth and speed to accommodate this growth. These anticipated advancements introduce new challenges to incorporating the DAS architecture in 3D-dynamic wireless networks. In the following, we discuss the possible challenges and open problems in DAS-assisted 3D-dynamic wireless networks.
\begin{itemize}
\item \textbf{Integration models:}  Having terrestrial BSs, high altitude platform systems, UAVs, and satellite nodes raises several research questions such as how these different segments would integrate to serve UEs? Many papers have studied and investigated different integration models \cite{9610113}, however, very few papers have studied these integration possibilities within the DAS architecture. The DAS architecture can be applied to 3D-dynamic wireless networks in three ways: 1) Deploying terrestrial DAS and dedicating non-terrestrial nodes to support the terrestrial DAS by improving the scalability and serving blocked, remote, or highly interfered UEs, 2) incorporating the DAS architecture in both terrestrial and non-terrestrial networks and associating the users based on their required QoS or based on the available data rates, and 3) deploying terrestrial DAS and dedicating non-terrestrial nodes for backhauling. In the first two models, the total resources of both segments can be virtualized and assigned to different network slices. For example, the nodes (whether they are terrestrial or not) that can provide low-latency services can be used for URLLC, while the nodes that have vast footprints (e.g., satellites or high-altitude platform systems) can be used to serve massive mMTC devices. In the last model of integration, non-terrestrial networks can mitigate scalability and backhaul traffic issues in terrestrial CF-mMIMO systems through enhanced data rate access, efficient data offloading, scalable network operation, and reduced backhaul overhead. 
\item \textbf{Synchronization: }  Synchronization is a critical challenge in DASs because it ensures that the signals transmitted by different antennas arrive at the receiver at the same time. Since the antennas are distributed in different locations, the propagation time for each UE will be different, which requires careful synchronization at the transmitters to ensure that the signals arrive at the receiver at the right time and in the right order. In 3D-dynamic wireless networks, the synchronization issue is a big challenge due to the large difference in distances between the transmitters and the receivers and the mobility of the non-terrestrial nodes. In addition, having wireless backhaul links increases the complexity of synchronizing the transmitted signals in the 3D-dynamic wireless networks. 
\item \textbf{Mobility management and handover: } Following the user-centric concept, each user is served by several APs that should be selected carefully to maximize the achievable rates. If users are mobile, it is required to re-assign these APs frequently. This poses a frequent handover challenge leading to significant overhead, and a potential temporary loss of service since when a user moves at high speed, the handover process can be delayed. In non-terrestrial networks, mobility management will be more critical since both UEs and APs are moving. In addition, the effect of the Doppler shift will be significant, especially in estimating the channels.  One approach to handling mobility is to use deep learning to predict the future locations of UEs and APs, which helps to predict the handovers. In general, mobility in 3D-dynamic wireless networks can introduce several challenges to wireless communication systems, including Doppler shift, interference, handover delays, capacity strain, and poor signal quality. Addressing these challenges requires advanced signal processing techniques, handover optimization techniques, and load balancing algorithms to ensure that the air-ground integrated DAS can efficiently handle the fast-moving users and APs and provide reliable communication. 
\item \textbf{Fronthaul Issue: } Numerous studies have suggested various strategies to enhance the constrained fronthaul capacity, ranging from link-level techniques like data compression to system-level methods such as user-centric designs.  However, as wireless communication systems continue to evolve, accommodating an ever-growing number of connected devices and the increasing demand for higher data rates, these existing approaches may not be sufficient. In addition, having mobile BSs such as satellites and UAVs brings new challenges to the DAS. For example, one challenge that may arise is the variable fronthaul capacity that changes due to the channel changes between the non-terrestrial nodes. The use of wireless fronthaul presents different problems compared to a fixed-capacity fronthaul. The wireless fronthaul capacity is affected by different factors such as the APs location and their velocity. As a result, studying the performance of wireless fronthaul can lead to novel analyses and conclusions for DASs. Therefore,  a truly transformative solution is needed to tackle the fronthaul challenge and ensure the scalability and sustainability of future DAS networks.
\end{itemize}

\subsection{Physical-layer Processing for DASs}
Physical-layer processing in DAS can be a significant source of overhead and delay, leading to reduced network performance and user satisfaction. To address these challenges, advanced relaying strategies, coupled with network automation, SDN, cloud-native networking, and network slicing are recently emerging as key technologies for improving the physical layer processing and relaxing the communications overhead over the fronthaul network in DAS.

Network automation involves the use of software tools and technologies to automate network management and configuration tasks, reducing the need for manual intervention and improving network agility and efficiency. In DAS, network automation can be used to optimize the fronthaul network by dynamically adjusting the allocation of resources, such as bandwidth and processing power, based on changing network conditions and user demand. This can enable more efficient transmission of data over the fronthaul network, reducing the signaling overhead and improving network performance. One of the enabler techniques for network automation is using SDN to enable centralized control and management of network resources, making it easier to configure and optimize the network to meet changing user requirements. Additionally, SDN can be used to dynamically allocate transport network resources based on user demand and network conditions, improving network performance and reducing overhead. Moreover, SDN can also be used to implement network slicing, which involves creating virtual network instances within a physical network, each with its own set of resources and policies. This can enable flexible fronthaul resource allocation based on the specific requirements of each network slice. Another technology to support network automation is exploiting cloud-native networking capabilities to build and deploy network infrastructure that leverages cloud computing technologies, such as Kubernetes, containers, and microservices. In particular, cloud-native networking can provide a flexible and scalable approach to building DAS, enabling more efficient use of fronthaul resources and reducing overhead. For instance, cloud-native networking can enable the deployment of lightweight network functions, such as baseband processing, on edge servers located closer to the RUs, reducing the need for dedicated high-capacity fronthaul links.

On the other side, advanced transmission strategies can play a pivotal role in optimizing the physical-layer processing by using advanced coding and signal processing techniques to improve the efficiency and reliability of codewords transmissions. These strategies can include cooperative relaying, interference mitigation, and dynamic resource allocation, among different RUs and users. Cooperative relaying involves the use of multiple relays to improve transmission range, reduce power consumption, and increase network capacity through using interference mitigation techniques to reduce interference from neighboring cells and dynamic resource allocation through allocating resources such as power, bandwidth, and frequency spectrum based on user demand and network conditions, further improving network efficiency.

\subsection{DASs and RIS Integration}
From the discussion provided in Section \ref{Sec:RIS-DAS}, we can learn that the main benefit of RIS in a cell-free system is the energy efficiency and the deployment cost, while the main challenge is optimizing the number of RISs, the number of reflecting elements, and  the number of APs and antennas. In addition, channel estimation in such systems is a big challenge, especially if the users are mobile. 

For RIS and AP density optimization, the problem should be studied to maximize the energy efficiency by taking into account the cost, the quality-of-service, and the fronthaul capacity under a given user density. Stochastic geometry methods can be used to determine the optimal density of both the APs and the RISs. Moreover, in dense systems especially ones with mobility, fast optimization of RIS and AP beamforming becomes paramount to deal with CSI aging, which is an interesting direction for future research. 

For the problem of channel estimation, it is known that DAS are designed to serve a large number of users with a large number of APs. In addition, the number of RIS reflecting elements is required to be huge to achieve a significant improvement in the system performance \cite{9756313}.   Since the channel estimation overhead is proportional to the number of users and the number of reflecting elements, the integration of RIS and DAS exacerbates the problem of channel estimation overhead, especially in environments with high-mobility. To tackle such a problem, efficient channel estimation schemes must be proposed that consider reducing both the overhead and the mean-squared error.

\subsection{DASs in Hybrid OWC/RF Systems}
As discussed above, using the DAS architecture in OWC systems promises to improve the coverage and the capacity significantly. However, the conducted work on DAS and OWC integration is not sufficient, and several open problems are required to be investigated. An important problem in this area is to optimize a  DAS that consists of RF and OWC APs in outdoor scenarios (e.g., vehicular systems). This problem can be challenging for the following reasons.
\begin{itemize}
\item First, the users are expected to be highly mobile, which means that the channel aging impact is significant. Hence, robust channel estimation schemes that consider the overhead and the accuracy of the estimates are required. 
\item Second, since there are two networks, the problem of load balancing needs to be investigated. In other words, the problem of associating the users to RF and OWC APs needs to be formulated and tackled.
\item Additionally, the density of OWC and RF APs needs to be optimized. In particular, for a given typical user density and distribution, finding the optimal number of OWC and RF APs that balances between coverage and capacity is an interesting research direction.  
\end{itemize}

\subsection{Cross-Layer Optimization for DASs}
Cross-layer optimization gained a strong reputation in efficiently reducing the transmission duration of files while preserving its low-computation characteristics in terms of coding/decoding complexity. The area of cross-layer optimization in DAS with instantaneous NC presents many opportunities for future research as follows. 
\begin{itemize}

\item First, powerful cross-layer optimization techniques are needed  to jointly deal with the scheduling and coding, e.g., network coding, power optimization, and beamforming, in DAS. As another example, for D2D-enabled systems, studying the overhead introduced by the DAS and methods to optimize the overhead and the system performance using cross-layer optimization techniques is a promising research direction. In general, these problems are characterized by a joint optimization of discrete and continuous variables, which makes them challenging. Machine learning-based optimization methods can be helpful in obtaining solutions for these problems. 

\item Second, many works considered cross-layer optimization with the assumption that transmitting nodes ideally know users' side information and keep a record of such information in their memory, which is impractical. A few works \cite{8255832, 6809217} begin to address general cross-layer optimization with partial knowledge of users' side information, which is called blind NC.  In practical scenarios of DAS, feedback signaling from users are prone for failure, thus powerful cross-layer optimization techniques can be used jointly fully blind coding.

\item Recently, new concepts such as heterogeneous networks, C-RANs, F-RANs, and D2D communications emerged. The network architecture changed from a conventional single high-powered BS serving all users in its coverage area into a mass deployment of low-power APs connected through high-speed fronthaul links and network routers, which leads to DAS. Meanwhile, the increasing number of users makes these distributed systems dense. Therefore, proposing distributed solutions over the increased numbers of low-power APs and users to accommodate network densification is crucial. All aforementioned works on cross-layer optimization used centralized graph theory and classical iterative algorithms, e.g., \cite{6588350, 8027131, 7342981, 6811617, RAIDNC1, RAIDNC2, RAIDNC3, RAIDNC4, RAIDNC5, CLNC1, CLNC2,CLNC3,CLNC4,CLNC6, CLNC5a, 9519548}. To align with the emerging distributed systems, one may use machine learning to propose distributed and decentralized solutions to overcome the burden on cloud/fog processors. For example, developing a distributed file placement and delivery algorithm in DAS is an interesting research direction, where the caching decision depends on the resource allocation at the physical layer. Generally, this joint optimization is complex, but reinforcement learning can be used to simplify this problem. 

\item Additionally, all the previous works on cross-layer optimization  assumed perfect channel estimation which can be considered practical for static networks, where users are not moving or moving slowly. However, dense DAS with mobility, e.g., D2D and vehicle-to-vehicle communications scenarios, considering channel state information to be fixed during transmission is not practical. Therefore, considering  channel estimation errors is an area with potential opportunities for future work  in  cross-layer optimization for DASs. 
\end{itemize}

\section{Conclusion}\label{C}
 
In this paper, we surveyed existing literature related to DAS architectures for 5G mobile networks, namely C-RAN, CF-MIMO, cell-free, F-RAN, and D2D communications. We also discussed DAS architectures from various layer perspectives, including physical and network layers, and different performance metrics, such as fronthaul capacity, energy efficiency, coverage probability, sum-rate, and delay, were investigated. In addition, we have investigated the key enabling solutions for DAS architectures such as network slicing, functional split, uplink and downlink coding, cross-layer network coding, and channel estimation techniques. All these techniques play a crucial role in the design of different efficient DAS architectures. Finally, we highlighted some major research challenges in the field of DAS for 6G mobile networks and recommended several future research directions. 

\vspace{-.2cm}
\bibliographystyle{IEEEtran}
\bibliography{Double_Column_Archive}

\begin{thebibliography}{100}
\providecommand{\url}[1]{#1}
\csname url@samestyle\endcsname
\providecommand{\newblock}{\relax}
\providecommand{\bibinfo}[2]{#2}
\providecommand{\BIBentrySTDinterwordspacing}{\spaceskip=0pt\relax}
\providecommand{\BIBentryALTinterwordstretchfactor}{4}
\providecommand{\BIBentryALTinterwordspacing}{\spaceskip=\fontdimen2\font plus
\BIBentryALTinterwordstretchfactor\fontdimen3\font minus \fontdimen4\font\relax}
\providecommand{\BIBforeignlanguage}[2]{{%
\expandafter\ifx\csname l@#1\endcsname\relax
\typeout{** WARNING: IEEEtran.bst: No hyphenation pattern has been}%
\typeout{** loaded for the language `#1'. Using the pattern for}%
\typeout{** the default language instead.}%
\else
\language=\csname l@#1\endcsname
\fi
#2}}
\providecommand{\BIBdecl}{\relax}
\BIBdecl

\bibitem{ITUI}
{ITU-R}, \emph{{IMT} Traffic Estimates for the Years 2020 to 2030}, July 2015.

\bibitem{Eric}
{Ericsson}, \emph{Ericsson Mobility Report}, November 2022.

\bibitem{Eric2}
------, \emph{Ericsson Breaking the Energy Curve Report 2022: {5G} network success can be achieved in an energy-efficient way}, October 2022.

\bibitem{8826541}
A.~Ghosh, A.~Maeder, M.~Baker, and D.~Chandramouli, ``{5G} evolution: {A} view on {5G} cellular technology beyond {3GPP} release 15,'' \emph{IEEE Access}, vol.~7, pp. 127\,639--127\,651, 2019.

\bibitem{6824752}
J.~G. Andrews, S.~Buzzi, W.~Choi, S.~V. Hanly, A.~Lozano, A.~C.~K. Soong, and J.~C. Zhang, ``What will {5G} be?'' \emph{IEEE Journal on Selected Areas in Communications}, vol.~32, no.~6, pp. 1065--1082, 2014.

\bibitem{7858136}
D.~Chitimalla, K.~Kondepu, L.~Valcarenghi, M.~Tornatore, and B.~Mukherjee, ``{5G} fronthaul-latency and jitter studies of {CPRI} over ethernet,'' \emph{Journal of Optical Communications and Networking}, vol.~9, no.~2, pp. 172--182, 2017.

\bibitem{R5}
H.~Q. {Ngo}, A.~{Ashikhmin}, H.~{Yang}, E.~G. {Larsson}, and T.~L. {Marzetta}, ``Cell-free massive {MIMO}: {U}niformly great service for everyone,'' in \emph{2015 IEEE 16th International Workshop on Signal Processing Advances in Wireless Communications (SPAWC)}, 2015, pp. 201--205.

\bibitem{R7}
E.~{Nayebi}, A.~{Ashikhmin}, T.~L. {Marzetta}, and H.~{Yang}, ``Cell-free massive {MIMO} systems,'' in \emph{2015 49th Asilomar Conference on Signals, Systems and Computers}, 2015, pp. 695--699.

\bibitem{R13}
A.~A.~M. {Saleh}, A.~{Rustako}, and R.~{Roman}, ``Distributed antennas for indoor radio communications,'' \emph{IEEE Transactions on Communications}, vol.~35, no.~12, pp. 1245--1251, 1987.

\bibitem{8642354}
F.~Giannone, K.~Kondepu, H.~Gupta, F.~Civerchia, P.~Castoldi, A.~Antony~Franklin, and L.~Valcarenghi, ``Impact of virtualization technologies on virtualized {RAN} midhaul latency budget: {A} quantitative experimental evaluation,'' \emph{IEEE Communications Letters}, vol.~23, no.~4, pp. 604--607, 2019.

\bibitem{9075181}
M.~Masoudi, S.~S. Lisi, and C.~Cavdar, ``Cost-effective migration toward virtualized {C-RAN} with scalable fronthaul design,'' \emph{IEEE Systems Journal}, vol.~14, no.~4, pp. 5100--5110, 2020.

\bibitem{7555389}
M.~Peng and K.~Zhang, ``Recent advances in fog radio access networks: {P}erformance analysis and radio resource allocation,'' \emph{IEEE Access}, vol.~4, pp. 5003--5009, 2016.

\bibitem{7558153}
S.-H. Park, O.~Simeone, and S.~Shamai~Shitz, ``Joint optimization of cloud and edge processing for fog radio access networks,'' \emph{IEEE Transactions on Wireless Communications}, vol.~15, no.~11, pp. 7621--7632, 2016.

\bibitem{7537176}
R.~Tandon and O.~Simeone, ``Harnessing cloud and edge synergies: {T}oward an information theory of fog radio access networks,'' \emph{IEEE Communications Magazine}, vol.~54, no.~8, pp. 44--50, 2016.

\bibitem{9861780}
M.~S. Al-Abiad, M.~Z. Hassan, and M.~J. Hossain, ``Task offloading optimization in {NOMA}-enabled dual-hop mobile edge computing system using conflict graph,'' \emph{IEEE Transactions on Wireless Communications}, vol.~22, no.~2, pp. 761--777, 2023.

\bibitem{9824969}
S.~Bahanshal, M.~S. Al–Abiad, and M.~J. Hossain, ``Minimizing energy consumption for mobile edge computing with non-orthogonal multiple access,'' in \emph{2022 International Wireless Communications and Mobile Computing (IWCMC)}, 2022, pp. 1088--1093.

\bibitem{10024837}
M.~Polese, L.~Bonati, S.~D’Oro, S.~Basagni, and T.~Melodia, ``Understanding {O-RAN}: {A}rchitecture, interfaces, algorithms, security, and research challenges,'' \emph{IEEE Communications Surveys and Tutorials}, vol.~25, no.~2, pp. 1376--1411, 2023.

\bibitem{10071958}
F.~Kavehmadavani, V.-D. Nguyen, T.~X. Vu, and S.~Chatzinotas, ``Intelligent traffic steering in beyond {5G} open {RAN} based on {LSTM} traffic prediction,'' \emph{IEEE Transactions on Wireless Communications}, pp. 1--1, 2023.

\bibitem{CPRI1}
{CPRI}, \emph{{CPRI} Interface Specification, V. 7.0}, October 2015.

\bibitem{tt}
M.~{Paolini}, ``{C-RAN} and {5G} take transport to new capacity and latency levels: {T}rends in backhaul, fronthaul, xhaul and mm{W},'' \emph{Senza Fili, Sammamish, WA, USA, Tech. Rep.}, March 2016.

\bibitem{ORI}
{European Telecommunications Standards Initiative}, \emph{Open Radio Interface, Industrial Standardization Group ({ORI ISG})}, 2014.

\bibitem{8384342}
G.~O. Perez, J.~A. Hernandez, and D.~Larrabeiti, ``Fronthaul network modeling and dimensioning meeting ultra-low latency requirements for {5G},'' \emph{Journal of Optical Communications and Networking}, vol.~10, no.~6, pp. 573--581, 2018.

\bibitem{3GPP1}
{3GPP TR 38.801 V2.0.0 (R14)}, \emph{Technical Specification Group Radio Access Network; Study on New Radio Access Technology; Radio Access Architecture and Interfaces}, 2017.

\bibitem{eCPRI1}
{CPRI}, \emph{Common Public Radio Interface: eCPRI Interface Specification V1.2}, June 2016.

\bibitem{SCF}
S.~C. Forum, \emph{Small Cell Virtualization Functional Splits and Use Cases, Release 7.0}.\hskip 1em plus 0.5em minus 0.4em\relax Small Cell Forum, January 2016.

\bibitem{ORAN-FH}
O.-R. Alliance, \emph{Control, User and Synchronization Plane Specification}.\hskip 1em plus 0.5em minus 0.4em\relax WG4.CUS.0-v03.00, April 2020.

\bibitem{quek_peng_simeone_yu_2017}
\emph{Cloud Radio Access Networks: {P}rinciples, technologies, and applications}.\hskip 1em plus 0.5em minus 0.4em\relax Cambridge University Press, 2017.

\bibitem{6342931}
S.~{Park}, O.~{Simeone}, O.~{Sahin}, and S.~{Shamai}, ``Robust and efficient distributed compression for cloud radio access networks,'' \emph{IEEE Transactions on Vehicular Technology}, vol.~62, no.~2, pp. 692--703, 2013.

\bibitem{6522158}
S.-N. Hong and G.~Caire, ``Compute-and-forward strategies for cooperative distributed antenna systems,'' \emph{IEEE Transactions on Information Theory}, vol.~59, no.~9, pp. 5227--5243, 2013.

\bibitem{5752460}
S.~H. Lim, Y.-H. Kim, A.~El~Gamal, and S.-Y. Chung, ``Noisy network coding,'' \emph{IEEE Transactions on Information Theory}, vol.~57, no.~5, pp. 3132--3152, 2011.

\bibitem{6920005}
B.~{Dai} and W.~{Yu}, ``Sparse beamforming and user-centric clustering for downlink cloud radio access network,'' \emph{IEEE Access}, vol.~2, pp. 1326--1339, 2014.

\bibitem{6588350}
S.~{Park}, O.~{Simeone}, O.~{Sahin}, and S.~{Shamai}, ``Joint precoding and multivariate backhaul compression for the downlink of cloud radio access networks,'' \emph{IEEE Transactions on Signal Processing}, vol.~61, no.~22, pp. 5646--5658, 2013.

\bibitem{6283033}
S.~{Hong} and G.~{Caire}, ``Reverse compute and forward: {A} low-complexity architecture for downlink distributed antenna systems,'' in \emph{2012 IEEE International Symposium on Information Theory Proceedings}, 2012, pp. 1147--1151.

\bibitem{R21}
E.~Bj{\"o}rnson, J.~Hoydis, L.~Sanguinetti \emph{et~al.}, ``Massive {MIMO} networks: {S}pectral, energy, and hardware efficiency,'' \emph{Foundations and Trends{\textregistered} in Signal Processing}, vol.~11, no. 3-4, pp. 154--655, 2017.

\bibitem{R20}
E.~Björnson and L.~Sanguinetti, ``Scalable cell-free massive {MIMO} systems,'' \emph{IEEE Transactions on Communications}, vol.~68, no.~7, pp. 4247--4261, 2020.

\bibitem{R32}
H.~Q. {Ngo}, L.~{Tran}, T.~Q. {Duong}, M.~{Matthaiou}, and E.~G. {Larsson}, ``On the total energy efficiency of cell-free massive {MIMO},'' \emph{IEEE Transactions on Green Communications and Networking}, vol.~2, no.~1, pp. 25--39, 2018.

\bibitem{R33}
L.~D. {Nguyen}, T.~Q. {Duong}, H.~Q. {Ngo}, and K.~{Tourki}, ``Energy efficiency in cell-free massive {MIMO} with zero-forcing precoding design,'' \emph{IEEE Communications Letters}, vol.~21, no.~8, pp. 1871--1874, 2017.

\bibitem{8972478}
A.~Papazafeiropoulos, P.~Kourtessis, M.~D. Renzo, S.~Chatzinotas, and J.~M. Senior, ``Performance analysis of cell-free massive {MIMO} systems: {A} stochastic geometry approach,'' \emph{IEEE Transactions on Vehicular Technology}, vol.~69, no.~4, pp. 3523--3537, 2020.

\bibitem{R3}
H.~Q. {Ngo}, A.~{Ashikhmin}, H.~{Yang}, E.~G. {Larsson}, and T.~L. {Marzetta}, ``Cell-free massive {MIMO} versus small cells,'' \emph{IEEE Transactions on Wireless Communications}, vol.~16, no.~3, pp. 1834--1850, 2017.

\bibitem{R45}
E.~{Björnson} and L.~{Sanguinetti}, ``Making cell-free massive {MIMO} competitive with {MMSE} processing and centralized implementation,'' \emph{IEEE Transactions on Wireless Communications}, vol.~19, no.~1, pp. 77--90, 2020.

\bibitem{R6}
G.~Interdonato, E.~Bj{\"o}rnson, H.~Q. Ngo, P.~Frenger, and E.~G. Larsson, ``Ubiquitous cell-free massive {MIMO} communications,'' \emph{EURASIP Journal on Wireless Communications and Networking}, vol. 2019, no.~1, p. 197, 2019.

\bibitem{CLNC1}
M.~S. Al-Abiad, A.~Douik, S.~Sorour, and M.~J. Hossain, ``Throughput maximization in cloud-radio access networks using cross-layer network coding,'' \emph{IEEE Transactions on Mobile Computing}, vol.~21, no.~2, pp. 696--711, 2022.

\bibitem{4068001}
J.~Huang, V.~Subramanian, R.~Agrawal, and R.~Berry, ``Downlink scheduling and resource allocation for {OFDM} systems,'' in \emph{2006 40th Annual Conference on Information Sciences and Systems}, 2006, pp. 1272--1279.

\bibitem{4769397}
J.~Huang, V.~G. Subramanian, R.~Agrawal, and R.~Berry, ``Joint scheduling and resource allocation in uplink {OFDM} systems for broadband wireless access networks,'' \emph{IEEE Journal on Selected Areas in Communications}, vol.~27, no.~2, pp. 226--234, 2009.

\bibitem{5464889}
W.~Yu, T.~Kwon, and C.~Shin, ``Joint scheduling and dynamic power spectrum optimization for wireless multicell networks,'' in \emph{2010 44th Annual Conference on Information Sciences and Systems (CISS)}, 2010, pp. 1--6.

\bibitem{6525475}
------, ``Multicell coordination via joint scheduling, beamforming, and power spectrum adaptation,'' \emph{IEEE Transactions on Wireless Communications}, vol.~12, no.~7, pp. 1--14, 2013.

\bibitem{5062043}
A.~L. Stolyar and H.~Viswanathan, ``Self-organizing dynamic fractional frequency reuse for best-effort traffic through distributed inter-cell coordination,'' in \emph{IEEE INFOCOM 2009}, 2009, pp. 1287--1295.

\bibitem{6662474}
X.~Wang, C.~Yuen, and Y.~Xu, ``Coding-based data broadcasting for time-critical applications with rate adaptation,'' \emph{IEEE Transactions on Vehicular Technology}, vol.~63, no.~5, pp. 2429--2442, 2014.

\bibitem{CLNC2}
M.~S. Al-Abiad, A.~Douik, S.~Sorour, and M.~D.~J. Hossain, ``Throughput maximization in cloud radio access networks using network coding,'' in \emph{2018 IEEE International Conference on Communications Workshops (ICC Workshops)}, 2018, pp. 1--6.

\bibitem{CLNC3}
M.~S. Al-Abiad, M.~Z. Hassan, A.~Douik, and M.~J. Hossain, ``Low-complexity power allocation for network-coded user scheduling in fog-{RAN}s,'' \emph{IEEE Communications Letters}, vol.~25, no.~4, pp. 1318--1322, 2021.

\bibitem{6692154}
X.~Wang, C.~Yuen, and S.~H. Dau, ``Delay minimization for network coded cooperative data exchange with rate adaptation,'' in \emph{2013 IEEE 78th Vehicular Technology Conference (VTC Fall)}, 2013, pp. 1--5.

\bibitem{CLNC4}
M.~S. Al-Abiad, S.~Sorour, and M.~J. Hossain, ``Cloud offloading with {Q}o{S} provisioning using cross-layer network coding,'' in \emph{2018 IEEE Global Communications Conference (GLOBECOM)}, 2018, pp. 1--7.

\bibitem{5462127}
T.-S. Kim, S.~Vural, I.~Broustis, D.~Syrivelis, S.~V. Krishnamurthy, and T.~F. La~Porta, ``A framework for joint network coding and transmission rate control in wireless networks,'' in \emph{2010 Proceedings IEEE INFOCOM}, 2010, pp. 1--9.

\bibitem{CLNC5}
M.~S. Al-Abiad, M.~J. Hossain, and S.~Sorour, ``Cross-layer cloud offloading with quality of service guarantees in fog-{RAN}s,'' \emph{IEEE Transactions on Communications}, vol.~67, no.~12, pp. 8435--8449, 2019.

\bibitem{CLNC6}
M.~S. Al-Abiad and M.~J. Hossain, ``Completion time minimization in fog-{RAN}s using {D2D} communications and rate-aware network coding,'' \emph{IEEE Transactions on Wireless Communications}, vol.~20, no.~6, pp. 3831--3846, 2021.

\bibitem{8288205}
Y.~N. Shnaiwer, S.~Sorour, P.~Sadeghi, and T.~Y. Al-Naffouri, ``Online cloud offloading using heterogeneous enhanced remote radio heads,'' in \emph{2017 IEEE 86th Vehicular Technology Conference (VTC-Fall)}, 2017, pp. 1--5.

\bibitem{NC1}
R.~Ahlswede, N.~Cai, S.-Y. Li, and R.~Yeung, ``Network information flow,'' \emph{IEEE Transactions on Information Theory}, vol.~46, no.~4, pp. 1204--1216, 2000.

\bibitem{NC2}
J.~K. Sundararajan, D.~Shah, and M.~Medard, ``Online network coding for optimal throughput and delay - the three-receiver case,'' in \emph{2008 International Symposium on Information Theory and Its Applications}, 2008, pp. 1--6.

\bibitem{NC3}
D.~Nguyen, T.~Nguyen, and X.~Yang, ``Multimedia wireless transmission with network coding,'' in \emph{Packet Video 2007}, 2007, pp. 326--335.

\bibitem{NC4}
D.~Nguyen and T.~Nguyen, ``Network coding-based wireless media transmission using {POMDP},'' in \emph{2009 17th International Packet Video Workshop}, 2009, pp. 1--9.

\bibitem{7845689}
A.~Douik, S.~Sorour, T.~Y. Al-Naffouri, and M.-S. Alouini, ``Instantly decodable network coding: {F}rom centralized to device-to-device communications,'' \emph{IEEE Communications Surveys and Tutorials}, vol.~19, no.~2, pp. 1201--1224, 2017.

\bibitem{8794557}
A.~Douik, M.~S. Al-Abiad, and M.~J. Hossain, ``An improved weight design for unwanted packets in multicast instantly decodable network coding,'' \emph{IEEE Communications Letters}, vol.~23, no.~11, pp. 2122--2125, 2019.

\bibitem{9512273}
M.~S. Al-Abiad and M.~J. Hossain, ``A low-complexity multicast scheduling for {D2D}-aided {F-RAN}s using network coding,'' \emph{IEEE Wireless Communications Letters}, vol.~10, no.~11, pp. 2484--2488, 2021.

\bibitem{9180294}
M.~S. Al-Abiad, A.~Douik, and M.~J. Hossain, ``Coalition formation game for cooperative content delivery in network coding assisted {D2D} communications,'' \emph{IEEE Access}, vol.~8, pp. 158\,152--158\,168, 2020.

\bibitem{8267199}
A.~Douik and S.~Sorour, ``Data dissemination using instantly decodable binary codes in fog-radio access networks,'' \emph{IEEE Transactions on Communications}, vol.~66, no.~5, pp. 2052--2064, 2018.

\bibitem{8444481}
A.~Douik, S.~Sorour, T.~Y. Al-Naffouri, H.-C. Yang, and M.-S. Alouini, ``Delay reduction in multi-hop device-to-device communication using network coding,'' \emph{IEEE Transactions on Wireless Communications}, vol.~17, no.~10, pp. 7040--7053, 2018.

\bibitem{6882208}
S.~Sorour and S.~Valaee, ``Completion delay minimization for instantly decodable network codes,'' \emph{IEEE/ACM Transactions on Networking}, vol.~23, no.~5, pp. 1553--1567, 2015.

\bibitem{1176612}
S.-Y. Li, R.~Yeung, and N.~Cai, ``Linear network coding,'' \emph{IEEE Transactions on Information Theory}, vol.~49, no.~2, pp. 371--381, 2003.

\bibitem{4557282}
L.~Lima, M.~Medard, and J.~Barros, ``Random linear network coding: {A} free cipher?'' in \emph{2007 IEEE International Symposium on Information Theory}, 2007, pp. 546--550.

\bibitem{1705002}
T.~Ho, M.~Medard, R.~Koetter, D.~Karger, M.~Effros, J.~Shi, and B.~Leong, ``A random linear network coding approach to multicast,'' \emph{IEEE Transactions on Information Theory}, vol.~52, no.~10, pp. 4413--4430, 2006.

\bibitem{4675714}
A.~Eryilmaz, A.~Ozdaglar, M.~Medard, and E.~Ahmed, ``On the delay and throughput gains of coding in unreliable networks,'' \emph{IEEE Transactions on Information Theory}, vol.~54, no.~12, pp. 5511--5524, 2008.

\bibitem{7511238}
M.~S. Karim, A.~Douik, S.~Sorour, and P.~Sadeghi, ``Rate-aware network codes for completion time reduction in device-to-device communications,'' in \emph{2016 IEEE International Conference on Communications (ICC)}, 2016, pp. 1--7.

\bibitem{4567144}
B.~Ni, N.~Santhapuri, C.~Gray, and S.~Nelakuditi, ``Selection of bit-rate for wireless network coding,'' in \emph{2008 5th IEEE Annual Communications Society Conference on Sensor, Mesh and Ad Hoc Communications and Networks Workshops}, 2008, pp. 1--6.

\bibitem{RAIDNC1}
A.~Douik, S.~Sorour, T.~Y. Al-Naffouri, and M.-S. Alouini, ``Rate aware instantly decodable network codes,'' \emph{IEEE Transactions on Wireless Communications}, vol.~16, no.~2, pp. 998--1011, 2017.

\bibitem{RAIDNC2}
M.~S. Karim, A.~Douik, and S.~Sorour, ``Rate-aware network codes for video distortion reduction in point-to-multipoint networks,'' \emph{IEEE Transactions on Vehicular Technology}, vol.~66, no.~8, pp. 7446--7460, 2017.

\bibitem{RAIDNC3}
M.~Saif, A.~Douik, and S.~Sorour, ``Rate aware network codes for coordinated multi base-station networks,'' in \emph{2016 IEEE International Conference on Communications (ICC)}, 2016, pp. 1--7.

\bibitem{RAIDNC4}
X.~Wang, C.~Yuen, and Y.~Xu, ``Coding-based data broadcasting for time-critical applications with rate adaptation,'' \emph{IEEE Transactions on Vehicular Technology}, vol.~63, no.~5, pp. 2429--2442, 2014.

\bibitem{RAIDNC5}
M.~S. Al-Abiad, A.~Douik, and S.~Sorour, ``Rate aware network codes for cloud radio access networks,'' \emph{IEEE Transactions on Mobile Computing}, vol.~18, no.~8, pp. 1898--1910, 2019.

\bibitem{8113473}
I.~A. Alimi, A.~L. Teixeira, and P.~P. Monteiro, ``Toward an efficient {C-RAN} optical fronthaul for the future networks: {A} tutorial on technologies, requirements, challenges, and solutions,'' \emph{IEEE Communications Surveys and Tutorials}, vol.~20, no.~1, pp. 708--769, 2018.

\bibitem{8479363}
L.~M.~P. Larsen, A.~Checko, and H.~L. Christiansen, ``A survey of the functional splits proposed for {5G} mobile crosshaul networks,'' \emph{IEEE Communications Surveys and Tutorials}, vol.~21, no.~1, pp. 146--172, 2019.

\bibitem{9650567}
H.~A. Ammar, R.~Adve, S.~Shahbazpanahi, G.~Boudreau, and K.~V. Srinivas, ``User-centric cell-free massive {MIMO} networks: {A} survey of opportunities, challenges and solutions,'' \emph{IEEE Communications Surveys and Tutorials}, vol.~24, no.~1, pp. 611--652, 2022.

\bibitem{chen2021survey}
S.~Chen, J.~Zhang, J.~Zhang, E.~Bj{\"o}rnson, and B.~Ai, ``A survey on user-centric cell-free massive {MIMO} systems,'' \emph{Digital Communications and Networks}, 2021.

\bibitem{elhoushy2021cell}
S.~Elhoushy, M.~Ibrahim, and W.~Hamouda, ``Cell-free massive {MIMO}: {A} survey,'' \emph{IEEE Communications Surveys and Tutorials}, vol.~24, no.~1, pp. 492--523, 2021.

\bibitem{bassoy2017coordinated}
S.~Bassoy, H.~Farooq, M.~A. Imran, and A.~Imran, ``Coordinated multi-point clustering schemes: {A} survey,'' \emph{IEEE Communications Surveys and Tutorials}, vol.~19, no.~2, pp. 743--764, 2017.

\bibitem{8327582}
L.~Li, G.~Zhao, and R.~S. Blum, ``A survey of caching techniques in cellular networks: {R}esearch issues and challenges in content placement and delivery strategies,'' \emph{IEEE Communications Surveys and Tutorials}, vol.~20, no.~3, pp. 1710--1732, 2018.

\bibitem{8367785}
I.~Parvez, A.~Rahmati, I.~Guvenc, A.~I. Sarwat, and H.~Dai, ``A survey on low latency towards {5G}: {RAN}, core network and caching solutions,'' \emph{IEEE Communications Surveys Tutorials}, vol.~20, no.~4, pp. 3098--3130, 2018.

\bibitem{Fogcomputing}
M.~Mukherjee, L.~Shu, and D.~Wang, ``Survey of fog computing: {F}undamental, network applications, and research challenges,'' \emph{IEEE Communications Surveys and Tutorials}, vol.~20, no.~3, pp. 1826--1857, 2018.

\bibitem{7009970}
A.~Pizzinat, P.~Chanclou, F.~Saliou, and T.~Diallo, ``Things you should know about fronthaul,'' \emph{Journal of Lightwave Technology}, vol.~33, no.~5, pp. 1077--1083, 2015.

\bibitem{7096298}
M.~{Peng}, C.~{Wang}, V.~{Lau}, and H.~V. {Poor}, ``Fronthaul-constrained cloud radio access networks: {I}nsights and challenges,'' \emph{IEEE Wireless Communications}, vol.~22, no.~2, pp. 152--160, 2015.

\bibitem{7331128}
T.~Pfeiffer, ``Next generation mobile fronthaul and midhaul architectures [invited],'' \emph{Journal of Optical Communications and Networking}, vol.~7, no.~11, pp. B38--B45, 2015.

\bibitem{9489948}
W.~Hu, Y.~Zhu, L.~Li, K.~Zhang, H.~Xin, Y.~Fu, and X.~Miao, ``Enabling technologies for comprehensive optical mobile fronthaul access network,'' in \emph{2021 Optical Fiber Communications Conference and Exhibition (OFC)}, 2021, pp. 1--3.

\bibitem{7402275}
A.~de~la Oliva, J.~A. Hernandez, D.~Larrabeiti, and A.~Azcorra, ``An overview of the {CPRI} specification and its application to {C-RAN}-based {LTE} scenarios,'' \emph{IEEE Communications Magazine}, vol.~54, no.~2, pp. 152--159, 2016.

\bibitem{8527624}
A.~Kukushkin, \emph{Annex: {B}ase‐Station Site Solutions}, 2018, pp. 367--375.

\bibitem{inbook1}
C.-L. I., J.~Huang, Y.~Yuan, and S.~Ma, \emph{{5G RAN} architecture: {C-RAN} with {NGFI}}, 10 2017, pp. 431--455.

\bibitem{7414147}
I.~Chih-Lin, J.~Huang, Y.~Yuan, S.~Ma, and R.~Duan, ``{NGFI}, the xhaul,'' in \emph{2015 IEEE Globecom Workshops (GC Wkshps)}, 2015, pp. 1--6.

\bibitem{ITU}
{TELECOMMUNICATION STANDARDIZATION SECTOR OF ITU}, \emph{{5G} wireless fronthaul requirements in a passive optical network context}.\hskip 1em plus 0.5em minus 0.4em\relax ITU-T G-series Recommendations – Supplement 66, 2020.

\bibitem{Pfeiffer:15}
\BIBentryALTinterwordspacing
T.~Pfeiffer, ``Next generation mobile fronthaul architectures,'' in \emph{Optical Fiber Communication Conference}.\hskip 1em plus 0.5em minus 0.4em\relax Optica Publishing Group, 2015, p. M2J.7. [Online]. Available: \url{http://opg.optica.org/abstract.cfm?URI=OFC-2015-M2J.7}
\BIBentrySTDinterwordspacing

\bibitem{Chanclou2013OpticalFS}
P.~Chanclou, A.~Pizzinat, F.~L. Clech, T.-L. Reedeker, Y.~Lagadec, F.~Saliou, B.~L. Guyader, L.~Guillo, Q.~Deniel, S.~Gosselin, S.~D. Le, T.~A. Diallo, R.~Brenot, F.~Lelarge, L.~Marazzi, P.~Parolari, M.~Martinelli, S.~P. O'Duill, S.~A. Gebrewold, D.~Hillerkuss, J.~Leuthold, G.~Gavioli, and P.~Galli, ``Optical fiber solution for mobile fronthaul to achieve cloud radio access network,'' \emph{2013 Future Network \& Mobile Summit}, pp. 1--11, 2013.

\bibitem{Oliveira2014AnalysisOT}
R.~S. Oliveira, C.~R.~L. Franc{\^e}s, J.~C. W.~A. Costa, D.~Viana, M.~Lima, and A.~L.~J. Teixeira, ``Analysis of the cost-effective digital radio over fiber system in the ng-pon2 context,'' \emph{2014 16th International Telecommunications Network Strategy and Planning Symposium (Networks)}, pp. 1--6, 2014.

\bibitem{LTE1}
H.~Holma and A.~Toskala, \emph{{LTE} Advanced: {3GPP} Solution for {IMT}-Advanced}.\hskip 1em plus 0.5em minus 0.4em\relax Hoboken, NJ, USA: Wiley, 2012, vol. 2012.

\bibitem{6897914}
A.~{Checko}, H.~L. {Christiansen}, Y.~{Yan}, L.~{Scolari}, G.~{Kardaras}, M.~S. {Berger}, and L.~{Dittmann}, ``Cloud {RAN} for mobile networks— {A} technology overview,'' \emph{IEEE Communications Surveys Tutorials}, vol.~17, no.~1, pp. 405--426, 2015.

\bibitem{7550569}
L.~Valcarenghi, K.~Kondepu, F.~Giannone, and P.~Castoldi, ``Requirements for {5G} fronthaul,'' in \emph{2016 18th International Conference on Transparent Optical Networks (ICTON)}, 2016, pp. 1--5.

\bibitem{7980777}
P.~Arnold, N.~Bayer, J.~Belschner, and G.~Zimmermann, ``{5G} radio access network architecture based on flexible functional control/user plane splits,'' in \emph{2017 European Conference on Networks and Communications (EuCNC)}, 2017, pp. 1--5.

\bibitem{9107209}
S.~Matoussi, I.~Fajjari, S.~Costanzo, N.~Aitsaadi, and R.~Langar, ``{5G RAN}: {F}unctional split orchestration optimization,'' \emph{IEEE Journal on Selected Areas in Communications}, vol.~38, no.~7, pp. 1448--1463, 2020.

\bibitem{9433522}
Z.~Becvar, P.~Mach, M.~Elfiky, and M.~Sakamoto, ``Hierarchical scheduling for suppression of fronthaul delay in {C-RAN} with dynamic functional split,'' \emph{IEEE Communications Magazine}, vol.~59, no.~4, pp. 95--101, 2021.

\bibitem{9120828}
S.~Matoussi, I.~Fajjari, N.~Aitsaadi, and R.~Langar, ``User slicing scheme with functional split selection in {5G} cloud-{RAN},'' in \emph{2020 IEEE Wireless Communications and Networking Conference (WCNC)}, 2020, pp. 1--8.

\bibitem{9684900}
E.~Zeydan, J.~Mangues-Bafalluy, J.~Baranda, M.~Requena, and Y.~Turk, ``Service based virtual {RAN} architecture for next generation cellular systems,'' \emph{IEEE Access}, vol.~10, pp. 9455--9470, 2022.

\bibitem{6771075}
U.~Dötsch, M.~Doll, H.-P. Mayer, F.~Schaich, J.~Segel, and P.~Sehier, ``Quantitative analysis of split base station processing and determination of advantageous architectures for {LTE},'' \emph{Bell Labs Technical Journal}, vol.~18, no.~1, pp. 105--128, 2013.

\bibitem{7996493}
N.~Makris, P.~Basaras, T.~Korakis, N.~Nikaein, and L.~Tassiulas, ``Experimental evaluation of functional splits for {5G} cloud-{RAN}s,'' in \emph{2017 IEEE International Conference on Communications (ICC)}, 2017, pp. 1--6.

\bibitem{long}
R.~Agrawal, A.~Bedekar, T.~Kolding, and V.~Ram, ``Cloud {RAN} challenges and solutions,'' \emph{Annales des Telecommunications/Annals of Telecommunications}, vol.~72, pp. 1--14, 06 2017.

\bibitem{10.5555/3181071}
M.~A. Imran, S.~A.~R. Zaidi, and M.~Z. Shakir, \emph{Access, Fronthaul and Backhaul Networks for 5G and Beyond}.\hskip 1em plus 0.5em minus 0.4em\relax Institution of Engineering and Technology, 2017.

\bibitem{8320765}
I.~Afolabi, T.~Taleb, K.~Samdanis, A.~Ksentini, and H.~Flinck, ``Network slicing and softwarization: {A} survey on principles, enabling technologies, and solutions,'' \emph{IEEE Communications Surveys Tutorials}, vol.~20, no.~3, pp. 2429--2453, 2018.

\bibitem{8736299}
Y.~Tsukamoto, R.~K. Saha, S.~Nanba, and K.~Nishimura, ``Experimental evaluation of {RAN} slicing architecture with flexibly located functional components of base station according to diverse {5G} services,'' \emph{IEEE Access}, vol.~7, pp. 76\,470--76\,479, 2019.

\bibitem{9594852}
L.~Diez, A.~M. Alba, W.~Kellerer, and R.~Agüero, ``Flexible functional split and fronthaul delay: {A} queuing-based model,'' \emph{IEEE Access}, vol.~9, pp. 151\,049--151\,066, 2021.

\bibitem{9300210}
L.~Diez, C.~Hervella, and R.~Agüero, ``Understanding the performance of flexible functional split in {5G} v{RAN} controllers: {A} markov chain-based model,'' \emph{IEEE Transactions on Network and Service Management}, vol.~18, no.~1, pp. 456--468, 2021.

\bibitem{8985486}
B.~H. Prananto, Iskandar, and A.~Kurniawan, ``Low split cloud {RAN} opportunities and challenges,'' in \emph{2019 IEEE 13th International Conference on Telecommunication Systems, Services, and Applications (TSSA)}, 2019, pp. 119--123.

\bibitem{8419201}
J.~Baranda, J.~Mangues-Bafalluy, I.~Pascual, J.~Nunez-Martinez, J.~L. De~La~Cruz, R.~Casellas, R.~Vilalta, J.~X. Salvat, and C.~Turyagyenda, ``Orchestration of end-to-end network services in the {5G}-crosshaul multi-domain multi-technology transport network,'' \emph{IEEE Communications Magazine}, vol.~56, no.~7, pp. 184--191, 2018.

\bibitem{9322543}
A.~M. Alba, S.~Janardhanan, and W.~Kellerer, ``Dynamics of the flexible functional split selection in {5G} networks,'' in \emph{GLOBECOM 2020 - 2020 IEEE Global Communications Conference}, 2020, pp. 1--6.

\bibitem{9013336}
A.~Martinez~Alba and W.~Kellerer, ``A dynamic functional split in {5G} radio access networks,'' in \emph{2019 IEEE Global Communications Conference (GLOBECOM)}, 2019, pp. 1--6.

\bibitem{8845147}
A.~M. Alba, J.~H.~G. Velásquez, and W.~Kellerer, ``An adaptive functional split in {5G} networks,'' in \emph{IEEE INFOCOM 2019 - IEEE Conference on Computer Communications Workshops (INFOCOM WKSHPS)}, 2019, pp. 410--416.

\bibitem{9627736}
W.~da~Silva~Coelho, A.~Benhamiche, N.~Perrot, and S.~Secci, ``Function splitting, isolation, and placement trade-offs in network slicing,'' \emph{IEEE Transactions on Network and Service Management}, vol.~19, no.~2, pp. 1920--1936, 2022.

\bibitem{9373366}
E.~Datsika, J.~S. Vardakas, K.~Ramantas, P.-V. Mekikis, I.~T. Monroy, L.~A. Neto, and C.~Verikoukis, ``{SDN}-enabled resource management for converged fi-wi {5G} fronthaul,'' \emph{IEEE Journal on Selected Areas in Communications}, vol.~39, no.~9, pp. 2772--2788, 2021.

\bibitem{BORROMEO2022108931}
\BIBentryALTinterwordspacing
J.~C. Borromeo, K.~Kondepu, N.~Andriolli, and L.~Valcarenghi, ``{FPGA}-accelerated smartnic for supporting {5G} virtualized radio access network,'' \emph{Computer Networks}, p. 108931, 2022. [Online]. Available: \url{https://www.sciencedirect.com/science/article/pii/S1389128622001189}
\BIBentrySTDinterwordspacing

\bibitem{inbook2}
M.~Maule, B.~Ojaghi~Kahjogh, and F.~Rezazadeh, \emph{Advanced Cloud-Based Network Management for {5G C-RAN}}, 01 2022, pp. 371--397.

\bibitem{YOUNIS2022107}
\BIBentryALTinterwordspacing
A.~Younis, B.~Qiu, and D.~Pompili, ``Latency and quality-aware task offloading in multi-node next generation {RAN}s,'' \emph{Computer Communications}, vol. 184, pp. 107--117, 2022. [Online]. Available: \url{https://www.sciencedirect.com/science/article/pii/S0140366421004655}
\BIBentrySTDinterwordspacing

\bibitem{9717286}
L.~M.~M. Zorello, M.~Sodano, S.~Troia, and G.~Maier, ``Power-efficient baseband-function placement in latency-constrained {5G} metro access,'' \emph{IEEE Transactions on Green Communications and Networking}, pp. 1--1, 2022.

\bibitem{8761941}
S.~Matoussi, I.~Fajjari, N.~Aitsaadi, R.~Langar, and S.~Costanzo, ``Joint functional split and resource allocation in {5G} cloud-{RAN},'' in \emph{ICC 2019 - 2019 IEEE International Conference on Communications (ICC)}, 2019, pp. 1--7.

\bibitem{7306542}
J.~Liu, S.~Xu, S.~Zhou, and Z.~Niu, ``Redesigning fronthaul for next-generation networks: {B}eyond baseband samples and point-to-point links,'' \emph{IEEE Wireless Communications}, vol.~22, no.~5, pp. 90--97, 2015.

\bibitem{7444125}
M.~Peng, Y.~Sun, X.~Li, Z.~Mao, and C.~Wang, ``Recent advances in cloud radio access networks: System architectures, key techniques, and open issues,'' \emph{IEEE Communications Surveys Tutorials}, vol.~18, no.~3, pp. 2282--2308, 2016.

\bibitem{7504140}
R.~Agrawal, A.~Bedekar, S.~Kalyanasundaram, T.~Kolding, H.~Kroener, and V.~Ram, ``Architecture principles for cloud {RAN},'' in \emph{2016 IEEE 83rd Vehicular Technology Conference (VTC Spring)}, 2016, pp. 1--5.

\bibitem{6666}
J.~Teran, N.~Maletic, D.~Camps-Mur, E.~Garcia-Villegas, I.~Berberana, M.~Anastasopoulos, A.~Tzanakaki, V.~Kalokidou, P.~Flegkas, D.~Syrivelis, T.~Korakis, P.~Legg, D.~Markovic, G.~Lyberopoulos, J.~Bartelt, J.~Chaudhary, M.~Grieger, N.~Vucic, J.~Zou, and E.~Grass, ``{5G}-xhaul: {A} converged optical and wireless solution for {5G} transport networks,'' \emph{Transactions on Emerging Telecommunications Technologies}, vol.~27, 07 2016.

\bibitem{4444}
J.~Bartelt, N.~Vucic, D.~Camps-Mur, E.~Garcia-Villegas, I.~Demirkol, A.~Fehske, M.~Grieger, A.~Tzanakaki, J.~Teran, E.~Grass, G.~Lyberopoulos, and G.~Fettweis, ``{5G} transport network requirements for the next generation fronthaul interface,'' \emph{EURASIP Journal on Wireless Communications and Networking}, vol. 2017, p.~89, 05 2017.

\bibitem{8346013}
P.~Assimakopoulos, G.~S. Birring, and N.~J. Gomes, ``Effects of contention and delay in a switched ethernet evolved fronthaul for future cloud-{RAN} applications,'' in \emph{2017 European Conference on Optical Communication (ECOC)}, 2017, pp. 1--3.

\bibitem{8254712}
G.~S. Birring, P.~Assimakopoulos, and N.~J. Gomes, ``An ethernet-based fronthaul implementation with {MAC/PHY} split {LTE} processing,'' in \emph{GLOBECOM 2017 - 2017 IEEE Global Communications Conference}, 2017, pp. 1--6.

\bibitem{8025034}
P.~Assimakopulos, G.~S. Birring, M.~K. Al-Hares, and N.~J. Gomes, ``Ethernet-based fronthauling for cloud-radio access networks,'' in \emph{2017 19th International Conference on Transparent Optical Networks (ICTON)}, 2017, pp. 1--4.

\bibitem{8378032}
M.~J. Roldan, P.~Leithead, and J.~Mack, ``Experiments and results of a mm{W} transport platform to enable {5G} cloud {RAN} lower layer splits,'' in \emph{2018 IEEE Long Island Systems, Applications and Technology Conference (LISAT)}, 2018, pp. 1--6.

\bibitem{9200734}
D.~A. Temesgene, M.~Miozzo, D.~Gündüz, and P.~Dini, ``Distributed deep reinforcement learning for functional split control in energy harvesting virtualized small cells,'' \emph{IEEE Transactions on Sustainable Computing}, vol.~6, no.~4, pp. 626--640, 2021.

\bibitem{Marotta:19}
\BIBentryALTinterwordspacing
A.~Marotta, D.~Cassioli, K.~Kondepu, C.~Antonelli, and L.~Valcarenghi, ``Exploiting flexible functional split in converged software defined access networks,'' \emph{J. Opt. Commun. Netw.}, vol.~11, no.~11, pp. 536--546, Nov 2019. [Online]. Available: \url{http://opg.optica.org/jocn/abstract.cfm?URI=jocn-11-11-536}
\BIBentrySTDinterwordspacing

\bibitem{9678321}
D.~David López-Pérez, A.~De~Domenico, N.~Piovesan, G.~Xinli, H.~Bao, S.~Qitao, and M.~Debbah, ``A survey on {5G} radio access network energy efficiency: {M}assive {MIMO}, lean carrier design, sleep modes, and machine learning,'' \emph{IEEE Communications Surveys Tutorials}, vol.~24, no.~1, pp. 653--697, 2022.

\bibitem{GARIMA2022100674}
\BIBentryALTinterwordspacing
Garima, V.~Jha, and R.~K. Singh, ``A novel dynamic bandwidth allocation scheme for {XGPON} based mobile fronthaul for small cell {CRAN},'' \emph{Optical Switching and Networking}, vol.~45, p. 100674, 2022. [Online]. Available: \url{https://www.sciencedirect.com/science/article/pii/S1573427722000108}
\BIBentrySTDinterwordspacing

\bibitem{8644108}
A.~Alabbasi, M.~Berg, and C.~Cavdar, ``Delay constrained hybrid {CRAN}: {A} functional split optimization framework,'' in \emph{2018 IEEE Globecom Workshops (GC Wkshps)}, 2018, pp. 1--7.

\bibitem{9249096}
V.~Q. Rodriguez, R.~Corbel, F.~Guillemin, and A.~Ferrieux, ``Cloud-{RAN} factory: {I}nstantiating virtualized mobile networks with {ONAP},'' in \emph{2020 11th International Conference on Network of the Future (NoF)}, 2020, pp. 120--122.

\bibitem{iet1}
\BIBentryALTinterwordspacing
J.~Li, ``Hybrid beamforming designs for {5G} new radio with fronthaul compression and functional splits,'' \emph{IET Communications}, vol.~14, pp. 3676--3685(9), December 2020. [Online]. Available: \url{https://digital-library.theiet.org/content/journals/10.1049/iet-com.2020.0188}
\BIBentrySTDinterwordspacing

\bibitem{9148093}
V.~Q. Rodriguez, F.~Guillemin, A.~Ferrieux, and L.~Thomas, ``Cloud-{RAN} functional split for an efficient fronthaul network,'' in \emph{2020 International Wireless Communications and Mobile Computing (IWCMC)}, 2020, pp. 245--250.

\bibitem{8959317}
J.~Francis, J.~K. Chaudhary, A.~N. Barreto, and G.~Fettweis, ``Uplink latency in massive {MIMO}-based {C-RAN} with intra-{PHY} functional split,'' \emph{IEEE Communications Letters}, vol.~24, no.~4, pp. 912--916, 2020.

\bibitem{8891191}
Y.~Huang, W.~Lei, C.~Lu, and M.~Berg, ``Fronthaul functional split of {IRC}-based beamforming for massive {MIMO} systems,'' in \emph{2019 IEEE 90th Vehicular Technology Conference (VTC2019-Fall)}, 2019, pp. 1--5.

\bibitem{MEI2020368}
\BIBentryALTinterwordspacing
H.~Mei and L.~Peng, ``Flexible functional split for cost-efficient {C-RAN},'' \emph{Computer Communications}, vol. 161, pp. 368--374, 2020. [Online]. Available: \url{https://www.sciencedirect.com/science/article/pii/S0140366420318417}
\BIBentrySTDinterwordspacing

\bibitem{9492647}
M.~Sander-Frigau, T.~Zhang, H.~Zhang, A.~E. Kamal, and A.~K. Somani, ``Physical wireless resource virtualization for software-defined whole-stack slicing,'' in \emph{2021 IEEE 7th International Conference on Network Softwarization (NetSoft)}, 2021, pp. 106--114.

\bibitem{3GPP2}
3GPP, \emph{Transport Requirement for {CU} \& {DU} Functional Splits Options}.\hskip 1em plus 0.5em minus 0.4em\relax Document TSG RAN WG3 Meeting \# 93 R3-161813, 3GPP, August 2016.

\bibitem{6934902}
H.~Niu, C.~Li, A.~Papathanassiou, and G.~Wu, ``{RAN} architecture options and performance for {5G} network evolution,'' in \emph{2014 IEEE Wireless Communications and Networking Conference Workshops (WCNCW)}, 2014, pp. 294--298.

\bibitem{6923535}
D.~Wubben, P.~Rost, J.~S. Bartelt, M.~Lalam, V.~Savin, M.~Gorgoglione, A.~Dekorsy, and G.~Fettweis, ``Benefits and impact of cloud computing on {5G} signal processing: {F}lexible centralization through cloud-{RAN},'' \emph{IEEE Signal Processing Magazine}, vol.~31, no.~6, pp. 35--44, 2014.

\bibitem{7306544}
J.~Bartelt, P.~Rost, D.~Wubben, J.~Lessmann, B.~Melis, and G.~Fettweis, ``Fronthaul and backhaul requirements of flexibly centralized radio access networks,'' \emph{IEEE Wireless Communications}, vol.~22, no.~5, pp. 105--111, 2015.

\bibitem{7145716}
D.~Sabella, P.~Rost, A.~Banchs, V.~Savin, M.~Consonni, M.~Di~Girolamo, M.~Lalam, A.~Maeder, and I.~Berberana, ``Benefits and challenges of cloud technologies for {5G} architecture,'' in \emph{2015 IEEE 81st Vehicular Technology Conference (VTC Spring)}, 2015, pp. 1--5.

\bibitem{7841885}
C.-Y. Chang, N.~Nikaein, and T.~Spyropoulos, ``Impact of packetization and scheduling on {C-RAN} fronthaul performance,'' in \emph{2016 IEEE Global Communications Conference (GLOBECOM)}, 2016, pp. 1--7.

\bibitem{7503792}
M.~Jaber, D.~Owens, M.~A. Imran, R.~Tafazolli, and A.~Tukmanov, ``A joint backhaul and {RAN} perspective on the benefits of centralised {RAN} functions,'' in \emph{2016 IEEE International Conference on Communications Workshops (ICC)}, 2016, pp. 226--231.

\bibitem{Lund}
A.~Lund, \emph{Signal Processing for Next Generation Fronthaul Interface ({NGFI})}.\hskip 1em plus 0.5em minus 0.4em\relax Tech. Univ. Denmark, Lyngby, Denmark, 2016.

\bibitem{Obaidi}
R.~Al-Obaidi, \emph{Next Generation Fronthaul Networks}.\hskip 1em plus 0.5em minus 0.4em\relax Tech. Univ. Denmark, Lyngby, Denmark, 2016.

\bibitem{10.1007/978-3-319-52171-8_3}
M.~Makhanbet, X.~Zhang, H.~Gao, and H.~A. Suraweera, ``An overview of cloud {RAN}: {A}rchitecture, issues and future directions,'' in \emph{Emerging Trends in Electrical, Electronic and Communications Engineering}, P.~Fleming, N.~Vyas, S.~Sanei, and K.~Deb, Eds.\hskip 1em plus 0.5em minus 0.4em\relax Cham: Springer International Publishing, 2017, pp. 44--60.

\bibitem{8250956}
G.~O. Perez, J.~A. Hernández, and D.~L. Lopez, ``Delay analysis of fronthaul traffic in {5G} transport networks,'' in \emph{2017 IEEE 17th International Conference on Ubiquitous Wireless Broadband (ICUWB)}, 2017, pp. 1--5.

\bibitem{8801953}
J.~K. Chaudhary, A.~Kumar, J.~Bartelt, and G.~Fettweis, ``{C-RAN} employing x{RAN} functional split: {C}omplexity analysis for {5G NR} remote radio unit,'' in \emph{2019 European Conference on Networks and Communications (EuCNC)}, 2019, pp. 580--585.

\bibitem{9439518}
C.~Nahum, L.~Ramalho, A.~Klautau, E.~Medeiros, I.~Almeida, and E.~Trojer, ``Functional split and frequency-domain processing for fronthaul traffic reduction,'' \emph{IEEE Communications Letters}, vol.~25, no.~8, pp. 2758--2762, 2021.

\bibitem{Das:20}
\BIBentryALTinterwordspacing
S.~Das, F.~Slyne, A.~Kaszubowska, and M.~Ruffini, ``Virtualized east--west {PON} architecture supporting low-latency communication for mobile functional split based on multiaccess edge computing,'' \emph{J. Opt. Commun. Netw.}, vol.~12, no.~10, pp. D109--D119, Oct 2020. [Online]. Available: \url{http://opg.optica.org/jocn/abstract.cfm?URI=jocn-12-10-D109}
\BIBentrySTDinterwordspacing

\bibitem{6886880}
S.-H. Cho, H.~Park, H.~S. Chung, K.~H. Doo, S.~Lee, and J.~H. Lee, ``Cost-effective next generation mobile fronthaul architecture with multi-if carrier transmission scheme,'' in \emph{OFC 2014}, 2014, pp. 1--3.

\bibitem{7511579}
C.-Y. Chang, R.~Schiavi, N.~Nikaein, T.~Spyropoulos, and C.~Bonnet, ``Impact of packetization and functional split on {C-RAN} fronthaul performance,'' in \emph{2016 IEEE International Conference on Communications (ICC)}, 2016, pp. 1--7.

\bibitem{7504410}
J.~Duan, X.~Lagrange, and F.~Guilloud, ``Performance analysis of several functional splits in {C-RAN},'' in \emph{2016 IEEE 83rd Vehicular Technology Conference (VTC Spring)}, 2016, pp. 1--5.

\bibitem{8025007}
H.~Wang, M.~Aftab~Hossain, and C.~Cavdar, ``Cloud {RAN} architectures with optical and mm-{W}ave transport technologies,'' in \emph{2017 19th International Conference on Transparent Optical Networks (ICTON)}, 2017, pp. 1--4.

\bibitem{7996632}
C.-Y. Chang, N.~Nikaein, R.~Knopp, T.~Spyropoulos, and S.~S. Kumar, ``Flex{CRAN}: {A} flexible functional split framework over ethernet fronthaul in cloud-{RAN},'' in \emph{2017 IEEE International Conference on Communications (ICC)}, 2017, pp. 1--7.

\bibitem{8025037}
L.~Valcarenghi, F.~Giannone, D.~Manicone, and P.~Castoldi, ``Virtualized e{NB} latency limits,'' in \emph{2017 19th International Conference on Transparent Optical Networks (ICTON)}, 2017, pp. 1--4.

\bibitem{8169875}
H.~Gupta, D.~Manicone, F.~Giannone, K.~Kondepu, A.~Franklin, P.~Castoldi, and L.~Valcarenghi, ``How much is fronthaul latency budget impacted by {RAN} virtualisation ?'' in \emph{2017 IEEE Conference on Network Function Virtualization and Software Defined Networks (NFV-SDN)}, 2017, pp. 315--320.

\bibitem{8252720}
L.~Ramalho, I.~Freire, C.~Lu, M.~Berg, and A.~Klautau, ``Improved {LPC}-based fronthaul compression with high rate adaptation resolution,'' \emph{IEEE Communications Letters}, vol.~22, no.~3, pp. 458--461, 2018.

\bibitem{8252881}
A.~S. Thyagaturu, Z.~Alharbi, and M.~Reisslein, ``{R-FFT}: Function split at {IFFT/FFT} in unified {LTE} {CRAN} and cable access network,'' \emph{IEEE Transactions on Broadcasting}, vol.~64, no.~3, pp. 648--665, 2018.

\bibitem{9690182}
C.-Y. Chang, N.~Nikaein, T.~Spyropoulos, and K.~De~Schepper, ``Flex{DRAN}: {F}lexible centralization in disaggregated radio access networks,'' \emph{IEEE Access}, vol.~10, pp. 11\,789--11\,808, 2022.

\bibitem{6882182}
C.-L. I, J.~Huang, R.~Duan, C.~Cui, J.~Jiang, and L.~Li, ``Recent progress on {C-RAN} centralization and cloudification,'' \emph{IEEE Access}, vol.~2, pp. 1030--1039, 2014.

\bibitem{7950267}
D.~Wang, Y.~Wang, S.~Meng, and X.~Zhang, ``Game based wireless fronthaul {C-RAN} baseband function splitting and placement,'' in \emph{2016 12th International Conference on Mobile Ad-Hoc and Sensor Networks (MSN)}, 2016, pp. 400--404.

\bibitem{7247156}
N.~P. Anthapadmanabhan, A.~Walid, and T.~Pfeiffer, ``Mobile fronthaul over latency-optimized time division multiplexed passive optical networks,'' in \emph{2015 IEEE International Conference on Communication Workshop (ICCW)}, 2015, pp. 62--67.

\bibitem{KANI201542}
\BIBentryALTinterwordspacing
J.~ichi Kani, S.~Kuwano, and J.~Terada, ``Options for future mobile backhaul and fronthaul,'' \emph{Optical Fiber Technology}, vol.~26, pp. 42--49, 2015, next Generation Access Networks. [Online]. Available: \url{https://www.sciencedirect.com/science/article/pii/S1068520015000954}
\BIBentrySTDinterwordspacing

\bibitem{8045624}
K.~M.~S. Huq and J.~Rodriguez, \emph{Fronthaul for a Flexible Centralization in Cloud Radio Access Networks}, 2016, pp. 55--84.

\bibitem{GOMES201550}
\BIBentryALTinterwordspacing
N.~J. Gomes, P.~Chanclou, P.~Turnbull, A.~Magee, and V.~Jungnickel, ``Fronthaul evolution: {F}rom {CPRI} to ethernet,'' \emph{Optical Fiber Technology}, vol.~26, pp. 50--58, 2015, next Generation Access Networks. [Online]. Available: \url{https://www.sciencedirect.com/science/article/pii/S1068520015000942}
\BIBentrySTDinterwordspacing

\bibitem{7958544}
C.~Ranaweera, E.~Wong, A.~Nirmalathas, C.~Jayasundara, and C.~Lim, ``{5G} {C-RAN} architecture: {A} comparison of multiple optical fronthaul networks,'' in \emph{2017 International Conference on Optical Network Design and Modeling (ONDM)}, 2017, pp. 1--6.

\bibitem{7894280}
M.~Shafi, A.~F. Molisch, P.~J. Smith, T.~Haustein, P.~Zhu, P.~De~Silva, F.~Tufvesson, A.~Benjebbour, and G.~Wunder, ``{5G}: {A} tutorial overview of standards, trials, challenges, deployment, and practice,'' \emph{IEEE Journal on Selected Areas in Communications}, vol.~35, no.~6, pp. 1201--1221, 2017.

\bibitem{7925770}
G.~Mountaser, M.~L. Rosas, T.~Mahmoodi, and M.~Dohler, ``On the feasibility of {MAC} and {PHY} split in cloud {RAN},'' in \emph{2017 IEEE Wireless Communications and Networking Conference (WCNC)}, 2017, pp. 1--6.

\bibitem{555}
Z.~Guizani and N.~Hamdi, ``{CRAN}, {H-CRAN}, and {F-RAN} for {5G} systems: {K}ey capabilities and recent advances,'' \emph{International Journal of Network Management}, vol.~27, p. e1973, 09 2017.

\bibitem{8584159}
M.~A. {Hasabelnaby}, H.~A.~I. {Selmy}, and M.~I. {Dessouky}, ``Joint optimal transceiver placement and resource allocation schemes for redirected cooperative hybrid {FSO}/mm{W} {5G} fronthaul networks,'' \emph{IEEE/OSA Journal of Optical Communications and Networking}, vol.~10, no.~12, pp. 975--990, 2018.

\bibitem{9724184}
M.~Ahsan, A.~Ahmed, A.~Al-Dweik, and A.~Ahmad, ``Functional split-aware optimal {BBU} placement for {5G} cloud-{RAN} over {WDM} access/aggregation network,'' \emph{IEEE Systems Journal}, pp. 1--12, 2022.

\bibitem{7487973}
A.~Checko, A.~P. Avramova, M.~S. Berger, and H.~L. Christiansen, ``Evaluating {C-RAN} fronthaul functional splits in terms of network level energy and cost savings,'' \emph{Journal of Communications and Networks}, vol.~18, no.~2, pp. 162--172, 2016.

\bibitem{7343513}
N.~Shibata, K.~Miyamoto, S.~Kuwano, J.~Terada, and A.~Otaka, ``System level performance of uplink transmission in split-{PHY} processing architecture with joint reception for future radio access,'' in \emph{2015 IEEE 26th Annual International Symposium on Personal, Indoor, and Mobile Radio Communications (PIMRC)}, 2015, pp. 1375--1379.

\bibitem{NGMN}
NGMN, \emph{Further study on critical {C-RAN} technologies, version 1.0}.\hskip 1em plus 0.5em minus 0.4em\relax White Paper, NGMN, Frankfurt, Germany, March 2015.

\bibitem{3GPP3}
3GPP, \emph{R3-173402-Overall of Proposed {L}1 Processing Diagram, Rev3}.\hskip 1em plus 0.5em minus 0.4em\relax Tech. Univ. Denmark, Lyngby, Denmark, 2017.

\bibitem{Jin:21}
\BIBentryALTinterwordspacing
W.~Jin, Z.~Q. Zhong, S.~Jiang, J.~X. He, S.~H. Hu, D.~Chang, R.~P. Giddings, Y.~H. Hong, X.~Q. Jin, M.~O'Sullivan, T.~Durrant, J.~Trewern, G.~Mariani, and J.~M. Tang, ``Experimental demonstrations of {DSP}-enabled flexibility, adaptability and elasticity of multi-channel \&gt;72gb/s over 25\&\#x2005;km imdd transmission systems,'' \emph{Opt. Express}, vol.~29, no.~25, pp. 41\,363--41\,377, Dec 2021. [Online]. Available: \url{http://opg.optica.org/oe/abstract.cfm?URI=oe-29-25-41363}
\BIBentrySTDinterwordspacing

\bibitem{7121511}
K.~Miyamoto, S.~Kuwano, J.~Terada, and A.~Otaka, ``Split-{PHY} processing architecture to realize base station coordination and transmission bandwidth reduction in mobile fronthaul,'' in \emph{2015 Optical Fiber Communications Conference and Exhibition (OFC)}, 2015, pp. 1--3.

\bibitem{Miyamoto:16}
\BIBentryALTinterwordspacing
------, ``Analysis of mobile fronthaul bandwidth and wireless transmission performance in split-{PHY} processing architecture,'' \emph{Opt. Express}, vol.~24, no.~2, pp. 1261--1268, Jan 2016. [Online]. Available: \url{http://opg.optica.org/oe/abstract.cfm?URI=oe-24-2-1261}
\BIBentrySTDinterwordspacing

\bibitem{7830261}
K.~Miyamoto, S.~Kuwano, T.~Shimizu, J.~Terada, and A.~Otaka, ``Performance evaluation of ethernet-based mobile fronthaul and wireless comp in split-{PHY} processing,'' \emph{Journal of Optical Communications and Networking}, vol.~9, no.~1, pp. A46--A54, 2017.

\bibitem{8255992}
D.~Harutyunyan and R.~Riggio, ``Flexible functional split in {5G} networks,'' in \emph{2017 13th International Conference on Network and Service Management (CNSM)}, 2017, pp. 1--9.

\bibitem{9678343}
B.~Kharel, O.~L.~A. López, N.~H. Mahmood, H.~Alves, and M.~Latva-Aho, ``Fog-{RAN} enabled multi-connectivity and multi-cell scheduling framework for ultra-reliable low latency communication,'' \emph{IEEE Access}, vol.~10, pp. 7059--7072, 2022.

\bibitem{9221322}
C.~S. Shinde, ``A pragmatic industrial road map for shifting the existing fronthaul from {CPRI} to {5G} compatible e{CPRI},'' in \emph{2020 IEEE 3rd 5G World Forum (5GWF)}, 2020, pp. 297--302.

\bibitem{9838846}
O.~T. Demir, M.~Masoudi, E.~E.~Bjornson, and C.~Cavdar, ``Cell-free massive {MIMO} in virtualized {CRAN}: {H}ow to minimize the total network power?'' in \emph{ICC 2022 - IEEE International Conference on Communications}, 2022, pp. 159--164.

\bibitem{arxiv1}
\BIBentryALTinterwordspacing
Y.~Cao, P.~Wang, K.~Zheng, X.~Liang, D.~Liu, M.~Lou, J.~Jin, Q.~Wang, D.~Wang, Y.~Huang, X.~You, and J.~Wang, ``Experimental performance evaluation of cell-free massive {MIMO} systems using {COTS RRU} with {OTA} reciprocity calibration and phase synchronization,'' 2022. [Online]. Available: \url{https://arxiv.org/abs/2208.14048}
\BIBentrySTDinterwordspacing

\bibitem{arxiv2}
\BIBentryALTinterwordspacing
X.~You, Y.~Huang, S.~Liu, D.~Wang, J.~Ma, W.~Xu, C.~Zhang, H.~Zhan, C.~Zhang, J.~Zhang, J.~Li, M.~Zhu, J.~You, D.~Liu, S.~He, G.~He, F.~Yang, Y.~Liu, J.~Wu, J.~Lu, G.~Li, X.~Chen, W.~Chen, and W.~Gao, ``Toward {6G} tk$μ$ extreme connectivity: {A}rchitecture, key technologies and experiments,'' 2022. [Online]. Available: \url{https://arxiv.org/abs/2208.01190}
\BIBentrySTDinterwordspacing

\bibitem{ALBREEM202279}
\BIBentryALTinterwordspacing
M.~A. Albreem, A.~Alhabbash, A.~M. Abu-Hudrouss, and T.~A. Almohamad, ``Data detection in decentralized and distributed massive {MIMO} networks,'' \emph{Computer Communications}, vol. 189, pp. 79--99, 2022. [Online]. Available: \url{https://www.sciencedirect.com/science/article/pii/S0140366422000901}
\BIBentrySTDinterwordspacing

\bibitem{8960442}
J.~Rodríguez~Sánchez, F.~Rusek, O.~Edfors, M.~Sarajlić, and L.~Liu, ``Decentralized massive {MIMO} processing exploring daisy-chain architecture and recursive algorithms,'' \emph{IEEE Transactions on Signal Processing}, vol.~68, pp. 687--700, 2020.

\bibitem{9319703}
S.~Shahabuddin, A.~Mämmelä, M.~Juntti, and O.~Silvén, ``Asip for {5G} and beyond: {O}pportunities and vision,'' \emph{IEEE Transactions on Circuits and Systems II: Express Briefs}, vol.~68, no.~3, pp. 851--857, 2021.

\bibitem{9748751}
D.~Larrabeiti, G.~Otero, J.~P. Fernandez-Palacios, L.~M. Contreras, and J.~A. Hernandez, ``Latency-aware network architectures for {5G} backhaul and fronthaul,'' in \emph{2022 Optical Fiber Communications Conference and Exhibition (OFC)}, 2022, pp. 1--3.

\bibitem{9492417}
J.~A. Hernandez, G.~Otero, D.~Larrabeiti, and O.~G. de~Dios, ``Dimensioning flex ethernet groups for the transport of {5G NR} fronthaul traffic in {C-RAN} scenarios,'' in \emph{2021 International Conference on Optical Network Design and Modeling (ONDM)}, 2021, pp. 1--3.

\bibitem{8736787}
G.~Otero~Perez, D.~Larrabeiti~Lopez, and J.~A. Hernandez, ``{5G} new radio fronthaul network design for e{CPRI}-ieee 802.1cm and extreme latency percentiles,'' \emph{IEEE Access}, vol.~7, pp. 82\,218--82\,230, 2019.

\bibitem{9013123}
G.~Kalfas, M.~Agus, A.~Pagano, L.~A. Neto, A.~Mesodiakaki, C.~Vagionas, J.~Vardakas, E.~Datsika, C.~Verikoukis, and N.~Pleros, ``Converged analog fiber-wireless point-to-multipoint architecture for e{CPRI} {5G} fronthaul networks,'' in \emph{2019 IEEE Global Communications Conference (GLOBECOM)}, 2019, pp. 1--6.

\bibitem{9045398}
H.~Touati, H.~Castel-Taleb, B.~Jouaber, and S.~Akbarzadeh, ``Split analysis and fronthaul dimensioning in {5G} {C-RAN} to guarantee ultra low latency,'' in \emph{2020 IEEE 17th Annual Consumer Communications Networking Conference (CCNC)}, 2020, pp. 1--4.

\bibitem{8385866}
A.~Di~Giglio, A.~Tzanakaki, and D.~Simeonidou, ``Scenarios and economic analysis of fronthaul,'' in \emph{2018 Optical Fiber Communications Conference and Exposition (OFC)}, 2018, pp. 1--3.

\bibitem{8526294}
A.~D. Giglio and A.~Pagano, ``Scenarios and economic analysis of fronthaul in {5G} optical networks,'' \emph{Journal of Lightwave Technology}, vol.~37, no.~2, pp. 585--591, 2019.

\bibitem{8723481}
M.~A. Habibi, M.~Nasimi, B.~Han, and H.~D. Schotten, ``A comprehensive survey of {RAN} architectures toward {5G} mobile communication system,'' \emph{IEEE Access}, vol.~7, pp. 70\,371--70\,421, 2019.

\bibitem{9798822}
A.~Arnaz, J.~Lipman, M.~Abolhasan, and M.~Hiltunen, ``Toward integrating intelligence and programmability in open radio access networks: {A} comprehensive survey,'' \emph{IEEE Access}, vol.~10, pp. 67\,747--67\,770, 2022.

\bibitem{TS}
{3GPP Release 15 Technical specification TS 38.470}, \emph{{NG-RAN}; F1 interface}, June 2017.

\bibitem{TR}
{3GPP Release 15 Technical report TR 38.816}, \emph{Study on {CU-DU} lower layer split for {NR}}, September 2017.

\bibitem{77777}
N.~Mahmood, N.~Marchenko, M.~Gidlund, and P.~Popovski, \emph{Wireless Networks and Industrial IoT Applications, Challenges and Enablers: {A}pplications, Challenges and Enablers}, 01 2021.

\bibitem{9014264}
A.~Martinez~Alba, J.~H. Gomez~Velasquez, and W.~Kellerer, ``Traffic characterization of the {MAC-PHY} split in {5G} networks,'' in \emph{2019 IEEE Global Communications Conference (GLOBECOM)}, 2019, pp. 1--6.

\bibitem{TakahiroKubo20172017XBL0041}
T.~Kubo, S.~Nishihara, K.~Miyamoto, T.~Shimizu, J.~Terada, and A.~Ootaka, ``Downlink bandwidth in {C-RAN} with mapping {FAPI} over {R}o{E},'' \emph{IEICE Communications Express}, vol.~6, no.~6, pp. 417--422, 2017.

\bibitem{Yajima2019OverviewOO}
A.~U.~T. Yajima, T.~Uchino, and S.~Okuyama, ``Overview of {O-RAN} fronthaul specifications,'' in \emph{\it{NTT DOCOMO} Technical Journal}, vol.~21, no.~1, 2019, pp. 46--59.

\bibitem{9685992}
S.~Lagén, X.~Gelabert, L.~Giupponi, and A.~Hansson, ``Fronthaul-aware scheduling strategies for next generation {RAN}s,'' in \emph{2021 IEEE Global Communications Conference (GLOBECOM)}, 2021, pp. 1--6.

\bibitem{9887515}
A.~Ahmed, A.~Aburas, K.~Al-Mashouq, and A.~A. Aburas, ``{5G} small cell and open {RAN}: {D}esign parameter perspectives and analysis for indoor coverage,'' in \emph{2022 IEEE USNC-URSI Radio Science Meeting (Joint with AP-S Symposium)}, 2022, pp. 112--113.

\bibitem{Silva_Ramalho_Almeida_Medeiros_Berg_Klautau_2022}
\BIBentryALTinterwordspacing
M.~D. Silva, L.~Ramalho, I.~Almeida, E.~Medeiros, M.~Berg, and A.~Klautau, ``A new {O-RAN} compression approach for improved performance on uplink signals,'' \emph{Journal of Communication and Information Systems}, vol.~37, no.~1, pp. 30--41, Feb. 2022. [Online]. Available: \url{https://jcis.sbrt.org.br/jcis/article/view/808}
\BIBentrySTDinterwordspacing

\bibitem{8408566}
D.~Harutyunyan and R.~Riggio, ``Flex{5G}: {F}lexible functional split in {5G} networks,'' \emph{IEEE Transactions on Network and Service Management}, vol.~15, no.~3, pp. 961--975, 2018.

\bibitem{7534741}
J.~Gil~Herrera and J.~F. Botero, ``Resource allocation in {NFV}: {A} comprehensive survey,'' \emph{IEEE Transactions on Network and Service Management}, vol.~13, no.~3, pp. 518--532, 2016.

\bibitem{Frauendorf2023}
\BIBentryALTinterwordspacing
J.~L. Frauendorf and {\'E}.~Almeida~de Souza, \emph{The Evolution of {RAN} (Radio Access Network), {D-RAN}, {C-RAN}, {V-RAN}, and {O-RAN}}.\hskip 1em plus 0.5em minus 0.4em\relax Cham: Springer International Publishing, 2023, pp. 139--154. [Online]. Available: \url{https://doi.org/10.1007/978-3-031-10650-7_10}
\BIBentrySTDinterwordspacing

\bibitem{9500721}
T.~Pamuklu, M.~Erol-Kantarci, and C.~Ersoy, ``Reinforcement learning based dynamic function splitting in disaggregated green open {RAN}s,'' in \emph{ICC 2021 - IEEE International Conference on Communications}, 2021, pp. 1--6.

\bibitem{8432268}
T.~RABIA and O.~BRAHAM, ``A new {SDN}-based next generation fronthaul interface for a partially centralized {C-RAN},'' in \emph{2018 IEEE 32nd International Conference on Advanced Information Networking and Applications (AINA)}, 2018, pp. 393--398.

\bibitem{6898939}
P.~Rost, C.~J. Bernardos, A.~D. Domenico, M.~D. Girolamo, M.~Lalam, A.~Maeder, D.~Sabella, and D.~Wübben, ``Cloud technologies for flexible {5G} radio access networks,'' \emph{IEEE Communications Magazine}, vol.~52, no.~5, pp. 68--76, 2014.

\bibitem{7744804}
A.~Maeder, A.~Ali, A.~Bedekar, A.~F. Cattoni, D.~Chandramouli, S.~Chandrashekar, L.~Du, M.~Hesse, C.~Sartori, and S.~Turtinen, ``A scalable and flexible radio access network architecture for fifth generation mobile networks,'' \emph{IEEE Communications Magazine}, vol.~54, no.~11, pp. 16--23, 2016.

\bibitem{7929675}
O.~Chabbouh, S.~B. Rejeb, N.~Agoulmine, and Z.~Choukair, ``Cloud {RAN} architecture model based upon flexible {RAN} functionalities split for {5G} networks,'' in \emph{2017 31st International Conference on Advanced Information Networking and Applications Workshops (WAINA)}, 2017, pp. 184--188.

\bibitem{9844117}
H.~Gupta, A.~Antony~Franklin, M.~Kumar, and B.~R. Tamma, ``Traffic-aware dynamic functional split for {5G} cloud radio access networks,'' in \emph{2022 IEEE 8th International Conference on Network Softwarization (NetSoft)}, 2022, pp. 297--301.

\bibitem{9399108}
A.~M. Alba, S.~Janardhanan, and W.~Kellerer, ``Enabling dynamically centralized {RAN} architectures in {5G} and beyond,'' \emph{IEEE Transactions on Network and Service Management}, vol.~18, no.~3, pp. 3509--3526, 2021.

\bibitem{8283832}
A.~Garcia-Saavedra, J.~X. Salvat, X.~Li, and X.~Costa-Perez, ``Wizhaul: {O}n the centralization degree of cloud {RAN} next generation fronthaul,'' \emph{IEEE Transactions on Mobile Computing}, vol.~17, no.~10, pp. 2452--2466, 2018.

\bibitem{7248612}
J.~Liu, S.~Zhou, J.~Gong, Z.~Niu, and S.~Xu, ``Graph-based framework for flexible baseband function splitting and placement in {C-RAN},'' in \emph{2015 IEEE International Conference on Communications (ICC)}, 2015, pp. 1958--1963.

\bibitem{9316303}
R.~I. Rony, E.~Lopez-Aguilera, and E.~Garcia-Villegas, ``Cost analysis of {5G} fronthaul networks through functional splits at the {PHY} layer in a capacity and cost limited scenario,'' \emph{IEEE Access}, vol.~9, pp. 8733--8750, 2021.

\bibitem{9781611}
A.~Martínez~Alba and W.~Kellerer, ``Dynamic functional split adaptation in next-generation radio access networks,'' \emph{IEEE Transactions on Network and Service Management}, vol.~19, no.~3, pp. 3239--3263, 2022.

\bibitem{9946423}
R.~Joda, T.~Pamuklu, P.~E. Iturria-Rivera, and M.~Erol-Kantarci, ``Deep reinforcement learning-based joint user association and {CU-DU} placement in {O-RAN},'' \emph{IEEE Transactions on Network and Service Management}, pp. 1--1, 2022.

\bibitem{8486243}
A.~Garcia-Saavedra, X.~Costa-Perez, D.~J. Leith, and G.~Iosifidis, ``Fluidran: {O}ptimized v{RAN}/{MEC} orchestration,'' in \emph{IEEE INFOCOM 2018 - IEEE Conference on Computer Communications}, 2018, pp. 2366--2374.

\bibitem{8280524}
A.~Alabbasi, X.~Wang, and C.~Cavdar, ``Optimal processing allocation to minimize energy and bandwidth consumption in hybrid {CRAN},'' \emph{IEEE Transactions on Green Communications and Networking}, vol.~2, no.~2, pp. 545--555, 2018.

\bibitem{7997127}
X.~Wang, A.~Alabbasi, and C.~Cavdar, ``Interplay of energy and bandwidth consumption in {CRAN} with optimal function split,'' in \emph{2017 IEEE International Conference on Communications (ICC)}, 2017, pp. 1--6.

\bibitem{8386280}
K.~Kondepu, N.~Sambo, F.~Giannone, P.~Castoldi, and L.~Valcarenghi, ``Orchestrating lightpath adaptation and flexible functional split to recover virtualized {RAN} connectivity,'' in \emph{2018 Optical Fiber Communications Conference and Exposition (OFC)}, 2018, pp. 1--3.

\bibitem{8718107}
Z.~Cheng, Y.~Tang, and H.~Wu, ``Joint task offloading and flexible functional split in {5G} radio access network,'' in \emph{2019 International Conference on Information Networking (ICOIN)}, 2019, pp. 114--119.

\bibitem{9912981}
M.~R. Dos~Santos, M.~K.~R. Mei, A.~C. Oliveira, R.~A.~L. Rabelo, and G.~B. Figueiredo, ``A new soccer game optimization modeling for flexible functional splitting dimensioning in {CF-RAN} networks,'' in \emph{2022 IEEE Symposium on Computers and Communications (ISCC)}, 2022, pp. 1--6.

\bibitem{8612914}
R.~Kassab, O.~Simeone, P.~Popovski, and T.~Islam, ``Non-orthogonal multiplexing of ultra-reliable and broadband services in fog-radio architectures,'' \emph{IEEE Access}, vol.~7, pp. 13\,035--13\,049, 2019.

\bibitem{8790771}
J.~Moon, O.~Simeone, S.-H. Park, and I.~Lee, ``Online reinforcement learning of x-haul content delivery mode in fog radio access networks,'' \emph{IEEE Signal Processing Letters}, vol.~26, no.~10, pp. 1451--1455, 2019.

\bibitem{8422754}
S.~Matoussi, I.~Fajjari, S.~Costanzo, N.~Aitsaadi, and R.~Langar, ``A user centric virtual network function orchestration for agile {5G} cloud-{RAN},'' in \emph{2018 IEEE International Conference on Communications (ICC)}, 2018, pp. 1--7.

\bibitem{8403571}
L.~Wang and S.~Zhou, ``Flexible functional split in {C-RAN} with renewable energy powered remote radio units,'' in \emph{2018 IEEE International Conference on Communications Workshops (ICC Workshops)}, 2018, pp. 1--6.

\bibitem{8917915}
------, ``Flexible functional split and power control for energy harvesting cloud radio access networks,'' \emph{IEEE Transactions on Wireless Communications}, vol.~19, no.~3, pp. 1535--1548, 2020.

\bibitem{8877874}
H.~Gupta, M.~Sharma, A.~Franklin~A., and B.~R. Tamma, ``{Apt-RAN}: {A} flexible split-based {5G} {RAN} to minimize energy consumption and handovers,'' \emph{IEEE Transactions on Network and Service Management}, vol.~17, no.~1, pp. 473--487, 2020.

\bibitem{8693874}
Y.~Zhou, J.~Li, Y.~Shi, and V.~W.~S. Wong, ``Flexible functional split design for downlink {C-RAN} with capacity-constrained fronthaul,'' \emph{IEEE Transactions on Vehicular Technology}, vol.~68, no.~6, pp. 6050--6063, 2019.

\bibitem{9751689}
F.~S. Vajd, M.~Hadi, C.~Bhar, M.~R. Pakravan, and E.~Agrell, ``Dynamic joint functional split and resource allocation optimization in elastic optical fronthaul,'' \emph{IEEE Transactions on Network and Service Management}, pp. 1--1, 2022.

\bibitem{8761081}
B.~Ojaghi, F.~Adelantado, E.~Kartsakli, A.~Antonopoulos, and C.~Verikoukis, ``Sliced-{RAN}: {J}oint slicing and functional split in future {5G} radio access networks,'' in \emph{ICC 2019 - 2019 IEEE International Conference on Communications (ICC)}, 2019, pp. 1--6.

\bibitem{8762089}
A.~De~Domenico, Y.-F. Liu, and W.~Yu, ``Optimal computational resource allocation and network slicing deployment in {5G} hybrid {C-RAN},'' in \emph{ICC 2019 - 2019 IEEE International Conference on Communications (ICC)}, 2019, pp. 1--6.

\bibitem{9615541}
E.~Sarikaya and E.~Onur, ``Placement of {5G} {RAN} slices in multi-tier {O-RAN} {5G} networks with flexible functional splits,'' in \emph{2021 17th International Conference on Network and Service Management (CNSM)}, 2021, pp. 274--282.

\bibitem{9617786}
M.~Dalgitsis, J.~S. Vardakas, and C.~Verikoukis, ``{5G} {RAN} resource slicing with flexible functional splits over multi-tenant environment,'' in \emph{2021 IEEE 26th International Workshop on Computer Aided Modeling and Design of Communication Links and Networks (CAMAD)}, 2021, pp. 1--6.

\bibitem{8361844}
Z.~Zaidi, V.~Friderikos, Z.~Yousaf, S.~Fletcher, M.~Dohler, and H.~Aghvami, ``Will {SDN} be part of {5G}?'' \emph{IEEE Communications Surveys Tutorials}, vol.~20, no.~4, pp. 3220--3258, 2018.

\bibitem{7010528}
M.~Y. Arslan, K.~Sundaresan, and S.~Rangarajan, ``Software-defined networking in cellular radio access networks: {P}otential and challenges,'' \emph{IEEE Communications Magazine}, vol.~53, no.~1, pp. 150--156, 2015.

\bibitem{8386082}
J.~Zou, A.~Magee, M.~Eiselt, A.~Straw, T.~Edwards, P.~Wright, and A.~Lord, ``Demonstration of x-haul architecture for {5G} over converged {SDN} fiber network,'' in \emph{2018 Optical Fiber Communications Conference and Exposition (OFC)}, 2018, pp. 1--3.

\bibitem{8428882}
G.~C. Valastro, D.~Panno, and S.~Riolo, ``A {SDN/NFV} based {C-RAN} architecture for {5G} mobile networks,'' in \emph{2018 International Conference on Selected Topics in Mobile and Wireless Networking (MoWNeT)}, 2018, pp. 1--8.

\bibitem{8422555}
A.~Marotta, D.~Cassioli, K.~Kondepu, C.~Antonelli, and L.~Valcarenghi, ``Efficient management of flexible functional split through software defined {5G} converged access,'' in \emph{2018 IEEE International Conference on Communications (ICC)}, 2018, pp. 1--6.

\bibitem{9061000}
Y.~Li, J.~Martensson, B.~Skubic, Y.~Zhao, J.~Zhang, L.~Wosinska, and P.~Monti, ``Flexible {RAN}: Combining dynamic baseband split selection and reconfigurable optical transport to optimize {RAN} performance,'' \emph{IEEE Network}, vol.~34, no.~4, pp. 180--187, 2020.

\bibitem{8386298}
Y.~Alfadhli, M.~Xu, S.~Liu, F.~Lu, P.-C. Peng, and G.-K. Chang, ``Real-time demonstration of adaptive functional split in {5G} flexible mobile fronthaul networks,'' in \emph{2018 Optical Fiber Communications Conference and Exposition (OFC)}, 2018, pp. 1--3.

\bibitem{Dalgitsis2022}
\BIBentryALTinterwordspacing
M.~Dalgitsis, M.~Mosahebfard, E.~Datsika, and J.~S. Vardakas, \emph{{SDN}-Based Resource Management for Optical-Wireless Fronthaul}.\hskip 1em plus 0.5em minus 0.4em\relax Cham: Springer International Publishing, 2022, pp. 467--500. [Online]. Available: \url{https://doi.org/10.1007/978-3-030-74648-3_14}
\BIBentrySTDinterwordspacing

\bibitem{8432465}
H.~Cheng, X.~Yuan, and Y.~Tan, ``Generalized compute-compress-and-forward,'' \emph{IEEE Transactions on Information Theory}, vol.~65, no.~1, pp. 462--481, 2019.

\bibitem{8114545}
Y.~{Li}, T.~{Jiang}, K.~{Luo}, and S.~{Mao}, ``Green heterogeneous cloud radio access networks: {P}otential techniques, performance trade-offs, and challenges,'' \emph{IEEE Communications Magazine}, vol.~55, no.~11, pp. 33--39, 2017.

\bibitem{1499041}
G.~Kramer, M.~Gastpar, and P.~Gupta, ``Cooperative strategies and capacity theorems for relay networks,'' \emph{IEEE Transactions on Information Theory}, vol.~51, no.~9, pp. 3037--3063, 2005.

\bibitem{7573000}
S.~{Park}, K.~{Lee}, C.~{Song}, and I.~{Lee}, ``Joint design of fronthaul and access links for {C-RAN} with wireless fronthauling,'' \emph{IEEE Signal Processing Letters}, vol.~23, no.~11, pp. 1657--1661, 2016.

\bibitem{4305394}
A.~El~Gamal, N.~Hassanpour, and J.~Mammen, ``Relay networks with delays,'' \emph{IEEE Transactions on Information Theory}, vol.~53, no.~10, pp. 3413--3431, 2007.

\bibitem{8644380}
M.~Raceala-Motoc, P.~Jung, Z.~Utkovski, and S.~Stańczak, ``{C-RAN}-assisted non-coherent grant-free random access based on compute-and-forward,'' in \emph{2018 IEEE Gl\textbf{}obecom Workshops (GC Wkshps)}, 2018, pp. 1--7.

\bibitem{9148793}
S.~Gelincik and G.~R.-B. Othman, ``Lattice codes for {C-RAN} based sectored cellular networks,'' in \emph{ICC 2020 - 2020 IEEE International Conference on Communications (ICC)}, 2020, pp. 1--7.

\bibitem{7962724}
Q.~Huang and A.~Burr, ``Compute-and-forward in cell-free massive {MIMO}: Great performance with low backhaul load,'' in \emph{2017 IEEE International Conference on Communications Workshops (ICC Workshops)}, 2017, pp. 601--606.

\bibitem{9755878}
A.~L.~P. Fernandes, D.~D. Souza, D.~B. da~Costa, A.~M. Cavalcante, and J.~C. W.~A. Costa, ``Cell-free massive {MIMO} with segmented fronthaul: {R}eliability and protection aspects,'' \emph{IEEE Wireless Communications Letters}, vol.~11, no.~8, pp. 1580--1584, 2022.

\bibitem{JIANG2023}
\BIBentryALTinterwordspacing
H.~Jiang, L.~Kong, and S.~Du, ``Compute-and-forward transmission scheme in cell-free massive {MIMO} systems,'' \emph{ICT Express}, 2023. [Online]. Available: \url{https://www.sciencedirect.com/science/article/pii/S2405959523000255}
\BIBentrySTDinterwordspacing

\bibitem{6034734}
B.~{Nazer} and M.~{Gastpar}, ``Compute-and-forward: {H}arnessing interference through structured codes,'' \emph{IEEE Transactions on Information Theory}, vol.~57, no.~10, pp. 6463--6486, 2011.

\bibitem{WPNC}
\emph{Wireless Physical Layer Network Coding}.\hskip 1em plus 0.5em minus 0.4em\relax Cambridge University Press, 2018.

\bibitem{8186899}
A.~Chaaban and A.~Sezgin, \emph{Multi-way Communications: {A}n information theoretic perspective}, 2015, vol.~12, no. 3-4.

\bibitem{5205840}
B.~{Nazer}, A.~{Sanderovich}, M.~{Gastpar}, and S.~{Shamai}, ``Structured superposition for backhaul constrained cellular uplink,'' in \emph{2009 IEEE International Symposium on Information Theory}, 2009, pp. 1530--1534.

\bibitem{7028519}
Y.~Tan, X.~Yuan, S.~C. Liew, and A.~Kavcic, ``Asymmetric compute-and-forward: {G}oing beyond one hop,'' in \emph{2014 52nd Annual Allerton Conference on Communication, Control, and Computing (Allerton)}, 2014, pp. 667--674.

\bibitem{6736658}
V.~Ntranos, V.~R. Cadambe, B.~Nazer, and G.~Caire, ``Asymmetric compute-and-forward,'' in \emph{2013 51st Annual Allerton Conference on Communication, Control, and Computing (Allerton)}, 2013, pp. 1174--1181.

\bibitem{7017572}
J.~Zhu and M.~Gastpar, ``Lattice codes for many-to-one interference channels with and without cognitive messages,'' \emph{IEEE Transactions on Information Theory}, vol.~61, no.~3, pp. 1309--1324, 2015.

\bibitem{7517355}
B.~Nazer, V.~R. Cadambe, V.~Ntranos, and G.~Caire, ``Expanding the compute-and-forward framework: {U}nequal powers, signal levels, and multiple linear combinations,'' \emph{IEEE Transactions on Information Theory}, vol.~62, no.~9, pp. 4879--4909, 2016.

\bibitem{8756704}
J.~Zhang, J.~Zhang, J.~Zheng, S.~Jin, and B.~Ai, ``Expanded compute-and-forward for backhaul-limited cell-free massive {MIMO},'' in \emph{2019 IEEE International Conference on Communications Workshops (ICC Workshops)}, 2019, pp. 1--6.

\bibitem{6736536}
O.~Ordentlich, U.~Erez, and B.~Nazer, ``Successive integer-forcing and its sum-rate optimality,'' in \emph{2013 51st Annual Allerton Conference on Communication, Control, and Computing (Allerton)}, 2013, pp. 282--292.

\bibitem{7349245}
M.~Hejazi, S.~M. Azimi-Abarghouyi, B.~Makki, M.~Nasiri-Kenari, and T.~Svensson, ``Robust successive compute-and-forward over multiuser multirelay networks,'' \emph{IEEE Transactions on Vehicular Technology}, vol.~65, no.~10, pp. 8112--8129, 2016.

\bibitem{9622183}
J.~Zhang, J.~Zhang, D.~W.~K. Ng, S.~Jin, and B.~Ai, ``Improving sum-rate of cell-free massive{MIMO} with expanded compute-and-forward,'' \emph{IEEE Transactions on Signal Processing}, vol.~70, pp. 202--215, 2022.

\bibitem{7275188}
Y.~Tan and X.~Yuan, ``Compute-compress-and-forward: {E}xploiting asymmetry of wireless relay networks,'' \emph{IEEE Transactions on Signal Processing}, vol.~64, no.~2, pp. 511--524, 2016.

\bibitem{7572049}
I.~E. {Aguerri} and A.~{Zaidi}, ``Lossy compression for compute-and-forward in limited backhaul uplink multicell processing,'' \emph{IEEE Transactions on Communications}, vol.~64, no.~12, pp. 5227--5238, 2016.

\bibitem{7465790}
Y.~{Zhou} and W.~{Yu}, ``Fronthaul compression and transmit beamforming optimization for multi-antenna uplink {C-RAN},'' \emph{IEEE Transactions on Signal Processing}, vol.~64, no.~16, pp. 4138--4151, 2016.

\bibitem{4305387}
S.~Borade, L.~Zheng, and R.~Gallager, ``Amplify-and-forward in wireless relay networks: {R}ate, diversity, and network size,'' \emph{IEEE Transactions on Information Theory}, vol.~53, no.~10, pp. 3302--3318, 2007.

\bibitem{6034234}
B.~Liu and N.~Cai, ``Analog network coding in the generalized high-{SNR} regime,'' in \emph{2011 IEEE International Symposium on Information Theory Proceedings}, 2011, pp. 74--78.

\bibitem{6824778}
Y.~{Zhou} and W.~{Yu}, ``Optimized backhaul compression for uplink cloud radio access network,'' \emph{IEEE Journal on Selected Areas in Communications}, vol.~32, no.~6, pp. 1295--1307, 2014.

\bibitem{5730555}
A.~S. Avestimehr, S.~N. Diggavi, and D.~N.~C. Tse, ``Wireless network information flow: {A} deterministic approach,'' \emph{IEEE Transactions on Information Theory}, vol.~57, no.~4, pp. 1872--1905, 2011.

\bibitem{5594708}
D.~{Gesbert}, S.~{Hanly}, H.~{Huang}, S.~{Shamai Shitz}, O.~{Simeone}, and W.~{Yu}, ``Multi-cell {MIMO} cooperative networks: {A} new look at interference,'' \emph{IEEE Journal on Selected Areas in Communications}, vol.~28, no.~9, pp. 1380--1408, 2010.

\bibitem{6924850}
S.-H. Park, O.~Simeone, O.~Sahin, and S.~Shamai~Shitz, ``Fronthaul compression for cloud radio access networks: {S}ignal processing advances inspired by network information theory,'' \emph{IEEE Signal Processing Magazine}, vol.~31, no.~6, pp. 69--79, 2014.

\bibitem{7590010}
Y.~{Zhou}, Y.~{Xu}, W.~{Yu}, and J.~{Chen}, ``On the optimal fronthaul compression and decoding strategies for uplink cloud radio access networks,'' \emph{IEEE Transactions on Information Theory}, vol.~62, no.~12, pp. 7402--7418, 2016.

\bibitem{1055508}
A.~Wyner and J.~Ziv, ``The rate-distortion function for source coding with side information at the decoder,'' \emph{IEEE Transactions on Information Theory}, vol.~22, no.~1, pp. 1--10, 1976.

\bibitem{6601765}
L.~Zhou and W.~Yu, ``Uplink multicell processing with limited backhaul via per-base-station successive interference cancellation,'' \emph{IEEE Journal on Selected Areas in Communications}, vol.~31, no.~10, pp. 1981--1993, 2013.

\bibitem{706450}
R.~{Zamir} and S.~{Shamai}, ``Nested linear/lattice codes for {W}yner-{Z}iv encoding,'' in \emph{1998 Information Theory Workshop (Cat. No.98EX131)}, 1998, pp. 92--93.

\bibitem{6089357}
Y.~{Song} and N.~{Devroye}, ``A lattice compress-and-forward scheme,'' in \emph{2011 IEEE Information Theory Workshop}, 2011, pp. 110--114.

\bibitem{5766129}
------, ``A lattice compress-and-forward strategy for canceling known interference in {G}aussian multi-hop channels,'' in \emph{2011 45th Annual Conference on Information Sciences and Systems}, 2011, pp. 1--5.

\bibitem{7745894}
O.~{Ordentlich} and U.~{Erez}, ``Integer-forcing source coding,'' \emph{IEEE Transactions on Information Theory}, vol.~63, no.~2, pp. 1253--1269, 2017.

\bibitem{8262720}
I.~{El Bakoury} and B.~{Nazer}, ``Integer-forcing architectures for uplink cloud radio access networks,'' in \emph{2017 55th Annual Allerton Conference on Communication, Control, and Computing (Allerton)}, 2017, pp. 67--75.

\bibitem{8422865}
M.~Bashar, K.~Cumanan, A.~G. Burr, H.~Q. Ngo, and M.~Debbah, ``Cell-free massive {MIMO} with limited backhaul,'' in \emph{2018 IEEE International Conference on Communications (ICC)}, 2018, pp. 1--7.

\bibitem{8645433}
M.~Bashar, H.~Q. Ngo, A.~G. Burr, D.~Maryopi, K.~Cumanan, and E.~G. Larsson, ``On the performance of backhaul constrained cell-free massive {MIMO} with linear receivers,'' in \emph{2018 52nd Asilomar Conference on Signals, Systems, and Computers}, 2018, pp. 624--628.

\bibitem{9110901}
M.~Bashar, A.~Akbari, K.~Cumanan, H.~Q. Ngo, A.~G. Burr, P.~Xiao, M.~Debbah, and J.~Kittler, ``Exploiting deep learning in limited-fronthaul cell-free massive {MIMO} uplink,'' \emph{IEEE Journal on Selected Areas in Communications}, vol.~38, no.~8, pp. 1678--1697, 2020.

\bibitem{8891922}
H.~Masoumi and M.~J. Emadi, ``Performance analysis of cell-free massive {MIMO} system with limited fronthaul capacity and hardware impairments,'' \emph{IEEE Transactions on Wireless Communications}, vol.~19, no.~2, pp. 1038--1053, 2020.

\bibitem{Bashar20201}
\BIBentryALTinterwordspacing
M.~Bashar, A.~Akbari, K.~Cumanan, H.~Quoc~Ngo, A.~Burr, P.~Xiao, and M.~Debbah, ``Deep learning-aided finite-capacity fronthaul cellfree massive {MIMO} with zero forcing,'' \emph{Proc. Icc}, p. 1 – 6, 2020, cited by: 2. [Online]. Available: \url{https://www.scopus.com/inward/record.uri?eid=2-s2.0-85124804924&partnerID=40&md5=a843286188d1d68061936e81798cf078}
\BIBentrySTDinterwordspacing

\bibitem{9123382}
G.~Femenias and F.~Riera-Palou, ``Fronthaul-constrained cell-free massive {MIMO} with low resolution {ADCs},'' \emph{IEEE Access}, vol.~8, pp. 116\,195--116\,215, 2020.

\bibitem{8761134}
M.~Bashar, K.~Cumanan, A.~G. Burr, H.~Q. Ngo, E.~G. Larsson, and P.~Xiao, ``On the energy efficiency of limited-backhaul cell-free massive {MIMO},'' in \emph{ICC 2019 - 2019 IEEE International Conference on Communications (ICC)}, 2019, pp. 1--7.

\bibitem{8756286}
M.~Bashar, K.~Cumanan, A.~G. Burr, H.~Q. Ngo, M.~Debbah, and P.~Xiao, ``Max–min rate of cell-free massive {MIMO} uplink with optimal uniform quantization,'' \emph{IEEE Transactions on Communications}, vol.~67, no.~10, pp. 6796--6815, 2019.

\bibitem{8359129}
C.~{Wang}, M.~{Wigger}, and A.~{Zaidi}, ``On achievability for downlink cloud radio access networks with base station cooperation,'' \emph{IEEE Transactions on Information Theory}, vol.~64, no.~8, pp. 5726--5742, 2018.

\bibitem{5997324}
R.~{Zakhour} and D.~{Gesbert}, ``Optimized data sharing in {MIMO} with finite backhaul capacity,'' \emph{IEEE Transactions on Signal Processing}, vol.~59, no.~12, pp. 6102--6111, 2011.

\bibitem{8732995}
A.~{Alameer Ahmad}, H.~{Dahrouj}, A.~{Chaaban}, A.~{Sezgin}, and M.~{Alouini}, ``Interference mitigation via rate-splitting and common message decoding in cloud radio access networks,'' \emph{IEEE Access}, vol.~7, pp. 80\,350--80\,365, 2019.

\bibitem{8690797}
P.~Parida, H.~S. Dhillon, and A.~F. Molisch, ``Downlink performance analysis of cell-free massive {MIMO} with finite fronthaul capacity,'' in \emph{2018 IEEE 88th Vehicular Technology Conference (VTC-Fall)}, 2018, pp. 1--6.

\bibitem{8902784}
G.~Femenias and F.~Riera-Palou, ``Reduced-complexity downlink cell-free mmwave massive {MIMO} systems with fronthaul constraints,'' in \emph{2019 27th European Signal Processing Conference (EUSIPCO)}, 2019, pp. 1--5.

\bibitem{8678745}
------, ``Cell-free millimeter-wave massive {MIMO} systems with limited fronthaul capacity,'' \emph{IEEE Access}, vol.~7, pp. 44\,596--44\,612, 2019.

\bibitem{10.1007/s11277-018-6038-1}
\BIBentryALTinterwordspacing
M.~N. Boroujerdi, A.~Abbasfar, and M.~Ghanbari, ``Cell free massive {MIMO} with limited capacity fronthaul,'' \emph{Wirel. Pers. Commun.}, vol. 104, no.~2, p. 633–648, jan 2019. [Online]. Available: \url{https://doi.org/10.1007/s11277-018-6038-1}
\BIBentrySTDinterwordspacing

\bibitem{7362826}
P.~{Patil}, B.~{Dai}, and W.~{Yu}, ``Performance comparison of data-sharing and compression strategies for cloud radio access networks,'' in \emph{2015 23rd European Signal Processing Conference (EUSIPCO)}, 2015, pp. 2456--2460.

\bibitem{7541570}
L.~{Liu}, P.~{Patil}, and W.~{Yu}, ``An uplink-downlink duality for cloud radio access network,'' in \emph{2016 IEEE International Symposium on Information Theory (ISIT)}, 2016, pp. 1606--1610.

\bibitem{8635883}
I.~E. {Bakoury} and B.~{Nazer}, ``Uplink-downlink duality for integer-forcing in cloud radio access networks,'' in \emph{2018 56th Annual Allerton Conference on Communication, Control, and Computing (Allerton)}, 2018, pp. 39--47.

\bibitem{9893884}
M.~A. Hasabelnaby and A.~Chaaban, ``Multi-pair computation for {C-RAN} with intra-cloud and inter-cloud communications,'' \emph{IEEE Wireless Communications Letters}, vol.~11, no.~12, pp. 2537--2541, 2022.

\bibitem{9687122}
------, ``Multi-pair computation for two-way intra cloud radio-access network communications,'' \emph{IEEE Transactions on Wireless Communications}, vol.~21, no.~7, pp. 5586--5599, 2022.

\bibitem{miretti2023uldl}
L.~Miretti, R.~L.~G. Cavalcante, E.~Björnson, and S.~Stańczak, ``{UL-DL} duality for cell-free massive {MIMO} with per-{AP} power and information constraints,'' 2023.

\bibitem{6449246}
X.~Hou and C.~Yang, ``Feedback overhead analysis for base station cooperative transmission,'' \emph{IEEE Transactions on Wireless Communications}, vol.~15, no.~7, pp. 4491--4504, 2016.

\bibitem{8387197}
C.~Pan, M.~Elkashlan, J.~Wang, J.~Yuan, and L.~Hanzo, ``User-centric {C-RAN} architecture for ultra-dense {5G} networks: {C}hallenges and methodologies,'' \emph{IEEE Communications Magazine}, vol.~56, no.~6, pp. 14--20, 2018.

\bibitem{8379438}
Z.~{Chen} and E.~{Björnson}, ``Channel hardening and favorable propagation in cell-free massive {MIMO} with stochastic geometry,'' \emph{IEEE Transactions on Communications}, vol.~66, no.~11, pp. 5205--5219, 2018.

\bibitem{8799031}
G.~Interdonato, H.~Q. Ngo, P.~Frenger, and E.~G. Larsson, ``Downlink training in cell-free massive {MIMO}: {A} blessing in disguise,'' \emph{IEEE Transactions on Wireless Communications}, vol.~18, no.~11, pp. 5153--5169, 2019.

\bibitem{6812159}
H.~{Yin}, D.~{Gesbert}, and L.~{Cottatellucci}, ``Dealing with interference in distributed large-scale {MIMO} systems: {A} statistical approach,'' \emph{IEEE Journal of Selected Topics in Signal Processing}, vol.~8, no.~5, pp. 942--953, 2014.

\bibitem{8780756}
Y.~Zhang, H.~Cao, P.~Zhong, C.~Qi, and L.~Yang, ``Location-based greedy pilot assignment for cell-free massive {MIMO} systems,'' in \emph{2018 IEEE 4th International Conference on Computer and Communications (ICCC)}, 2018, pp. 392--396.

\bibitem{R25}
M.~{Attarifar}, A.~{Abbasfar}, and A.~{Lozano}, ``Random vs structured pilot assignment in cell-free massive {MIMO} wireless networks,'' in \emph{2018 IEEE International Conference on Communications Workshops (ICC Workshops)}, 2018, pp. 1--6.

\bibitem{R26}
R.~{Sabbagh}, C.~{Pan}, and J.~{Wang}, ``Pilot allocation and sum-rate analysis in cell-free massive {MIMO} systems,'' in \emph{2018 IEEE International Conference on Communications (ICC)}, 2018, pp. 1--6.

\bibitem{8914726}
H.~{Liu}, J.~{Zhang}, X.~{Zhang}, A.~{Kurniawan}, T.~{Juhana}, and B.~{Ai}, ``Tabu-search-based pilot assignment for cell-free massive {MIMO} systems,'' \emph{IEEE Transactions on Vehicular Technology}, vol.~69, no.~2, pp. 2286--2290, 2020.

\bibitem{7080877}
Z.~Chen, X.~Hou, and C.~Yang, ``Training resource allocation for user-centric base station cooperation networks,'' \emph{IEEE Transactions on Vehicular Technology}, vol.~65, no.~4, pp. 2729--2735, 2016.

\bibitem{8247283}
C.~Pan, H.~Mehrpouyan, Y.~Liu, M.~Elkashlan, and N.~Arumugam, ``Joint pilot allocation and robust transmission design for ultra-dense user-centric {TDD C-RAN} with imperfect {CSI},'' \emph{IEEE Transactions on Wireless Communications}, vol.~17, no.~3, pp. 2038--2053, 2018.

\bibitem{9154310}
H.~Song, X.~You, C.~Zhang, O.~Tirkkonen, and C.~Studer, ``Minimizing pilot overhead in cell-free massive {MIMO} systems via joint estimation and detection,'' in \emph{2020 IEEE 21st International Workshop on Signal Processing Advances in Wireless Communications (SPAWC)}, 2020, pp. 1--5.

\bibitem{9201112}
Y.~Zhang, X.~Qiao, L.~Yang, and H.~Zhu, ``Superimposed pilots are beneficial for mitigating pilot contamination in cell-free massive {MIMO},'' \emph{IEEE Communications Letters}, vol.~25, no.~1, pp. 279--283, 2021.

\bibitem{8815888}
Y.~Jin, J.~Zhang, S.~Jin, and B.~Ai, ``Channel estimation for cell-free mmwave massive {MIMO} through deep learning,'' \emph{IEEE Transactions on Vehicular Technology}, vol.~68, no.~10, pp. 10\,325--10\,329, 2019.

\bibitem{obeed2022alternating}
M.~Obeed, Y.~Al-Eryani, and A.~Chaaban, ``Alternating channel estimation and prediction for cell-free {mMIMO} with channel aging: A deep learning based scheme,'' pp. 3590--3595, 2023.

\bibitem{7869024}
E.~Nayebi, A.~Ashikhmin, T.~L. Marzetta, and B.~D. Rao, ``Performance of cell-free massive {MIMO} systems with {MMSE} and {LSFD} receivers,'' in \emph{2016 50th Asilomar Conference on Signals, Systems and Computers}, 2016, pp. 203--207.

\bibitem{9739189}
O.~T. Demir, E.~Bjoernson, and L.~Sanguinetti, ``Cell-free massive {MIMO} with large-scale fading decoding and dynamic cooperation clustering,'' in \emph{WSA 2021; 25th International ITG Workshop on Smart Antennas}, 2021, pp. 1--6.

\bibitem{1683918}
H.~{Weingarten}, Y.~{Steinberg}, and S.~S. {Shamai}, ``The capacity region of the gaussian multiple-input multiple-output broadcast channel,'' \emph{IEEE Transactions on Information Theory}, vol.~52, no.~9, pp. 3936--3964, 2006.

\bibitem{R12}
J.~{Wang} and L.~{Dai}, ``Asymptotic rate analysis of downlink multi-user systems with co-located and distributed antennas,'' \emph{IEEE Transactions on Wireless Communications}, vol.~14, no.~6, pp. 3046--3058, 2015.

\bibitem{R34}
E.~{Nayebi}, A.~{Ashikhmin}, T.~L. {Marzetta}, H.~{Yang}, and B.~D. {Rao}, ``Precoding and power optimization in cell-free massive {MIMO} systems,'' \emph{IEEE Transactions on Wireless Communications}, vol.~16, no.~7, pp. 4445--4459, 2017.

\bibitem{8579566}
C.~Pan, H.~Ren, M.~Elkashlan, A.~Nallanathan, and L.~Hanzo, ``Robust beamforming design for ultra-dense user-centric {C-RAN} in the face of realistic pilot contamination and limited feedback,'' \emph{IEEE Transactions on Wireless Communications}, vol.~18, no.~2, pp. 780--795, 2019.

\bibitem{8660693}
J.~Kim, S.-H. Park, O.~Simeone, I.~Lee, and S.~Shamai~Shitz, ``Joint design of fronthauling and hybrid beamforming for downlink {C-RAN} systems,'' \emph{IEEE Transactions on Communications}, vol.~67, no.~6, pp. 4423--4434, 2019.

\bibitem{8599043}
M.~{Attarifar}, A.~{Abbasfar}, and A.~{Lozano}, ``Modified conjugate beamforming for cell-free massive {MIMO},'' \emph{IEEE Wireless Communications Letters}, vol.~8, no.~2, pp. 616--619, 2019.

\bibitem{9340389}
G.~Interdonato, H.~Q. Ngo, and E.~G. Larsson, ``Enhanced normalized conjugate beamforming for cell-free massive {MIMO},'' \emph{IEEE Transactions on Communications}, vol.~69, no.~5, pp. 2863--2877, 2021.

\bibitem{9108198}
A.~Zhou, J.~Wu, E.~G. Larsson, and P.~Fan, ``Max-min optimal beamforming for cell-free massive {MIMO},'' \emph{IEEE Communications Letters}, vol.~24, no.~10, pp. 2344--2348, 2020.

\bibitem{atzeni2020distributed}
I.~Atzeni, B.~Gouda, and A.~Tölli, ``Distributed precoding design via over-the-air signaling for cell-free massive {MIMO},'' \emph{IEEE Transactions on Wireless Communications}, vol.~20, no.~2, pp. 1201--1216, 2021.

\bibitem{alouiniR26}
E.~{Basar}, M.~{Di Renzo}, J.~{De Rosny}, M.~{Debbah}, M.~{Alouini}, and R.~{Zhang}, ``Wireless communications through reconfigurable intelligent surfaces,'' \emph{IEEE Access}, vol.~7, pp. 116\,753--116\,773, 2019.

\bibitem{9756313}
M.~Obeed and A.~Chaaban, ``Joint beamforming design for multiuser {MISO} downlink aided by a reconfigurable intelligent surface and a relay,'' \emph{IEEE Transactions on Wireless Communications}, vol.~21, no.~10, pp. 8216--8229, 2022.

\bibitem{10431714}
B.~Al-Nahhas, M.~Obeed, A.~Chaaban, and M.~J. Hossain, ``Performance of multi-{RIS}-aided cell-free massive {MIMO}: {D}o more {RISs} always help?'' \emph{IEEE Transactions on Communications}, pp. 1--1, 2024.

\bibitem{9473521}
------, ``{RIS}-aided cell-free massive {MIMO}: {P}erformance analysis and competitiveness,'' in \emph{2021 IEEE International Conference on Communications Workshops (ICC Workshops)}, 2021, pp. 1--6.

\bibitem{9298890}
Y.~Zheng, S.~Bi, Y.-J.~A. Zhang, X.~Lin, and H.~Wang, ``Joint beamforming and power control for throughput maximization in {IRS}-assisted {MISO WPCNs},'' \emph{IEEE Internet of Things Journal}, vol.~8, no.~10, pp. 8399--8410, 2021.

\bibitem{9352948}
Y.~Zhang, B.~Di, H.~Zhang, J.~Lin, C.~Xu, D.~Zhang, Y.~Li, and L.~Song, ``Beyond cell-free {MIMO}: {E}nergy efficient reconfigurable intelligent surface aided cell-free {MIMO} communications,'' \emph{IEEE Transactions on Cognitive Communications and Networking}, vol.~7, no.~2, pp. 412--426, 2021.

\bibitem{9448858}
K.~Liu and Z.~Zhang, ``On the energy-efficiency fairness of reconfigurable intelligent surface-aided cell-free network,'' in \emph{2021 IEEE 93rd Vehicular Technology Conference (VTC2021-Spring)}, 2021, pp. 1--6.

\bibitem{9665300}
T.~Van~Chien, H.~Q. Ngo, S.~Chatzinotas, M.~Di~Renzo, and B.~Ottersten, ``Reconfigurable intelligent surface-assisted cell-free massive {MIMO} systems over spatially-correlated channels,'' \emph{IEEE Transactions on Wireless Communications}, vol.~21, no.~7, pp. 5106--5128, 2022.

\bibitem{9911685}
W.~Hao, J.~Li, G.~Sun, M.~Zeng, and O.~A. Dobre, ``Securing reconfigurable intelligent surface-aided cell-free networks,'' \emph{IEEE Transactions on Information Forensics and Security}, vol.~17, pp. 3720--3733, 2022.

\bibitem{10640072}
Z.~Sui, H.~Q. Ngo, T.~V. Chien, M.~Matthaiou, and L.~Hanzo, ``{RIS}-assisted cell-free massive {MIMO} relying on reflection pattern modulation,'' \emph{IEEE Transactions on Communications}, pp. 1--1, 2024.

\bibitem{8723525}
M.~Cui, G.~Zhang, and R.~Zhang, ``Secure wireless communication via intelligent reflecting surface,'' \emph{IEEE Wireless Communications Letters}, vol.~8, no.~5, pp. 1410--1414, 2019.

\bibitem{9206080}
H.~Yang, Z.~Xiong, J.~Zhao, D.~Niyato, L.~Xiao, and Q.~Wu, ``Deep reinforcement learning-based intelligent reflecting surface for secure wireless communications,'' \emph{IEEE Transactions on Wireless Communications}, vol.~20, no.~1, pp. 375--388, 2021.

\bibitem{alabiad2023effectivenessreconfigurableintelligentsurfaces}
M.~Saif, M.~Javad-Kalbasi, and S.~Valaee, ``Effectiveness of reconfigurable intelligent surfaces to enhance connectivity in {UAV} networks,'' \emph{IEEE Transactions on Wireless Communications}, pp. 1--1, 2024.

\bibitem{10437661}
M.~S. Al-Abiad, M.~Javad-Kalbasi, and S.~Valaee, ``Maximizing network connectivity for {UAV} communications via reconfigurable intelligent surfaces,'' in \emph{GLOBECOM 2023 - 2023 IEEE Global Communications Conference}, 2023, pp. 6395--6400.

\bibitem{10437618}
M.~Javad-Kalbasi, M.~S. Al-Abiad, and S.~Valaee, ``Energy efficient communications in {RIS}-assisted {UAV} networks based on genetic algorithm,'' in \emph{GLOBECOM 2023 - 2023 IEEE Global Communications Conference}, 2023, pp. 5901--5906.

\bibitem{8113550}
R.~Jiang, Q.~Wang, H.~Haas, and Z.~Wang, ``Joint user association and power allocation for cell-free visible light communication networks,'' \emph{IEEE Journal on Selected Areas in Communications}, vol.~36, no.~1, pp. 136--148, 2018.

\bibitem{8779676}
J.~Chen, Z.~Wang, and R.~Jiang, ``Downlink interference management in cell-free {VLC} network,'' \emph{IEEE Transactions on Vehicular Technology}, vol.~68, no.~9, pp. 9007--9017, 2019.

\bibitem{8574984}
M.~Obeed, A.~M. Salhab, S.~A. Zummo, and M.-S. Alouini, ``New algorithms for energy-efficient {VLC} networks with user-centric cell formation,'' \emph{IEEE Transactions on Green Communications and Networking}, vol.~3, no.~1, pp. 108--121, 2019.

\bibitem{liu2022joint}
H.~Liu, X.~Gong, M.~Huang, Y.~Chen, X.~Yuan, K.~Chen, and S.~Yang, ``Joint {AP} grouping and user clustering for interference management in cell-free {VLC} network,'' \emph{Optics \& Laser Technology}, vol. 156, p. 108465, 2022.

\bibitem{almehdhar2022hybrid}
A.~Almehdhar, M.~Obeed, A.~Chaaban, and S.~Zummo, ``A hybrid {VLC/RF} cell-free massive {MIMO} system,'' in \emph{ICC 2022-IEEE International Conference on Communications}.\hskip 1em plus 0.5em minus 0.4em\relax IEEE, 2022, pp. 1871--1876.

\bibitem{almehdhar2022user}
A.~Almehdhar, M.~Obeed, A.~Chaaban, and S.~A. Zummo, ``User association in user-centric hybrid {VLC/RF} cell-free massive {MIMO} systems,'' \emph{Arabian Journal for Science and Engineering}, pp. 1--15, 2024.

\bibitem{giordani2020non}
M.~Giordani and M.~Zorzi, ``Non-terrestrial networks in the {6G} era: {C}hallenges and opportunities,'' \emph{IEEE Network}, vol.~35, no.~2, pp. 244--251, 2020.

\bibitem{riera2022scalable}
F.~Riera-Palou, G.~Femenias, M.~Caus, M.~Shaat, and A.~I. P{\'e}rez-Neira, ``Scalable cell-free massive {MIMO} networks with {LEO} satellite support,'' \emph{IEEE Access}, vol.~10, pp. 37\,557--37\,571, 2022.

\bibitem{abdelsadek2023broadband}
M.~Y. Abdelsadek, G.~Karabulut-Kurt, H.~Yanikomeroglu, P.~Hu, G.~Lamontagne, and K.~Ahmed, ``Broadband connectivity for handheld devices via {LEO} satellites: {I}s distributed massive {MIMO} the answer?'' \emph{IEEE Open Journal of the Communications Society}, 2023.

\bibitem{abdelsadek2021future}
M.~Y. Abdelsadek, H.~Yanikomeroglu, and G.~K. Kurt, ``Future ultra-dense {LEO} satellite networks: {A} cell-free massive {MIMO} approach,'' in \emph{2021 IEEE International Conference on Communications Workshops (ICC Workshops)}.\hskip 1em plus 0.5em minus 0.4em\relax IEEE, 2021, pp. 1--6.

\bibitem{10622597}
J.~Yu, C.~Hua, L.~Liu, and P.~Gu, ``Joint beamforming optimization for user-centric multi-satellite system,'' in \emph{ICC 2024 - IEEE International Conference on Communications}, 2024, pp. 1861--1866.

\bibitem{10634184}
T.-N. Tran, H.~Yu, and T.~Kim, ``Strategies for optimizing uplink spectrum efficiency in cell-free massive {MIMO} satellite-{UAV} network,'' \emph{IEEE Internet of Things Journal}, pp. 1--1, 2024.

\bibitem{RLNC_1}
A.~Eryilmaz, A.~Ozdaglar, M.~Medard, and E.~Ahmed, ``On the delay and throughput gains of coding in unreliable networks,'' \emph{IEEE Transactions on Information Theory}, vol.~54, no.~12, pp. 5511--5524, 2008.

\bibitem{RLNC_2}
S.-Y.~R. Li, Q.~T. Sun, and Z.~Shao, ``Linear network coding: {T}heory and algorithms,'' \emph{Proceedings of the IEEE}, vol.~99, no.~3, pp. 372--387, 2011.

\bibitem{RLNC_3}
T.~Ho, M.~Medard, R.~Koetter, D.~Karger, M.~Effros, J.~Shi, and B.~Leong, ``A random linear network coding approach to multicast,'' \emph{IEEE Transactions on Information Theory}, vol.~52, no.~10, pp. 4413--4430, 2006.

\bibitem{RLNC_4}
D.~E. Lucani, M.~Medard, and M.~Stojanovic, ``Broadcasting in time-division duplexing: {A} random linear network coding approach,'' in \emph{2009 Workshop on Network Coding, Theory, and Applications}, 2009, pp. 62--67.

\bibitem{9036050}
Y.~Li, J.~Zhou, J.~Wang, Z.~Bao, T.~Q.~S. Quek, and J.~Wang, ``On data dissemination enhanced by network coded device-to-device communications,'' \emph{IEEE Transactions on Wireless Communications}, vol.~19, no.~6, pp. 3963--3976, 2020.

\bibitem{6874566}
Z.~Dong, S.~H. Dau, C.~Yuen, Y.~Gu, and X.~Wang, ``Delay minimization for relay-based cooperative data exchange with network coding,'' \emph{IEEE/ACM Transactions on Networking}, vol.~23, no.~6, pp. 1890--1902, 2015.

\bibitem{9982443}
Z.~Mei, ``Average packet decoding delay minimization for rate-aware buffered instantly decodable network codes,'' \emph{IEEE Communications Letters}, vol.~27, no.~2, pp. 409--413, 2023.

\bibitem{10229924}
------, ``Delay reduction for instantly decodable network codes with lossy feedback channels,'' \emph{IEEE Transactions on Communications}, vol.~71, no.~12, pp. 6821--6833, 2023.

\bibitem{5683677}
S.~Sorour and S.~Valaee, ``Minimum broadcast decoding delay for generalized instantly decodable network coding,'' in \emph{2010 IEEE Global Telecommunications Conference GLOBECOM 2010}, 2010, pp. 1--5.

\bibitem{9391701}
M.~S. Al-Abiad, M.~J. Hossain, and A.~Douik, ``Low-complexity scheduling for delay minimization in {D2D} communications using network coding,'' \emph{IEEE Communications Letters}, vol.~25, no.~7, pp. 2430--2434, 2021.

\bibitem{10.5555/574848}
M.~R. Garey and D.~S. Johnson, \emph{Computers and Intractability; A Guide to the Theory of {NP}-Completeness}.\hskip 1em plus 0.5em minus 0.4em\relax USA: W. H. Freeman and Co., 1990.

\bibitem{9044331}
A.~Douik, H.~Dahrouj, T.~Y. Al-Naffouri, and M.-S. Alouini, ``A tutorial on clique problems in communications and signal processing,'' \emph{Proceedings of the IEEE}, vol. 108, no.~4, pp. 583--608, 2020.

\bibitem{7864466}
M.~S. Karim, A.~Douik, and S.~Sorour, ``Rate-aware network codes for video distortion reduction in point-to-multipoint networks,'' \emph{IEEE Transactions on Vehicular Technology}, vol.~66, no.~8, pp. 7446--7460, 2017.

\bibitem{5072351}
Y.~Kim and G.~De~Veciana, ``Is rate adaptation beneficial for inter-session network coding?'' \emph{IEEE Journal on Selected Areas in Communications}, vol.~27, no.~5, pp. 635--646, 2009.

\bibitem{5424026}
K.~Chi, X.~Jiang, and S.~Horiguchi, ``Joint design of network coding and transmission rate selection for multihop wireless networks,'' \emph{IEEE Transactions on Vehicular Technology}, vol.~59, no.~5, pp. 2435--2444, 2010.

\bibitem{9519548}
M.~S. Al-Abiad, M.~Z. Hassan, and M.~J. Hossain, ``Throughput maximization of network-coded and multi-level cache-enabled heterogeneous network,'' \emph{IEEE Transactions on Vehicular Technology}, vol.~70, no.~10, pp. 11\,039--11\,043, 2021.

\bibitem{CLNC5a}
------, ``A joint reinforcement-learning enabled caching and cross-layer network code in {F-RAN} with {D2D} communications,'' \emph{IEEE Transactions on Communications}, vol.~70, no.~7, pp. 4400--4416, 2022.

\bibitem{9791405}
------, ``Cross-layer network codes for completion time minimization in device-to-device networks,'' \emph{IEEE Access}, vol.~10, pp. 61\,567--61\,584, 2022.

\bibitem{8027131}
A.~Douik, H.~Dahrouj, T.~Y. Al-Naffouri, and M.-S. Alouini, ``Distributed hybrid scheduling in multi-cloud networks using conflict graphs,'' \emph{IEEE Transactions on Communications}, vol.~66, no.~1, pp. 209--224, 2018.

\bibitem{7342981}
------, ``Coordinated scheduling and power control in cloud-radio access networks,'' \emph{IEEE Transactions on Wireless Communications}, vol.~15, no.~4, pp. 2523--2536, 2016.

\bibitem{6811617}
H.~Dahrouj, W.~Yu, T.~Tang, J.~Chow, and R.~Selea, ``Coordinated scheduling for wireless backhaul networks with soft frequency reuse,'' in \emph{21st European Signal Processing Conference (EUSIPCO 2013)}, 2013, pp. 1--5.

\bibitem{10550002}
M.~Saif, M.~Z. Hassan, and M.~J. Hossain, ``Decentralized aggregation for energy-efficient federated learning in mm{W}ave aerial-terrestrial integrated networks,'' \emph{IEEE Transactions on Machine Learning in Communications and Networking}, pp. 1--1, 2024.

\bibitem{9844152}
M.~S. Al-Abiad, M.~Z. Hassan, and M.~J. Hossain, ``Energy-efficient resource allocation for federated learning in noma-enabled and relay-assisted internet of things networks,'' \emph{IEEE Internet of Things Journal}, vol.~9, no.~24, pp. 24\,736--24\,753, 2022.

\bibitem{9932020}
M.~S. Al-Abiad and M.~J. Hossain, ``Coordinated scheduling and decentralized federated learning using conflict clustering graphs in fog-assisted {IoD} networks,'' \emph{IEEE Transactions on Vehicular Technology}, vol.~72, no.~3, pp. 3455--3472, 2023.

\bibitem{10061474}
M.~S. Al-Abiad, M.~Obeed, M.~J. Hossain, and A.~Chaaban, ``Decentralized aggregation for energy-efficient federated learning via {D2D} communications,'' \emph{IEEE Transactions on Communications}, vol.~71, no.~6, pp. 3333--3351, 2023.

\bibitem{10199618}
M.~S. Al-Abiad and M.~J. Hossain, ``Minimizing energy consumption for decentralized federated learning using {D2D} communications,'' in \emph{2023 IEEE 97th Vehicular Technology Conference (VTC2023-Spring)}, 2023, pp. 1--6.

\bibitem{7248768}
A.~Douik, H.~Dahrouj, T.~Y. Al-Naffouri, and M.-S. Alouini, ``Coordinated scheduling for the downlink of cloud radio-access networks,'' in \emph{2015 IEEE International Conference on Communications (ICC)}, 2015, pp. 2906--2911.

\bibitem{6809217}
S.~Sorour, A.~Douik, S.~Valaee, T.~Y. Al-Naffouri, and M.-S. Alouini, ``Partially blind instantly decodable network codes for lossy feedback environment,'' \emph{IEEE Transactions on Wireless Communications}, vol.~13, no.~9, pp. 4871--4883, 2014.

\bibitem{9610113}
X.~Zhu and C.~Jiang, ``Integrated satellite-terrestrial networks toward {6G}: {A}rchitectures, applications, and challenges,'' \emph{IEEE Internet of Things Journal}, vol.~9, no.~1, pp. 437--461, 2022.

\bibitem{8255832}
A.~Arefi, M.~Khabbazian, M.~Ardakani, and G.~Bansal, ``Blind instantly decodable network codes for wireless broadcast of real-time multimedia,'' \emph{IEEE Transactions on Wireless Communications}, vol.~17, no.~4, pp. 2276--2288, 2018.

\end{thebibliography}

\end{document}